\documentclass[
aps,
prd,
amsmath,amssymb,
notitlepage,
reprint,
11pt,
onecolumn,
superscriptaddress,
floatfix,
nofootinbib,
a4paper
]{revtex4-2}

\usepackage{physics}
\usepackage[margin=2cm]{geometry}

\usepackage[english]{babel}
\usepackage[utf8]{luainputenc}
\usepackage[T1]{fontenc}
\usepackage{mathrsfs}
\usepackage{bbm}
\usepackage{bm}
\usepackage{enumerate}
\usepackage{graphicx}
\usepackage[dvipsnames]{xcolor}
\usepackage{float}
\usepackage{subdepth}
\usepackage{fnpct}

\usepackage{lipsum}
\usepackage{comment}

\usepackage{tikz}
\usepackage[compat=1.1.0]{tikz-feynman} 

\usepackage[colorlinks,
linkcolor=BrickRed,
citecolor=MidnightBlue,
urlcolor=MidnightBlue,
bookmarks=true,        
bookmarksopen=true,    
bookmarksnumbered=true,
]{hyperref}

\usepackage{setspace}
\usepackage[raggedright,small]{titlesec}

\renewcommand\thesubsection{\arabic{section}.\arabic{subsection}}

\renewcommand*\d{\mathop{}\!\mathrm{d}}

\newcommand{\av}[1]{\left\langle#1\right\rangle}

\renewcommand{\vec}[1]{\underline{#1}}
\renewcommand{\i}{\mathrm{i}}
\renewcommand{\(}{\left(}
\renewcommand{\)}{\right)}
\newcommand{\id}{\mathbbm{1}}

\renewcommand{\va}{{\vec{a}}}
\renewcommand{\vb}{{\vec{b}}}
\newcommand{\vc}{{\vec{c}}}
\newcommand{\vd}{{\vec{d}}}
\newcommand{\vast}{{\vec{a}^{\raisebox{-2.5pt}{$\scriptstyle\prime$}}}}
\newcommand{\vbst}{{\vec{b}^{\raisebox{-2.5pt}{$\scriptstyle\prime$}}}}
\newcommand{\must}{{\mu^{\raisebox{-2.5pt}{$\scriptstyle\prime$}}}}
\newcommand{\qb}{{\hat{q}}}
\newcommand{\sM}{\mathcal{M}}
\newcommand{\sS}{\mathcal{S}}
\newcommand{\sigmaH}{\sigma_\mathrm{H}}
\newcommand{\sigmaL}{\sigma_\mathrm{L}}
\newcommand{\EH}{E_{0\mathrm{H}}}
\newcommand{\EL}{E_{0\mathrm{L}}}

\allowdisplaybreaks

\begin{document}

\title{$Q$-Laguerre spectral density and quantum chaos in the Wishart-Sachdev-Ye-Kitaev model}

\author{Lucas S\'a}
\email{lucas.seara.sa@tecnico.ulisboa.pt}
\affiliation{CeFEMA, Instituto Superior T\'ecnico, Universidade de Lisboa, Av.\ Rovisco Pais, 1049-001 Lisboa, Portugal}

\author{Antonio M.\ Garc\'ia-Garc\'ia}
\email{amgg@sjtu.edu.cn}
\affiliation{Shanghai Center for Complex Physics,
	School of Physics and Astronomy, Shanghai Jiao Tong
	University, Shanghai 200240, China}

\begin{abstract}
  We study the Wishart-Sachdev-Ye-Kitaev (WSYK) model consisting of two $\hat{q}$-body Sachdev-Ye-Kitaev (SYK) models with general complex couplings, one the Hermitian conjugate of the other, living in off-diagonal blocks of a larger WSYK Hamiltonian. The spectrum is positive with a hard edge at zero energy.
  We employ diagrammatic and combinatorial techniques to compute analytically the low-order moments of the Hamiltonian. In the limit of large number $N$ of Majoranas, we have found striking similarities with the moments of the weight function of the Al-Salam-Chihara $Q$-Laguerre polynomials.
  For $\hat{q} = 3, 4$, the $Q$-Laguerre prediction, with $Q=Q(\hat{q},N)$ also computed analytically, agrees well with exact diagonalization results for $30 < N \leq 34$ while we observe some deviations for $\hat q =  2$. The most salient feature of the spectral density is that, for odd $\hat{q}$, low-energy excitations grow as a stretched exponential, with a functional form different from that of the supersymmetric SYK model. 
  For $\hat q = 4$, a detailed analysis of level statistics reveals quantum chaotic dynamics even for time scales substantially shorter than the Heisenberg time. More specifically, the spacing ratios in the bulk of the spectrum and the microscopic spectral density and the number variance close to the hard edge are very well approximated by that of an ensemble of random matrices that, depending on $N$, belong to the chiral or superconducting universality classes.  In particular, we report the first realization of level statistics belonging to the chGUE universality class, which completes the tenfold-way classification in the SYK model.
\end{abstract}

\maketitle
\newpage

{
\hypersetup{allcolors=Black}
\singlespacing
\tableofcontents
\addtocontents{toc}{\vspace{-3ex}}
}
	
\section{Introduction}
The Sachdev-Ye-Kitaev (SYK) model~\cite{kitaev2015,sachdev1993,sachdev2015,bohigas1971,bohigas1971a,french1970,french1971}, a model of $N$ Majorana, or Dirac, fermions in zero spatial dimensions with $q$-body infinite range interactions, is attracting a lot of attention in different research fields because, despite being analytically tractable, it can still reveal intriguing features of both strongly correlated systems and, through the use of holographic dualities, quantum gravity in near-AdS$_2$ backgrounds~\cite{almheiri2015,maldacena2016a}. 

An example of this is Kitaev's analytical calculation~\cite{kitaev2015} showing that, in the strong-coupling low-temperature limit, the Lyapunov exponent~\cite{larkin1969} of the SYK model saturates a universal bound on quantum chaos~\cite{maldacena2015}. This is an expected feature~\cite{maldacena2015} of field theories that have a gravity dual. Indeed, in the SYK case, the holographic dual in this infrared limit has been identified as Jackiw-Teitelboim (JT) gravity~\cite{maldacena2016a,almheiri2015,jensen2016}. 

It is also possible to compute analytically the spectral density, and, therefore, the free energy and other thermodynamic properties of the SYK model by using different techniques. In the strong-coupling, low-temperature limit, the density can be obtained by an exact evaluation~\cite{stanford2017,Belokurov:2017eit} of the path integral of the Schwarzian action that captures the spontaneous and explicit breaking of conformal symmetry to SL$(2,R)$ symmetry. This exact density shows an exponential growth close to the ground state, which is the expected result for quantum black holes as well~\cite{carlip2000,maldacena2016}.
The free energy has also been computed numerically, either from the solution of the large-$N$ Schwinger-Dyson equations or by exact diagonalization, and analytically, in the large-$q$, large-$N$ limit, or by the combined use of diagrammatic and combinatorial tools 
 ~\cite{maldacena2016,kitaev2015,cotler2016,garcia2017,garcia2016,jensen2016,jevicki2016,berkooz2018}.

These techniques have also been applied to many generalizations of the SYK model, including supersymmetric SYKs~\cite{li2017,kanazawa2017,garcia2018a,fu2018,gates2021} that can reproduce the tenfold way~\cite{altland1997} of random matrix theory, higher dimensional SYKs~\cite{berkooz2017}, one-body deformations leading to a quantum-chaos transition~\cite{garcia2018b}, and two-site SYKs dual to traversable~\cite{maldacena2018,garcia2019,ferrari2017} and Euclidean~\cite{garcia2020} wormholes. 
However, despite their power and versatility, it was not yet possible to use them to solve the SYK model for all temperatures or coupling strengths. 

At least for the spectral density, such an exact calculation is possible~\cite{erdos2014,feng2019} but only for $q \propto \sqrt{N}$. The procedure involves an explicit evaluation of the moments of the Hamiltonian by diagrammatic techniques. It turns out that the moments are given by a sum over the number of crossings of perfect-matching diagrams. Surprisingly, this sum can be evaluated explicitly and compactly by using the Touchard-Riordan formula~\cite{riordan1975,touchard1952} introduced in the context of combinatorial analysis. 
The resulting expression was later identified~\cite{ismail1987} with that of the moments of the weight function of $Q$-Hermite polynomials, which allows computing the spectral density exactly. For the more physically relevant case of $q=4$, these combinatorial techniques, though only approximate, still provide a quantitatively correct description of the spectral density~\cite{garcia2017,cotler2016,berkooz2018,garcia2018,berkooz2019,jia2020a} with $Q$ a function of $q$ and $N$ that is evaluated analytically. Still based on the role of perfect matchings and the mapping to the the $Q$-Hermite weight function, it has been possible to obtain analytical expressions for supersymmetric~\cite{garcia2018a,berkooz2020a,stanford2017}, sparse~\cite{garcia2020c}, and complex SYK models with a finite chemical potential~\cite{berkooz2020}. Moreover, it has been used to relate~\cite{jia2020} a variant of the SYK model to a $U(1)$ pure gauge theory in a hypercube proposed by Parisi~\cite{parisi1994} as a toy model of spin glasses.
 
In this paper, we introduce a generalized SYK model, termed Wishart~\cite{wishart1928}-Sachdev-Ye-Kitaev (WSYK) model, which can be represented by a matrix with two off-diagonal blocks, one the Hermitian conjugate of the other. In each of these blocks lives an SYK model with $\qb$ Majorana fermions and random complex couplings.

We will show that, unlike previous results, the spectral density in the scaling limit $\qb \propto \sqrt{N}$~\cite{erdos2014,feng2019} cannot be evaluated exactly because the combinatorial analysis is based only on a subset of perfect matchings. However, by identifying this subset as the set of permutations, we can establish an approximate mapping to the weight function of the Al-Salam-Chihara $Q$-Laguerre polynomials~\cite{al1976,ismail2005,koekoek2010,kasraoui2011AAM}, a variant of the $Q$-Laguerre polynomials~\cite{moak1981,ismail2005,koekoek2010}. For a fixed $\qb \geq 2$, the resulting spectral density, which is qualitatively different from the one in the $Q$-Hermite case, is in fair agreement with exact diagonalization results in the range of $N$ we can explore numerically. Moreover, we will identify a range of parameters where the WSYK model shares typical features expected in a field theory with a holographic dual such as an exponential growth of low energy excitations. The global symmetries of this generalized SYK model belong to the chiral and superconducting universality classes according to the random matrix classification~\cite{altland1997,verbaarschot1994}. Level statistics and the microscopic spectral density, which are believed to be universal, show good agreement with the random matrix predictions. Deviations from the random matrix results, for fixed $N$ and $\qb = 4$, are smaller than in the standard {\bf $q = 3$} supersymmetric SYK model~\cite{garcia2018a,kanazawa2017,li2017}. Our work includes the first observation of the chiral Gaussian Unitary Ensemble (chGUE) universality class in the context of the SYK model. Although it is possible to realize complex-fermion SYK models~\cite{kanazawa2017} with global symmetries consistent with those of chGUE, the number of zero modes of those models scales with system size and, thus, in the thermodynamic limit, one obtains GUE correlations.

Finally, we note that Iyoda \textit{et al.}~\cite{iyoda2018} studied a related Wishart extension of the complex-fermion Sachdev-Ye-Kitaev model with $\qb=2$ and found it to be integrable as it can be mapped to the Richardson-Gaudin model, which is known to be solved by the Bethe ansatz. 
 
\section{The Wishart-Sachdev-Ye-Kitaev model}
We start with the definition of the SYK Hamiltonian that describes $q$-body random interactions among $N$ Majorana fermions,
\begin{equation}\label{eq:def_SYK}
H=\i^{q(q-1)/2}\sum_{i_1<\dots<i_q=1}^N 
J_{i_1\cdots i_q}
\gamma_{i_1}\cdots\gamma_{i_q}
\equiv \i^{q(q-1)/2}\sum_{\va} J_\va \Gamma_\va,
\end{equation}
where $N$ and $q$ are even integers, $\va$ is a multi-index accounting for all $q$ indices $i_1,\dots,i_q=1,\dots,N$, $J_{i_1\cdots i_q}\equiv J_\va$ is a totally antisymmetric tensor with independent real Gaussian entries with zero mean, i.e.,
\begin{equation}
\av{J_\va}=0
\quad \text{and} \quad
\av{J_\va J_\vb}=\av{J^2}\delta_{\va,\vb},
\end{equation}
$\gamma_i$ are Majorana fermions satisfying the Clifford algebra $\acomm{\gamma_i}{\gamma_j}=2\delta_{ij}$ and, therefore, represented by $2^{N/2}$-dimensional (Hermitian) Dirac $\gamma$-matrices, and $\Gamma_\va=\gamma_{i_1}\cdots\gamma_{i_q}$ is a product of $q$ $\gamma$-matrices with all indices $i_1,\dots,i_q$ mutually different, satisfying
\begin{equation}\label{eq:relations_Gamma}
\Gamma_\va^2=\i^{q(q-1)}\id
\quad \text{and} \quad
\Gamma_\va\Gamma_\vb=(-1)^{q+\abs{\va\cap\vb}}
\Gamma_\vb\Gamma_\va,
\end{equation}
with $\id$ the $2^{N/2}$-dimensional identity and $\abs{\va\cap\vb}$ the number of $\gamma$-matrices that $\Gamma_\va$ and $\Gamma_\vb$ have in common.

Several restrictions in $H$ can be relaxed. Let us consider a set of $M$ independent \emph{non-Hermitian} $\qb$-body charges $L_\mu$ (with $\qb$ allowed to be odd):
\begin{equation}
L_\mu=\sum_{i_1,\dots,i_\qb=1}^N 
\ell_{\mu,i_1\cdots i_\qb}
\gamma_{i_1}\cdots\gamma_{i_\qb}
\equiv\sum_{\va} \ell_{\mu,\va} \Gamma_\va,
\end{equation}
where the $\ell_{\mu,i_1\cdots i_\qb}\equiv \ell_{\mu,\va}$ are antisymmetric (in $\va$) independent \emph{complex} Gaussian random variables (the exact variances are specified below). From the charges $L_\mu$, we can define a Hermitian and positive-definite Hamiltonian, which we dub the Wishart-Sachdev-Ye-Kitaev (WSYK) Hamiltonian:
\begin{equation}\label{eq:def_WSYK}
W=\sum_{\mu=1}^{M}L_\mu^\dagger L_\mu
=i^{\qb(\qb-1)}\sum_{\va,\vast}
\(\sum_{\mu=1}^{M} \ell_{\mu,\vast}^*\ell_{\mu,\va}\)
\Gamma_\vast \Gamma_\va,
\end{equation}
where the factor $i^{\qb(\qb-1)}$ arises from bringing $\Gamma_\vast^\dagger$ back to $\Gamma_\vast$ by commuting all the (Hermitian) $\gamma$-matrices in $\Gamma_\vast^\dagger$.
Note that while all $\gamma$-matrices in a single $\Gamma_\va$ are different (owing to the antisymmetry of $\ell_{\mu,\va}$), there could be overlaps between the $\gamma$-matrices in $\Gamma_\va$ and those in $\Gamma_\vast$ (which follows from the fact that the tensor $K_{\vast\va}=\sum_{\mu}\ell_{\mu,\vast}^{*}\ell_{\mu,\va}$
is \emph{not} totally antisymmetric). Setting $\qb=q/2$, the positive-definite Hamiltonian~$W$ describes $q$-body interactions like the original Hamiltonian $H$. Note that, for $M=1$, the eigenvalues of $W$ are exactly the squares of the eigenvalues of the off-diagonal block Hamiltonian 
\begin{equation}
\mathcal{H}=\begin{pmatrix}
0 & L \\
L^\dagger & 0
\end{pmatrix},
\end{equation}
which has a natural chiral structure.

We now address the exact distribution of the random couplings $\ell_{\mu,\va}$. We decompose $\ell_{\mu,\va}=\ell^{(1)}_{\mu,\va}+\i k \ell^{(2)}_{\mu,\va}$, $0\leq k\leq 1$. The real, $\ell^{(1)}_{\mu,\va}$, and imaginary, $\ell^{(2)}_{\mu,\va}$, parts of $\ell_{\mu,\va}$ are independent and identically distributed Gaussian random variables with
\begin{equation}
\av{\ell^{(1)}_{\mu,\va}}
=\av{\ell^{(2)}_{\mu,\va}}=0
\quad \text{and} \quad
\av{\ell^{(1)}_{\mu,\va} \ell^{(1)}_{\mu',\va'}}
=\av{\ell^{(2)}_{\mu,\va} \ell^{(2)}_{\mu',\va'}}
=\frac{1}{2}\av{\ell^2}\delta_{\mu,\mu'}\delta_{\va,\va'}.
\end{equation}
It then follows that $\av{\ell_{\mu,\va}}=\av{\ell_{\mu,\va}^*}=0$,
\begin{equation}\label{eq:variances_ell}
\av{\ell_{\must,\vast}^*\ell_{\mu,\va}}=
\frac{1+k^2}{2}\av{\ell^2}\delta_{\must\!,\mu}\delta_{\vast\!,\va},
\quad \text{and} \quad
\av{\ell_{\must,\vast}\ell_{\mu,\va}}=
\av{\ell_{\must,\vast}^*\ell_{\mu,\va}^*}=
\frac{1-k^2}{2}\av{\ell^2}\delta_{\must\!,\mu}\delta_{\vast\!,\va}.
\end{equation}
We name the three special cases $k=0$, $k=1$, and $0<k<1$ the linear, circular, and elliptic WSYK models, respectively (borrowing the nomenclature of standard random matrix theory, after the shapes of the support of the random variables $\ell_{\mu,\va}$ seen as random matrices).

Note that only for the $M=1$ linear WSYK model the eigenvalues of $W$ are the squares of the eigenvalues of $L$. When the charge is non-Hermitian, there is no relation between the eigenvalues of $L$ and its singular values (the eigenvalues of $W=L^\dagger L$), apart from some general inequalities. The same is true if there are $M>1$ independent (noncommuting) charges, irrespective of their Hermiticity. As a result, to compute the correct spectral density in those cases, one cannot rely on the known combinatorial expansion of the standard SYK moments (see Appendix~\ref{app:review_standard_SYK} for a review). The WSYK moments have to be computed afresh and a new combinatorial interpretation has to be given.

In this paper, we focus on the circular WSYK model. The $M=1$ linear WSYK model coincides with the $\mathcal{N}=1$ supersymmetric SYK model and is reviewed in Appendix~\ref{app:linear_WSYK}\footnote{Only in the linear model do the operators $L_\mu$ have the interpretation of a supercharge generating supersymmetry. For the other cases, we still use the nomenclature \emph{charge} to emphasize their role as the building block of the positive-definite Hamiltonian $W$.}. We compute the spectral density of the $M=1$ circular WSYK model in Secs.~\ref{sec:moments_WSYK}--\ref{sec:spectral_density_numerics} (with simple asymptotic formulas for different regimes derived in Appendix~\ref{app:asymptotics}) and also study its microscopic spectral density and level statistics in Sec.~\ref{sec:speden}. We propose an ansatz for the spectral density of the $M>1$ circular WSYK model in Appendix~\ref{app:M>1}. The elliptic WSYK model is more complicated so we leave it for future work.

\section{Moments of the circular WSYK model}
\label{sec:moments_WSYK}

Our aim is to derive the spectral density of the circular WSYK model using the method of moments. Let us briefly mention how this computation is carried out for the standard SYK model~\cite{garcia2016,garcia2017}, referring the reader to Appendix~\ref{app:review_standard_SYK} for a detailed review. One starts by an exact computation of the first few moments, which can be given a combinatorial interpretation in terms of perfect matchings. Arbitrarily high moments cannot be computed exactly in general, but one can find approximations to varying degrees of accuracy and exact results when $q\propto \sqrt{N}$. In particular, with this scaling, the moments can be expressed as a sum over the number of crossings of perfect matchings and, hence, identified with the moments of the weight function of the $Q$-Hermite polynomials~\cite{ismail1987}.

Let us apply this program to the circular WSYK model. Because the random variables $\ell_{\mu,\va}$ are Gaussian, the moments $\av{\Tr W^{p}}$ are evaluated by Wick contraction, i.e., by summing over all possible pair contractions of the indices $\va$, $\vast$, $\vb$, $\vbst$, etc. In contrast to the standard SYK model, the odd moments are nonvanishing and, more importantly, the non-Hermitian couplings suppress certain Wick contractions. For example, by performing all allowed contractions, the second moment of $W$ is explicitly found to be
\begin{equation}\label{eq:W_moment2_example}
\begin{split}
\av{\Tr W^2}&
=\sum_{\mu,\nu}\sum_{\va,\vb,\vc,\vd}
\av{\ell_{\mu,\va}^*\ell_{\mu,\vb}
	\ell_{\nu,\vc}^*\ell_{\nu,\vd}}
\Tr\(\Gamma_\va\Gamma_\vb\Gamma_\vc\Gamma_\vd\)\\
&=\av{\ell^2}^2\sum_{\mu,\nu}\sum_{\va,\vb}
\left[
\delta_{\mu\mu}\delta_{\nu\nu}
\Tr\(\Gamma_{\va}\Gamma_{\va}\Gamma_{\vb}\Gamma_{\vb}\)
+0\cdot
\Tr\(\Gamma_{\va}\Gamma_{\vb}\Gamma_{\va}\Gamma_{\vb}\)
+\delta_{\mu\nu}\delta_{\nu\mu}
\Tr\(\Gamma_{\va}\Gamma_{\vb}\Gamma_{\vb}\Gamma_{\va}\)
\right]\\
&=2^{N/2}\(\av{\ell^2}\binom{N}{\qb}\)^2\(M^2+M\),
\end{split}
\end{equation}
where the second term in the Wick expansion is identically zero because of Eq.~(\ref{eq:variances_ell}) with $k=1$ and we used Eq.~(\ref{eq:relations_Gamma}) to evaluate the traces.

As for the standard SYK model, we can introduce a combinatorial diagrammatic notation to simplify the representation and evaluation of the moments. First, we note that to avoid dealing with allowed and forbidden perfect-matching diagrams we can switch to an expansion in terms of permutation diagrams~\cite{corteel2007AAM,kasraoui2011AAM,josuat-verges2011DM,corteel2016BOOKCH,zeng2020BOOKCH}\footnote{It was noted in Ref.~\cite{berkooz2020a} that the moments $\mathcal{N}=2$ supersymmetric SYK model also admit an expansion in terms of permutations, but the connection to the $Q$-Laguerre polynomials was not made.}. Indeed, at $p$th order, the set of $\Gamma$-matrices in the trace $\Tr\(\Gamma_{\va_1}\Gamma_{\va_2}\cdots \Gamma_{\va_{2p}}\)$ is naturally bipartite: half of the $\Gamma$-matrices (more precisely, the even-numbered ones, $\{\Gamma_{\va_{2j}}\}$, $j=1,\dots,p$) come from insertions of matrices $L_\mu$, the other half ($\{\Gamma_{\va_{2j-1}}\}$) from insertions of $L_\mu^\dagger$, two odd-numbered (or two even-numbered) $\Gamma$-matrices cannot be coupled (according to Eq.~(\ref{eq:variances_ell}) with $k=1$), and each odd-numbered $\Gamma$-matrix is coupled to one and only one even-numbered one. We can, therefore, identify each Wick contraction with a permutation $\sigma\in\mathcal S_{p}$ such that if $\sigma(j)=k$, with $j,k=1,\dots,p$, then $\va_{2j-1}$ is contracted with $\va_{2k}$, leading to a factor $\langle\ell_{\mu,\va_{2j-1}}^*\ell_{\mu,\va_{2k}}\rangle$. The permutation diagram corresponding to $\sigma\in\mathcal{S}_p$ is then given by a set of $p$ dots, labeled $j=1,\dots,p$, with edges connecting all pairs of dots $(j,\sigma(j))$; if $j\leq\sigma(j)$ the edge is drawn above the dots, if $j>\sigma(j)$ it is drawn below. For example, the two allowed contractions in Eq.~(\ref{eq:W_moment2_example}) are represented diagrammatically as follows (we omit the labeling of the dots):
\begingroup
\allowdisplaybreaks
\begin{align}
\label{eq:W_moment2_diagram1}
\begin{tikzpicture}[baseline=(b)]
\begin{feynman}[inline=(a)]
\vertex[dot] (a) {};
\vertex[dot, right=0.6cm of a] (b) {};
\vertex[right=0.3cm of a] (x) {};
\vertex[below=0.55cm of x] {};
\end{feynman}
\draw [/tikzfeynman] (a) to[out=135, in=45, loop, min distance=0.9cm] (a);
\draw [/tikzfeynman] (b) to[out=135, in=45, loop, min distance=0.9cm] (b);
\end{tikzpicture}
\hspace{-0.4cm}
&=2^{-N/2}\binom{N}{\qb}^{-1}
\sum_{\mu,\nu}\sum_{\va,\vb}
\delta_{\mu\mu}\delta_{\nu\nu}
\Tr\(\Gamma_{\va}\Gamma_{\va}\Gamma_{\vb}\Gamma_{\vb}\)
\intertext{and}
\label{eq:W_moment2_diagram2}
\begin{tikzpicture}[inner sep=2pt,baseline=(b)]
\begin{feynman}[inline=(a)]
\vertex[dot] (a) {};
\vertex[dot, right=0.6cm of a] (b) {};
\diagram*{(a) --[half left] (b)};
\diagram*{(b) --[half left] (a)};
\vertex[right=0.3cm of a] (x) {};
\vertex[below=0.6cm of x] {};
\end{feynman}
\end{tikzpicture}
\hspace{+0.15cm}
&=2^{-N/2}\binom{N}{\qb}^{-1}
\sum_{\mu,\nu}\sum_{\va,\vb}
\delta_{\mu\nu}\delta_{\nu\mu}
\Tr\(\Gamma_{\va}\Gamma_{\vb}\Gamma_{\vb}\Gamma_{\va}\).
\end{align}
\endgroup
Note that each dot comes also equipped with an index $\mu_j$ related to which charge $L_\mu$ it belongs. Each closed loop in the diagram then contributes with a factor $M$. Equations~(\ref{eq:W_moment2_diagram1}) and (\ref{eq:W_moment2_diagram2}) thus give a factor of $M^2$ and $M$, respectively, in agreement with Eq.~(\ref{eq:W_moment2_example}).

In full generality, the (normalized) moments of the circular WSYK model can be written as a sum over permutations $\sigma\in\mathcal{S}_p$,
\begin{equation}\label{eq:moments_W_sum_permutations}
\frac{1}{\sigmaL^p}\frac{\av{\Tr W^p}}{\Tr \id}
=\sum_{\sigma\in\mathcal{S}_p}
t(\sigma)M^{\mathrm{cyc}(\sigma)},
\end{equation}
where $\sigmaL=\av{\ell^2}\binom{N}{\qb}$ is the energy scale of the WSYK model, $\Tr\id=2^{N/2}$, the weight~$t(\sigma)$ gives the normalized trace of each diagram~$\sigma$, and $\mathrm{cyc}(\sigma)$ is the number of cycles in the permutation $\sigma$ (number of closed loops in the respective diagram).

For example, the first four moments of $W$ are easily written down and organized in terms of permutation diagrams:
\begingroup
\allowdisplaybreaks
\begin{align}
\av{\Tr W^0}&=2^{N/2},
\\
\av{\Tr W^1}&=2^{N/2}\av{\ell^2} \binom{N}{\qb}
\begin{tikzpicture}[baseline=(a)]
\begin{feynman}[inline=(a)]
\vertex[dot] (a) {};
\vertex[below=0.35cm of a] {\footnotesize{$1\times M$}};
\end{feynman}
\draw [/tikzfeynman] (a) to[out=135, in=45, loop, min distance=0.9cm] (a);
\end{tikzpicture}
=2^{N/2}\av{\ell^2} \binom{N}{\qb}\, M,
\\
\label{eq:W_moment2}
\av{\Tr W^2}&=2^{N/2}\(\av{\ell^2} \binom{N}{\qb}\)^2
\(
\hspace{-0.3cm}
\begin{tikzpicture}[baseline=(b)]
\begin{feynman}[inline=(a)]
\vertex[dot] (a) {};
\vertex[dot, right=0.6cm of a] (b) {};
\vertex[right=0.3cm of a] (x) {};
\vertex[below=0.55cm of x] {\footnotesize{$1\times M^2$}};
\end{feynman}
\draw [/tikzfeynman] (a) to[out=135, in=45, loop, min distance=0.9cm] (a);
\draw [/tikzfeynman] (b) to[out=135, in=45, loop, min distance=0.9cm] (b);
\end{tikzpicture}
\hspace{-0.3cm}
+
\hspace{+0.15cm}
\begin{tikzpicture}[inner sep=2pt,baseline=(b)]
\begin{feynman}[inline=(a)]
\vertex[dot] (a) {};
\vertex[dot, right=0.6cm of a] (b) {};
\diagram*{(a) --[half left] (b)};
\diagram*{(b) --[half left] (a)};
\vertex[right=0.3cm of a] (x) {};
\vertex[below=0.6cm of x] {\footnotesize{$1\times M$}};
\end{feynman}
\end{tikzpicture}
\hspace{+0.15cm}
\)
=2^{N/2}\(\av{\ell^2} \binom{N}{\qb}\)^2
\(M^2+M\),
\\
\begin{split}
\label{eq:W_moment3}
\av{\Tr W^3}&=2^{N/2}\(\av{\ell^2} \binom{N}{\qb}\)^3\\
\times&\left(
\hspace{-0.3cm}
\begin{tikzpicture}[baseline=(b)]
\begin{feynman}[inline=(a)]
\vertex[dot] (a) {};
\vertex[dot, right=0.6cm of a] (b) {};
\vertex[dot, right=0.6cm of b] (c) {};
\vertex[below=0.635cm of b] {\footnotesize{$1\times M^3$}};
\end{feynman}
\draw [/tikzfeynman] (a) to[out=135, in=45, loop, min distance=0.9cm] (a);
\draw [/tikzfeynman] (b) to[out=135, in=45, loop, min distance=0.9cm] (b);
\draw [/tikzfeynman] (c) to[out=135, in=45, loop, min distance=0.9cm] (c);
\end{tikzpicture}
\hspace{-0.3cm}
+
\hspace{-0.3cm}
\begin{tikzpicture}[inner sep=2pt,baseline=(b)]
\begin{feynman}[inline=(a)]
\vertex[dot] (a) {};
\vertex[dot, right=0.6cm of a] (b) {};
\vertex[dot, right=0.6cm of b] (c) {};
\diagram*{(b) --[half left] (c)};
\diagram*{(c) --[half left] (b)};
\vertex[below=0.65cm of b] {\footnotesize{$1\times M^2$}};
\end{feynman}
\draw [/tikzfeynman] (a) to[out=135, in=45, loop, min distance=0.9cm] (a);
\end{tikzpicture}
\hspace{+0.15cm}
+
\hspace{+0.15cm}
\begin{tikzpicture}[inner sep=2pt,baseline=(b)]
\begin{feynman}[inline=(a)]
\vertex[dot] (a) {};
\vertex[dot, right=0.6cm of a] (b) {};
\vertex[dot, right=0.6cm of b] (c) {};
\diagram*{(a) --[half left] (b)};
\diagram*{(b) --[half left] (a)};
\vertex[below=0.65cm of b] {\footnotesize{$1\times M^2$}};
\end{feynman}
\draw [/tikzfeynman] (c) to[out=135, in=45, loop, min distance=0.9cm] (c);
\end{tikzpicture}
\hspace{-0.3cm}
+
\hspace{+0.15cm}
\begin{tikzpicture}[inner sep=2pt,baseline=(b)]
\begin{feynman}[inline=(a)]
\vertex[dot] (a) {};
\vertex[dot, right=0.6cm of a] (b) {};
\vertex[dot, right=0.6cm of b] (c) {};
\diagram*{(a) --[half left] (c)};
\diagram*{(c) --[out=260, in=280] (a)};
\vertex[below=0.65cm of b] {\footnotesize{$1\times M^2$}};
\end{feynman}
\draw [/tikzfeynman] (b) to[out=135, in=45, loop, min distance=0.8cm] (b);
\end{tikzpicture}
\hspace{+0.15cm}
+
\hspace{+0.15cm}
\begin{tikzpicture}[inner sep=2pt,baseline=(b)]
\begin{feynman}[inline=(a)]
\vertex[dot] (a) {};
\vertex[dot, right=0.6cm of a] (b) {};
\vertex[dot, right=0.6cm of b] (c) {};
\diagram*{(a) --[half left] (c)};
\diagram*{(c) --[half left] (b)};
\diagram*{(b) --[half left] (a)};
\vertex[below=0.71cm of b] {\footnotesize{$1\times M$}};
\end{feynman}
\end{tikzpicture}
\hspace{+0.15cm}
+
\hspace{+0.15cm}
\begin{tikzpicture}[inner sep=2pt,baseline=(b)]
\begin{feynman}[inline=(a)]
\vertex[dot] (a) {};
\vertex[dot, right=0.6cm of a] (b) {};
\vertex[dot, right=0.6cm of b] (c) {};
\diagram*{(a) --[half left] (b)};
\diagram*{(b) --[half left] (c)};
\diagram*{(c) --[out=260, in=280] (a)};
\vertex[below=0.73cm of b] {\footnotesize{$t_3\times M$}};
\end{feynman}
\end{tikzpicture}
\hspace{+0.15cm}
\right)
\\
&=2^{N/2}\(\av{\ell^2} \binom{N}{\qb}\)^3
\(M^3+3M^2+(t_3+1)M\),
\end{split}
\\
\begin{split}\nonumber
\label{eq:W_moment4}
\av{\Tr W^4}&=2^{N/2}\(\av{\ell^2} \binom{N}{\qb}\)^4
\end{split}
\\
\begin{split}\nonumber
\times&\left(
\hspace{-0.3cm}
\begin{tikzpicture}[baseline=(b)]
\begin{feynman}[inline=(a)]
\vertex[dot] (a) {};
\vertex[dot, right=0.6cm of a] (b) {};
\vertex[dot, right=0.6cm of b] (c) {};
\vertex[dot, right=0.6cm of c] (d) {};
\vertex[right=0.3cm of b] (x) {};
\vertex[below=0.62cm of x] {\footnotesize{$1\times M^4$}};
\end{feynman}
\draw [/tikzfeynman] (a) to[out=135, in=45, loop, min distance=0.9cm] (a);
\draw [/tikzfeynman] (b) to[out=135, in=45, loop, min distance=0.9cm] (b);
\draw [/tikzfeynman] (c) to[out=135, in=45, loop, min distance=0.9cm] (c);
\draw [/tikzfeynman] (d) to[out=135, in=45, loop, min distance=0.9cm] (d);
\end{tikzpicture}
\hspace{-0.3cm}
+
\hspace{-0.3cm}
\begin{tikzpicture}[inner sep=2pt,baseline=(b)]
\begin{feynman}[inline=(a)]
\vertex[dot] (a) {};
\vertex[dot, right=0.6cm of a] (b) {};
\vertex[dot, right=0.6cm of b] (c) {};
\vertex[dot, right=0.6cm of c] (d) {};
\vertex[right=0.3cm of b] (x) {};
\diagram*{(c) --[half left] (d)};
\diagram*{(d) --[half left] (c)};
\vertex[below=0.635cm of x] {\footnotesize{$1\times M^3$}};
\end{feynman}
\draw [/tikzfeynman] (a) to[out=135, in=45, loop, min distance=0.9cm] (a);
\draw [/tikzfeynman] (b) to[out=135, in=45, loop, min distance=0.9cm] (b);
\end{tikzpicture}
\hspace{+0.15cm}
+
\hspace{-0.30cm}
\begin{tikzpicture}[inner sep=2pt,baseline=(b)]
\begin{feynman}[inline=(a)]
\vertex[dot] (a) {};
\vertex[dot, right=0.6cm of a] (b) {};
\vertex[dot, right=0.6cm of b] (c) {};
\vertex[dot, right=0.6cm of c] (d) {};
\vertex[right=0.3cm of b] (x) {};
\diagram*{(b) --[half left] (c)};
\diagram*{(c) --[half left] (b)};
\vertex[below=0.635cm of x] {\footnotesize{$1\times M^3$}};
\end{feynman}
\draw [/tikzfeynman] (a) to[out=135, in=45, loop, min distance=0.9cm] (a);
\draw [/tikzfeynman] (d) to[out=135, in=45, loop, min distance=0.9cm] (d);
\end{tikzpicture}
\hspace{-0.3cm}
+
\hspace{+0.15cm}
\begin{tikzpicture}[inner sep=2pt,baseline=(b)]
\begin{feynman}[inline=(a)]
\vertex[dot] (a) {};
\vertex[dot, right=0.6cm of a] (b) {};
\vertex[dot, right=0.6cm of b] (c) {};
\vertex[dot, right=0.6cm of c] (d) {};
\vertex[right=0.3cm of b] (x) {};
\diagram*{(a) --[half left] (b)};
\diagram*{(b) --[half left] (a)};
\vertex[below=0.635cm of x] {\footnotesize{$1\times M^3$}};
\end{feynman}
\draw [/tikzfeynman] (c) to[out=135, in=45, loop, min distance=0.8cm] (c);
\draw [/tikzfeynman] (d) to[out=135, in=45, loop, min distance=0.8cm] (d);
\end{tikzpicture}
\hspace{-0.3cm}
+
\hspace{-0.3cm}
\begin{tikzpicture}[inner sep=2pt,baseline=(b)]
\begin{feynman}[inline=(a)]
\vertex[dot] (a) {};
\vertex[dot, right=0.6cm of a] (b) {};
\vertex[dot, right=0.6cm of b] (c) {};
\vertex[dot, right=0.6cm of c] (d) {};
\vertex[right=0.3cm of b] (x) {};
\diagram*{(b) --[half left] (d)};
\diagram*{(d) --[out=260, in=280] (b)};
\vertex[below=0.635cm of x] {\footnotesize{$1\times M^3$}};
\end{feynman}
\draw [/tikzfeynman] (a) to[out=135, in=45, loop, min distance=0.8cm] (a);
\draw [/tikzfeynman] (c) to[out=135, in=45, loop, min distance=0.8cm] (c);
\end{tikzpicture}
\right.
\end{split}
\\
\begin{split}\nonumber
&+
\hspace{+0.15cm}
\begin{tikzpicture}[inner sep=2pt,baseline=(b)]
\begin{feynman}[inline=(a)]
\vertex[dot] (a) {};
\vertex[dot, right=0.6cm of a] (b) {};
\vertex[dot, right=0.6cm of b] (c) {};
\vertex[dot, right=0.6cm of c] (d) {};
\vertex[right=0.3cm of b] (x) {};
\diagram*{(a) --[half left] (c)};
\diagram*{(c) --[out=260, in=280] (a)};
\vertex[below=0.8cm of x] {\footnotesize{$1\times M^3$}};
\end{feynman}
\draw [/tikzfeynman] (b) to[out=135, in=45, loop, min distance=0.8cm] (b);
\draw [/tikzfeynman] (d) to[out=135, in=45, loop, min distance=0.8cm] (d);
\end{tikzpicture}
\hspace{-0.3cm}
+
\hspace{+0.15cm}
\begin{tikzpicture}[inner sep=2pt,baseline=(b)]
\begin{feynman}[inline=(a)]
\vertex[dot] (a) {};
\vertex[dot, right=0.6cm of a] (b) {};
\vertex[dot, right=0.6cm of b] (c) {};
\vertex[dot, right=0.6cm of c] (d) {};
\vertex[right=0.3cm of b] (x) {};
\diagram*{(a) --[half left] (d)};
\diagram*{(d) --[out=240, in=300] (a)};
\vertex[below=0.8cm of x] {\footnotesize{$1\times M^3$}};
\end{feynman}
\draw [/tikzfeynman] (b) to[out=135, in=45, loop, min distance=0.8cm] (b);
\draw [/tikzfeynman] (c) to[out=135, in=45, loop, min distance=0.8cm] (c);
\end{tikzpicture}
\hspace{+0.15cm}
+
\hspace{-0.3cm}
\begin{tikzpicture}[inner sep=2pt,baseline=(b)]
\begin{feynman}[inline=(a)]
\vertex[dot] (a) {};
\vertex[dot, right=0.6cm of a] (b) {};
\vertex[dot, right=0.6cm of b] (c) {};
\vertex[dot, right=0.6cm of c] (d) {};
\vertex[right=0.3cm of b] (x) {};
\diagram*{(b) --[half left] (d)};
\diagram*{(d) --[half left] (c)};
\diagram*{(c) --[half left] (b)};
\vertex[below=0.785cm of x] {\footnotesize{$1\times M^2$}};
\end{feynman}
\draw [/tikzfeynman] (a) to[out=135, in=45, loop, min distance=0.8cm] (a);
\end{tikzpicture}
\hspace{+0.15cm}
+
\hspace{+0.15cm}
\begin{tikzpicture}[inner sep=2pt,baseline=(b)]
\begin{feynman}[inline=(a)]
\vertex[dot] (a) {};
\vertex[dot, right=0.6cm of a] (b) {};
\vertex[dot, right=0.6cm of b] (c) {};
\vertex[dot, right=0.6cm of c] (d) {};
\vertex[right=0.3cm of b] (x) {};
\diagram*{(a) --[half left] (c)};
\diagram*{(c) --[half left] (b)};
\diagram*{(b) --[half left] (a)};
\vertex[below=0.785cm of x] {\footnotesize{$1\times M^2$}};
\end{feynman}
\draw [/tikzfeynman] (d) to[out=135, in=45, loop, min distance=0.8cm] (d);
\end{tikzpicture}
\hspace{-0.30cm}
+
\hspace{+0.15cm}
\begin{tikzpicture}[inner sep=2pt,baseline=(b)]
\begin{feynman}[inline=(a)]
\vertex[dot] (a) {};
\vertex[dot, right=0.6cm of a] (b) {};
\vertex[dot, right=0.6cm of b] (c) {};
\vertex[dot, right=0.6cm of c] (d) {};
\vertex[right=0.3cm of b] (x) {};
\diagram*{(a) --[half left] (d)};
\diagram*{(d) --[half left] (c)};
\diagram*{(c) --[out=260, in=280] (a)};
\vertex[below=0.785cm of x] {\footnotesize{$1\times M^2$}};
\end{feynman}
\draw [/tikzfeynman] (b) to[out=135, in=45, loop, min distance=0.8cm] (b);
\end{tikzpicture}
\end{split}
\\
\begin{split}
&+
\hspace{+0.15cm}
\begin{tikzpicture}[inner sep=2pt,baseline=(b)]
\begin{feynman}[inline=(a)]
\vertex[dot] (a) {};
\vertex[dot, right=0.6cm of a] (b) {};
\vertex[dot, right=0.6cm of b] (c) {};
\vertex[dot, right=0.6cm of c] (d) {};
\vertex[right=0.3cm of b] (x) {};
\diagram*{(a) --[half left] (d)};
\diagram*{(d) --[out=260, in=280] (b)};
\diagram*{(b) --[half left] (a)};
\vertex[below=0.785cm of x] {\footnotesize{$1\times M^2$}};
\end{feynman}
\draw [/tikzfeynman] (c) to[out=135, in=45, loop, min distance=0.8cm] (c);
\end{tikzpicture}
\hspace{+0.15cm}
+
\hspace{+0.15cm}
\begin{tikzpicture}[inner sep=2pt,baseline=(b)]
\begin{feynman}[inline=(a)]
\vertex[dot] (a) {};
\vertex[dot, right=0.6cm of a] (b) {};
\vertex[dot, right=0.6cm of b] (c) {};
\vertex[dot, right=0.6cm of c] (d) {};
\vertex[right=0.3cm of b] (x) {};
\diagram*{(a) --[half left] (b)};
\diagram*{(b) --[half left] (a)};
\diagram*{(c) --[half left] (d)};
\diagram*{(d) --[half left] (c)};
\vertex[below=0.785cm of x] {\footnotesize{$1\times M^2$}};
\end{feynman}
\end{tikzpicture}
\hspace{+0.15cm}
+
\hspace{+0.15cm}
\begin{tikzpicture}[inner sep=2pt,baseline=(b)]
\begin{feynman}[inline=(a)]
\vertex[dot] (a) {};
\vertex[dot, right=0.6cm of a] (b) {};
\vertex[dot, right=0.6cm of b] (c) {};
\vertex[dot, right=0.6cm of c] (d) {};
\vertex[right=0.3cm of b] (x) {};
\diagram*{(a) --[out=80, in=100] (d)};
\diagram*{(d) --[out=260, in=280] (a)};
\diagram*{(b) --[half left] (c)};
\diagram*{(c) --[half left] (b)};
\vertex[below=0.785cm of x] {\footnotesize{$1\times M^2$}};
\end{feynman}
\end{tikzpicture}
\hspace{+0.15cm}
+
\hspace{-0.30cm}
\begin{tikzpicture}[inner sep=2pt,baseline=(b)]
\begin{feynman}[inline=(a)]
\vertex[dot] (a) {};
\vertex[dot, right=0.6cm of a] (b) {};
\vertex[dot, right=0.6cm of b] (c) {};
\vertex[dot, right=0.6cm of c] (d) {};
\vertex[right=0.3cm of b] (x) {};
\diagram*{(b) --[half left] (c)};
\diagram*{(c) --[half left] (d)};
\diagram*{(d) --[out=260, in=280] (b)};
\vertex[below=0.785cm of x] {\footnotesize{$t_3\times M^2$}};
\end{feynman}
\draw [/tikzfeynman] (a) to[out=135, in=45, loop, min distance=0.8cm] (a);
\end{tikzpicture}
\hspace{+0.15cm}
+
\hspace{+0.15cm}
\begin{tikzpicture}[inner sep=2pt,baseline=(b)]
\begin{feynman}[inline=(a)]
\vertex[dot] (a) {};
\vertex[dot, right=0.6cm of a] (b) {};
\vertex[dot, right=0.6cm of b] (c) {};
\vertex[dot, right=0.6cm of c] (d) {};
\vertex[right=0.3cm of b] (x) {};
\diagram*{(a) --[half left] (b)};
\diagram*{(b) --[half left] (c)};
\diagram*{(c) --[out=260, in=280] (a)};
\vertex[below=0.785cm of x] {\footnotesize{$t_3\times M^2$}};
\end{feynman}
\draw [/tikzfeynman] (d) to[out=135, in=45, loop, min distance=0.8cm] (d);
\end{tikzpicture}
\hspace{-0.3cm}
\end{split}
\\
\begin{split}\nonumber
&+
\hspace{+0.15cm}
\begin{tikzpicture}[inner sep=2pt,baseline=(b)]
\begin{feynman}[inline=(a)]
\vertex[dot] (a) {};
\vertex[dot, right=0.6cm of a] (b) {};
\vertex[dot, right=0.6cm of b] (c) {};
\vertex[dot, right=0.6cm of c] (d) {};
\vertex[right=0.3cm of b] (x) {};
\diagram*{(a) --[half left] (c)};
\diagram*{(c) --[half left] (d)};
\diagram*{(d) --[out=240, in=300] (a)};
\vertex[below=0.785cm of x] {\footnotesize{$t_3\times M^2$}};
\end{feynman}
\draw [/tikzfeynman] (b) to[out=135, in=45, loop, min distance=0.8cm] (b);
\end{tikzpicture}
\hspace{+0.15cm}
+
\hspace{+0.15cm}
\begin{tikzpicture}[inner sep=2pt,baseline=(b)]
\begin{feynman}[inline=(a)]
\vertex[dot] (a) {};
\vertex[dot, right=0.6cm of a] (b) {};
\vertex[dot, right=0.6cm of b] (c) {};
\vertex[dot, right=0.6cm of c] (d) {};
\vertex[right=0.3cm of b] (x) {};
\diagram*{(a) --[half left] (b)};
\diagram*{(b) --[half left] (d)};
\diagram*{(d) --[out=240, in=300] (a)};
\vertex[below=0.785cm of x] {\footnotesize{$t_3\times M^2$}};
\end{feynman}
\draw [/tikzfeynman] (c) to[out=135, in=45, loop, min distance=0.8cm] (c);
\end{tikzpicture}
\hspace{+0.15cm}
+
\hspace{+0.15cm}
\begin{tikzpicture}[inner sep=2pt,baseline=(b)]
\begin{feynman}[inline=(a)]
\vertex[dot] (a) {};
\vertex[dot, right=0.6cm of a] (b) {};
\vertex[dot, right=0.6cm of b] (c) {};
\vertex[dot, right=0.6cm of c] (d) {};
\vertex[right=0.3cm of b] (x) {};
\diagram*{(a) --[out=80, in=100] (c)};
\diagram*{(c) --[out=260, in=280] (a)};
\diagram*{(b) --[out=80, in=100] (d)};
\diagram*{(d) --[out=260, in=280] (b)};
\vertex[below=0.785cm of x] {\footnotesize{$t_4\times M^2$}};
\end{feynman}
\end{tikzpicture}
\hspace{+0.15cm}
+
\hspace{+0.15cm}
\begin{tikzpicture}[inner sep=2pt,baseline=(b)]
\begin{feynman}[inline=(a)]
\vertex[dot] (a) {};
\vertex[dot, right=0.6cm of a] (b) {};
\vertex[dot, right=0.6cm of b] (c) {};
\vertex[dot, right=0.6cm of c] (d) {};
\vertex[right=0.3cm of b] (x) {};
\diagram*{(a) --[out=80,in=100] (d)};
\diagram*{(d) --[half left] (c)};
\diagram*{(c) --[half left] (b)};
\diagram*{(b) --[half left] (a)};
\vertex[below=0.785cm of x] {\footnotesize{$1\times M$}};
\end{feynman}
\end{tikzpicture}
\hspace{+0.15cm}
+
\hspace{+0.15cm}
\begin{tikzpicture}[inner sep=2pt,baseline=(b)]
\begin{feynman}[inline=(a)]
\vertex[dot] (a) {};
\vertex[dot, right=0.6cm of a] (b) {};
\vertex[dot, right=0.6cm of b] (c) {};
\vertex[dot, right=0.6cm of c] (d) {};
\vertex[right=0.3cm of b] (x) {};
\diagram*{(a) --[half left] (b)};
\diagram*{(b) --[half left] (d)};
\diagram*{(d) --[half left] (c)};
\diagram*{(c) --[out=260, in=280] (a)};
\vertex[below=0.805cm of x] {\footnotesize{$t_3\times M$}};
\end{feynman}
\end{tikzpicture}
\hspace{+0.15cm}+
\end{split}
\\
\begin{split}\nonumber
&+\left.
\hspace{+0.15cm}
\begin{tikzpicture}[inner sep=2pt,baseline=(b)]
\begin{feynman}[inline=(a)]
\vertex[dot] (a) {};
\vertex[dot, right=0.6cm of a] (b) {};
\vertex[dot, right=0.6cm of b] (c) {};
\vertex[dot, right=0.6cm of c] (d) {};
\vertex[right=0.3cm of b] (x) {};
\diagram*{(a) --[half left] (c)};
\diagram*{(c) --[half left] (d)};
\diagram*{(d) --[out=260, in=280] (b)};
\diagram*{(b) --[half left] (a)};
\vertex[below=0.805cm of x] {\footnotesize{$t_3\times M$}};
\end{feynman}
\end{tikzpicture}
\hspace{+0.15cm}
+
\hspace{+0.15cm}
\begin{tikzpicture}[inner sep=2pt,baseline=(b)]
\begin{feynman}[inline=(a)]
\vertex[dot] (a) {};
\vertex[dot, right=0.6cm of a] (b) {};
\vertex[dot, right=0.6cm of b] (c) {};
\vertex[dot, right=0.6cm of c] (d) {};
\vertex[right=0.3cm of b] (x) {};
\diagram*{(a) --[half left] (c)};
\diagram*{(c) --[half left] (b)};
\diagram*{(b) --[half left] (d)};
\diagram*{(d) --[out=240, in=300] (a)};
\vertex[below=0.805cm of x] {\footnotesize{$t_3\times M$}};
\end{feynman}
\end{tikzpicture}
\hspace{+0.15cm}
+
\hspace{+0.15cm}
\begin{tikzpicture}[inner sep=2pt,baseline=(b)]
\begin{feynman}[inline=(a)]
\vertex[dot] (a) {};
\vertex[dot, right=0.6cm of a] (b) {};
\vertex[dot, right=0.6cm of b] (c) {};
\vertex[dot, right=0.6cm of c] (d) {};
\vertex[right=0.3cm of b] (x) {};
\diagram*{(a) --[out=800, in=100] (d)};
\diagram*{(d) --[out=260, in=280] (b)};
\diagram*{(b) --[half left] (c)};
\diagram*{(c) --[out=260, in=280] (a)};
\vertex[below=0.805cm of x] {\footnotesize{$t_3\times M$}};
\end{feynman}
\end{tikzpicture}
\hspace{+0.15cm}
+
\hspace{+0.15cm}
\begin{tikzpicture}[inner sep=2pt,baseline=(b)]
\begin{feynman}[inline=(a)]
\vertex[dot] (a) {};
\vertex[dot, right=0.6cm of a] (b) {};
\vertex[dot, right=0.6cm of b] (c) {};
\vertex[dot, right=0.6cm of c] (d) {};
\vertex[right=0.3cm of b] (x) {};
\diagram*{(a) --[half left] (b)};
\diagram*{(b) --[half left] (c)};
\diagram*{(c) --[half left] (d)};
\diagram*{(d) --[out=240, in=300] (a)};
\vertex[below=0.805cm of x] {\footnotesize{$t_4\times M$}};
\end{feynman}
\end{tikzpicture}
\hspace{+0.15cm}
\right)
\end{split}
\\
\begin{split}\nonumber
&=2^{N/2}\(\av{\ell^2} \binom{N}{\qb}\)^4
\(
M^4+6M^3+(6+4t_3+t_4)M^2+(1+4t_3+t_4)M
\).
\end{split}
\end{align}
\endgroup
Below each diagram we wrote (i) a factor of $M$ for each closed loop and (ii) the weight~$t(\sigma)$ coming from the trace of $\Gamma$-matrices. Different diagrams can correspond to the same trace because of the latter's cyclic property. 

The first nontrivial diagram, $t_3$, arises in the third moment of $W$ (sixth order in $L$ and $L^\dagger$) and it is explicitly given by~\cite{garcia2016,garcia2018}:
\begin{equation}\label{eq:diagram_t3}
\begin{split}
t_3(\qb,N)&\equiv
\begin{tikzpicture}[inner sep=2pt,baseline=(b)]
\begin{feynman}[inline=(a)]
\vertex[dot] (a) {};
\vertex[dot, right=0.6cm of a] (b) {};
\vertex[dot, right=0.6cm of b] (c) {};
\diagram*{(a) --[half left] (b)};
\diagram*{(b) --[half left] (c)};
\diagram*{(c) --[out=260, in=280] (a)};;
\end{feynman}
\end{tikzpicture}
= 2^{-N/2} \binom{N}{\qb}^{-3} \sum_{\va,\vb,\vc} \Tr\(
\Gamma_\va \Gamma_\vb \Gamma_\vc \Gamma_\va \Gamma_\vb \Gamma_\vc \)
\\
&=\binom{N}{\qb}^{-2} 
\sum_{s=0}^\qb \sum_{r=0}^\qb \sum_{m=0}^r 
(-1)^{\qb+s+m}
\binom{\qb}{s}\binom{N-\qb}{\qb-s}\binom{s}{r-m}
\binom{2(\qb-s)}{m}\binom{N-2\qb+s}{\qb-r}.
\end{split}
\end{equation}
To evaluate the trace, we have to first commute $\Gamma_\va$ with the product $\Gamma_\vb\Gamma_\vc$ and then commute $\Gamma_\vb$ with $\Gamma_\vc$, using Eq.~(\ref{eq:relations_Gamma}). Diagram $t_3$ also arises in the standard SYK model at sixth order in an expansion in powers of $H$; for its perfect-matching representation and a detailed account on how to compute it, see Appendix~\ref{app:review_standard_SYK}. However, the nontrivial diagram 
\begin{equation}\label{eq:diagram_t2}
t_2(\qb,N)
=2^{-N/2}\binom{N}{\qb}^{-2}\sum_{\va,\vb}\Tr\(\Gamma_\va \Gamma_\vb \Gamma_\va \Gamma_\vb\)
=\binom{N}{\qb}^{-1}\sum_{s=0}^\qb 
(-1)^{\qb+s}\binom{\qb}{s}\binom{N-\qb}{\qb-s}
\end{equation}
that arises at fourth order in an expansion in powers of $H$ [see Eqs.~(\ref{eq:H_moment4_example}) and (\ref{eq:diagram_t2_perfect_matching})] does not arise at fourth order in an expansion in powers of $L$ and $L^\dagger$. This is also true for higher orders: no diagrams related to $t_2$ (e.g., its powers) arise, as $t_2$ is a forbidden contraction when $k=1$ (i.e., it corresponds to a perfect matching that is not a permutation). In the context of the standard SYK model, the first nontrivial diagram (i.e., the single-crossing diagram, in that case $t_2$) has the combinatorial interpretation of the deformation parameter $Q$. For the circular WSYK model, we thus anticipate that it is diagram $t_3$ that assumes this role. 

Another nontrivial diagram, $t_4$, appears in the fourth moment. It cannot be reduced to a power of the lower-order diagram $t_3$ and it is evaluated analogously to before~\cite{garcia2018}:
\begin{equation}\label{eq:diagram_t4}
\begin{split}
t_4(\qb,N)&\equiv
\begin{tikzpicture}[inner sep=2pt,baseline=(b)]
\begin{feynman}[inline=(a)]
\vertex[dot] (a) {};
\vertex[dot, right=0.6cm of a] (b) {};
\vertex[dot, right=0.6cm of b] (c) {};
\vertex[dot, right=0.6cm of c] (d) {};
\diagram*{(a) --[half left] (b)};
\diagram*{(b) --[half left] (c)};
\diagram*{(c) --[half left] (d)};
\diagram*{(d) --[out=240, in=300] (a)};
\end{feynman}
\end{tikzpicture}
=2^{-N/2}\binom{N}{\qb}^{-4}\sum_{\va,\vb,\vc,\vd}
\Tr\(
\Gamma_\va \Gamma_\vb \Gamma_\vc \Gamma_\va
\Gamma_\vd \Gamma_\vc \Gamma_\vb \Gamma_\vd
 \)
\\
&=
\binom{N}{\qb}^{-3} 
\sum_{s=0}^\qb \binom{\qb}{s}\binom{N-\qb}{\qb-s}
\left\{\sum_{r=0}^\qb \sum_{m=0}^r 
(-1)^{m} \binom{s}{r-m} \binom{2(\qb-s)}{m}\binom{N-2\qb+s}{\qb-r}\right\}^2.
\end{split}
\end{equation}
To arrive at the explicit expression, we performed independent commutations of $\Gamma_\vb$ and $\Gamma_\vc$ over the product $\Gamma_\va\Gamma_\vd$, using again Eq.~(\ref{eq:relations_Gamma}).

In principle, one can compute $t(\sigma)$ exactly for every diagram and for all $p$ by explicitly computing traces. However, the computations quickly become intractable. Alternatively, when $\qb\propto\sqrt{N}$ and $M=1$, the weights $t(\sigma)$ can be computed exactly.

\section{Analytic spectral density of the circular WSYK model with \texorpdfstring{$M=1$}{M=1}}
\label{sec:spectral_density_WSYK}

When $\qb\propto \sqrt{N}$, the weight $t(\sigma)$ is fully characterized by the numbers of commutations required to bring the $\Gamma$-matrices to a trivial ordering. The number of commutations in a trace corresponds to the number of crossings in the corresponding perfect-matching, and not permutation, diagram. In this scaling limit, it is possible to show~\cite{erdos2014,feng2019} that, for a perfect matching $\pi$ with $\mathrm{cross}(\pi)$ crossings, $t(\pi)=Q^{\mathrm{cross}(\pi)}$, where $Q=(-1)^\qb\exp(-2\alpha)$, $\alpha=\qb^2/N$ is $N$ independent and $N\to\infty$. To use this exact scaling-limit result, the sum in Eq.~(\ref{eq:moments_W_sum_permutations}) would have to be performed over a subset of perfect matchings. Unfortunately, such a restricted sum cannot be performed in closed form. Instead, we noticed that if we approximate the number of commutations in the trace by the number of crossings in the permutation diagram, the sum in Eq.~(\ref{eq:moments_W_sum_permutations}) is feasible and the moments correspond to the moments of the weight function of certain $Q$-Laguerre polynomials, where $-1 < Q(\qb,N) < 1$ becomes independent of $N$ in the scaling limit. When $\qb$ is fixed and finite, the same combinatorial arguments still hold to order $1/N$~\cite{garcia2018}. Hence, for any $\qb$, we can compute the spectral density to leading and next-to-leading order in $1/N$.

For fixed $\qb$ and in the limit $N\to\infty$, there are, to leading order in $1/N$, no $\gamma$-matrices common to different $\Gamma$-matrices and, thus the commutations of the latter can be ignored, i.e., $t(\sigma)=1$ for all $\sigma\in\mathcal{S}_p$. The leading-order moments simply count the number of allowed diagrams at each order, which for permutations $\sigma\in\mathcal{S}_p$ are
\begin{equation}
\frac{1}{\sigmaL^p}\frac{\av{\Tr W^p}}{\Tr \id}=p!.
\end{equation}
These are the moments of the exponential distribution and, hence, to leading order, the spectral density of the $M=1$ circular WSYK model is $\varrho(E)=\exp(-E)$.

To next-to-leading order, we take the commutations of $\Gamma$-matrices into account but ignore correlations between different commutations. Therefore, we completely characterize a diagram~$\sigma$ simply by its number of crossings, $\mathrm{cross}(\sigma)$. Some additional care is required when counting crossings diagrammatically, compared to the perfect-matching case: $\mathrm{cross}(\sigma)$ is the number of pairs of edges above the line of dots that cross
(~$
\begin{tikzpicture}[inner sep=2pt,baseline=(b)]
\begin{feynman}[inline=(a)]
\vertex[dot] (a) {};
\vertex[dot,right=0.4cm of a] (b) {};
\vertex[dot,right=0.4cm of b] (c) {};
\vertex[dot,right=0.4cm of c] (d) {};
\diagram*{(a) --[half left,min distance=0.4cm] (c)};
\diagram*{(b) --[half left,min distance=0.4cm] (d)};
\end{feynman}
\end{tikzpicture}
$~) \emph{or touch} 
(~$
\begin{tikzpicture}[inner sep=2pt,baseline=(b)]
\begin{feynman}[inline=(a)]
\vertex[dot] (a) {};
\vertex[dot,right=0.4cm of a] (b) {};
\vertex[dot,right=0.4cm of b] (c) {};
\diagram*{(a) --[half left,min distance=0.4cm] (b)};
\diagram*{(b) --[half left,min distance=0.4cm] (c)};
\end{feynman}
\end{tikzpicture}
$~), plus the number of pairs of edges below the line of dots that cross
(~$
\begin{tikzpicture}[inner sep=2pt,baseline=(l)]
\begin{feynman}[inline=(a)]
\vertex[dot] (a) {};
\vertex[dot,right=0.4cm of a] (b) {};
\vertex[dot,right=0.4cm of b] (c) {};
\vertex[dot,right=0.4cm of c] (d) {};
\vertex[right=0.2cm of b, below=0.3cm of b] (l) {};
\diagram*{(a) --[half right,min distance=0.4cm] (c)};
\diagram*{(b) --[half right,min distance=0.4cm] (d)};
\end{feynman}
\end{tikzpicture}
$~)~\cite{corteel2007AAM}\footnote{This is the reason why the permutation $\sigma(1,2,3)=(2,3,1)$ has a crossing despite no edges actually crossing in its diagram, see Eq.~(\ref{eq:diagram_t3}).}.
The elementary single-crossing diagram is $t_3(\qb,N)$ and, as for the standard SYK model, we approximate more complicated diagrams with $\mathrm{cross}(\sigma)$ crossings by ascribing a factor $t_3$ to each crossing, i.e., we approximate $t(\sigma)=t_3^{\mathrm{cross}(\sigma)}$\footnote{The number of permutations in $\mathcal{S}_p$ with $k$ crossings can be explicitly computed for arbitrary $(p,k)$ and is tabulated as sequence A263776 in the Online Encyclopedia of Integer Sequences (OEIS)~\cite{OEIS_A263776}.}. As an example, we have $t_4\approx t_3^2$. As mentioned before, this identification does not become exact when $\qb\propto\sqrt{N}$. 

Summing over all diagrams, we arrive at
\begin{equation}
\frac{1}{\sigmaL^p}\frac{\av{\Tr W^p}}{\Tr \id}
=\sum_{\sigma\in\sS_p} t_3^{\mathrm{cross(\sigma)}},
\end{equation}
which are recognized as the moments of the orthogonality weight function of the Al-Salam-Chihara $Q$-Laguerre polynomials~\cite{kasraoui2011AAM} with $Q=t_3(\qb,N)$ and $y=1$:
\begin{equation}\label{eq:spectral_density_QLaguerre}
\varrho_{\mathrm{QL}}(E;Q)=
\frac{(Q;Q)_\infty^2(-Q;Q)_\infty^2}{(-Q^2;Q^2)_\infty^2}
\frac{2}{\pi \EL}\sqrt{\frac{1-E/\EL}{E/\EL}}
\prod_{k=1}^\infty 
\frac{
	1-\frac{4\(1-2E/\EL\)^2}{2+Q^{k}+Q^{-k}}}{
	\(1-\frac{2\(1-2E/\EL\)}{Q^k+Q^{-k}}\)^2},
\end{equation}
supported on $0\leq E\leq \EL$, where the (dimensionless) spectral edge of the WSYK model is
\begin{equation}\label{eq:E0_QLaguerre}
\EL=\frac{4}{1-Q},
\end{equation}
and $(a;Q)_\infty=\prod_{k=0}^\infty \(1-aQ^k\)$ is the $Q$-Pochhammer symbol. Note that the spectral density is of the form of a single-channel Marchenko-Pastur distribution~\cite{marchenko1967,forrester2010},
\begin{equation}\label{eq:Marchenko-Pastur}
\varrho_{\mathrm{MP}}(E)=
\frac{2}{\pi \EL}\sqrt{\frac{1-E/\EL}{E/\EL}},
\end{equation}
times a $Q$-dependant multiplicative correction\footnote{The Marchenko-Pastur distribution depends itself on $Q$ through its endpoint. In the random-matrix limit, $Q\to0$, the endpoint becomes $\EL\to4$.}.

Let us briefly comment on the nature of the $Q$-Laguerre approximation. It is exact to linear order in $Q$ or, alternatively, to next-to-leading order in $1/N$. The first correction is diagram $t_4$ that appears in the fourth moment. In the scaling limit $\qb^2=\alpha N$, we have the exact results $Q=t_3=-\exp[-6\alpha]$ and $t_4=\exp[-8\alpha]$~\cite{erdos2014,feng2019}; the $Q$-Laguerre approximation is, instead, $t_4\approx Q^2=\exp[-12\alpha]$; higher-order corrections proceed similarly. 
We thus see that, contrary to standard SYK case for which the $Q$-Hermite spectral density becomes exact in the scaling limit, the $Q$-Laguerre density is only an approximation of the true WSYK density both for fixed $\qb$ and in the scaling limit $\qb\propto \sqrt{N}$. Its accuracy has to be checked on a case-by-case basis, which we do for $\qb=2$, $3$, $4$ and $N\leq34$ in Sec.~\ref{sec:spectral_density_numerics}.

Notwithstanding the preceding considerations, it is important to emphasize that the $Q$-Laguerre mapping has two key features necessary to accurately predict the WSYK spectral density, namely, it accounts for the correct number of diagrams and it correctly identifies the deformation parameter as $Q=t_3$ (i.e., the minimal number of commutations in a trace is three). The other natural candidates for describing the spectral density ($Q$-Hermite and random matrix theory) do not possess either of these properties.

It is also important to note that the $Q$-Laguerre density~(\ref{eq:spectral_density_QLaguerre}) is \emph{not equal} to the $Q$-Hermite density after the change of variables $E\to E^2$, as it is the case for the linear WSYK model (equivalently, the supersymmetric SYK model), see Appendix~\ref{app:linear_WSYK}. First, $Q$ depends exclusively on the symmetries of the model (i.e., Hermiticity) and appropriate combinatorics are required to obtain its correct value: $Q=t_2$ [Eq.~(\ref{eq:diagram_t2})] for the linear WSYK model, while $Q=t_3$ [Eq.~(\ref{eq:diagram_t3})] for the circular WSYK model. In particular, the spectral edge, $\EL$, of both models is formally the same [see Eqs.~(\ref{eq:E0_QLaguerre}) and (\ref{eq:E0_linear_WSYK})] but it is evaluated at different $Q$. Second, even if we put in the correct value of $Q$ by hand, for $Q>0$ the multiplicative corrections are different for the circular and linear WSYK models, as is clear by inspection of Eqs.~(\ref{eq:spectral_density_QLaguerre}) and (\ref{eq:spectral_density_linearWSYK}), respectively. Specifically, as $Q\to1$, in the circular model $\varrho(E)\to\exp\(-E\)$, while in the linear model $\varrho(E)\to(1/\sqrt{2\pi E})\exp\(-E/2\)$. On the other hand, as $Q\to0$ (the random-matrix limit) both spectral densities go to the Marchenko-Pastur distribution. Indeed, the relevant diagrams for random matrix theory are noncrossing and because all noncrossing perfect matchings are also noncrossing permutations, the allowed diagrams for the Gaussian and Wishart-Laguerre ensembles coincide.

To conclude this section, we note that the $Q$-Laguerre density~(\ref{eq:spectral_density_QLaguerre}) has a simple asymptotic form in the bulk (i.e., for $0\ll E\ll \EL$), close to the hard edge $E\approx 0$, and near the soft edge $E\approx \EL$. These asymptotic formulas are obtained in the large-$N$ limit after a Poisson resummation of the $Q$-Laguerre spectral density; see Appendix~\ref{app:asymptotics} for their derivation. 

The asymptotic densities depend on the sign of $Q$. For a finite but large enough $N$ (the limit we are mostly interested in), $Q=t_3(\qb,N)$ is positive for even $\qb$ and negative for odd $\qb$. 

For positive $Q$ (even $\qb$), the asymptotic bulk density is given by
\begin{equation}\label{eq:asympt_initial}
\varrho_{Q>0}^{(\mathrm{bulk})}(E;Q)=C''_Q 
\exp\left[
\frac{2\arcsin^2\(1-\frac{2E}{\EL}\)-\arccos^2\(\frac{2E}{\EL}-1\)}{\log Q}\right],
\end{equation}
where the constant $C''_Q$ is
\begin{equation}
C''_Q
=\frac{(Q;Q)_\infty^2(-Q;Q)_\infty^2}{(-Q^2;Q^2)_\infty^2}\frac{\exp\left[\pi^2/\(4\log Q\)\right]}{\(1+\exp[\pi^2/\log Q]\)^2}
\frac{2}{\pi \EL}.
\end{equation}

Near the hard edge, $E\approx 0$, the asymptotic density is
\begin{equation}
\varrho_{Q>0}^{(\mathrm{hard})}(E;Q)=
C''_Q 
\exp\left[-\frac{\pi^2}{2\log Q}\right]
\coth\left[-\frac{2\pi}{\log Q}\sqrt{\frac{E}{\EL}} \right],
\end{equation}
which has the expected $1/\sqrt{E}$ divergence. Intriguingly, unlike the standard and supersymmetric SYK models, the spectral density in this low energy region does not increase exponentially. As was mentioned previously, a density of low-energy excitations proportional to $\exp[\gamma\sqrt{E-E_0}]$, with $\gamma > 0$ and $E_0$ the ground-state energy, is typical of both quantum black holes~\cite{maldacena2016} and nuclear matter~\cite{bethe1936,garcia2017}. Its absence in the WSYK model, for $Q(\qb,N) > 0$, is a strong indication that no gravity dual interpretation exists in this range of parameters. We shall see shortly that for $Q(\qb,N) < 0$ the situation is different. 

Near the soft edge $E\approx \EL$ corresponding to the high-energy region, the asymptotic density is found to be
\begin{equation}
\varrho_{Q>0}^{(\mathrm{soft})}(E;Q)= 
C''_Q\exp\left[\frac{\pi^2}{2\log Q}\right]
\,2\sinh\left[-\frac{4\pi}{\log Q}\sqrt{1-\frac{E}{\EL}}\right],
\end{equation}
Note that, apart from some multiplicative constants, this is the same density found near the edges of the standard SYK model. This is not surprising as, far away from the origin, the positive-definiteness of the spectrum is irrelevant and one does not expect the density to be very sensitive to the exact distribution of matrix elements of the Hamiltonian. Since $E$ is smaller than, or comparable to, $\EL$, there is no exponential growth of excitations in this region which in any case would not be expected to be related to gravity systems as holographic relations in the context of the SYK model are restricted to the low-energy, strong-coupling region.

For negative $Q$ (odd $\qb$), the bulk spectral density instead reads as:
\begin{equation}
\begin{split}
\varrho_{Q<0}^{\mathrm{(bulk)}}=
C'_\abs{Q}
&\cosh\left[
\frac{\pi}{\log\abs{Q}}
\abs{\arcsin\(1-\frac{2E}{\EL}\)}
\right]
\\
\times&\exp\left[
\frac{2\arcsin^2\(1-\frac{2E}{\EL}\)-\frac{1}{2}\arccos^2\(1-\frac{2E}{\EL}\)-\frac{1}{2}\arccos^2\(\frac{2E}{\EL}-1\)}{\log \abs{Q}}
\right],
\end{split}
\end{equation}
where the global constant is
\begin{equation}
C'_\abs{Q}
=\frac{(Q;Q)_\infty^2(-Q;Q)_\infty^2}{(-Q^2;Q^2)_\infty^2}
\exp\left[\frac{\pi^2}{4\log\abs{Q}}\right]
\frac{2}{\pi \EL}.
\end{equation}

\begin{figure}[tbp]
	\centering
	\includegraphics[width=\textwidth]{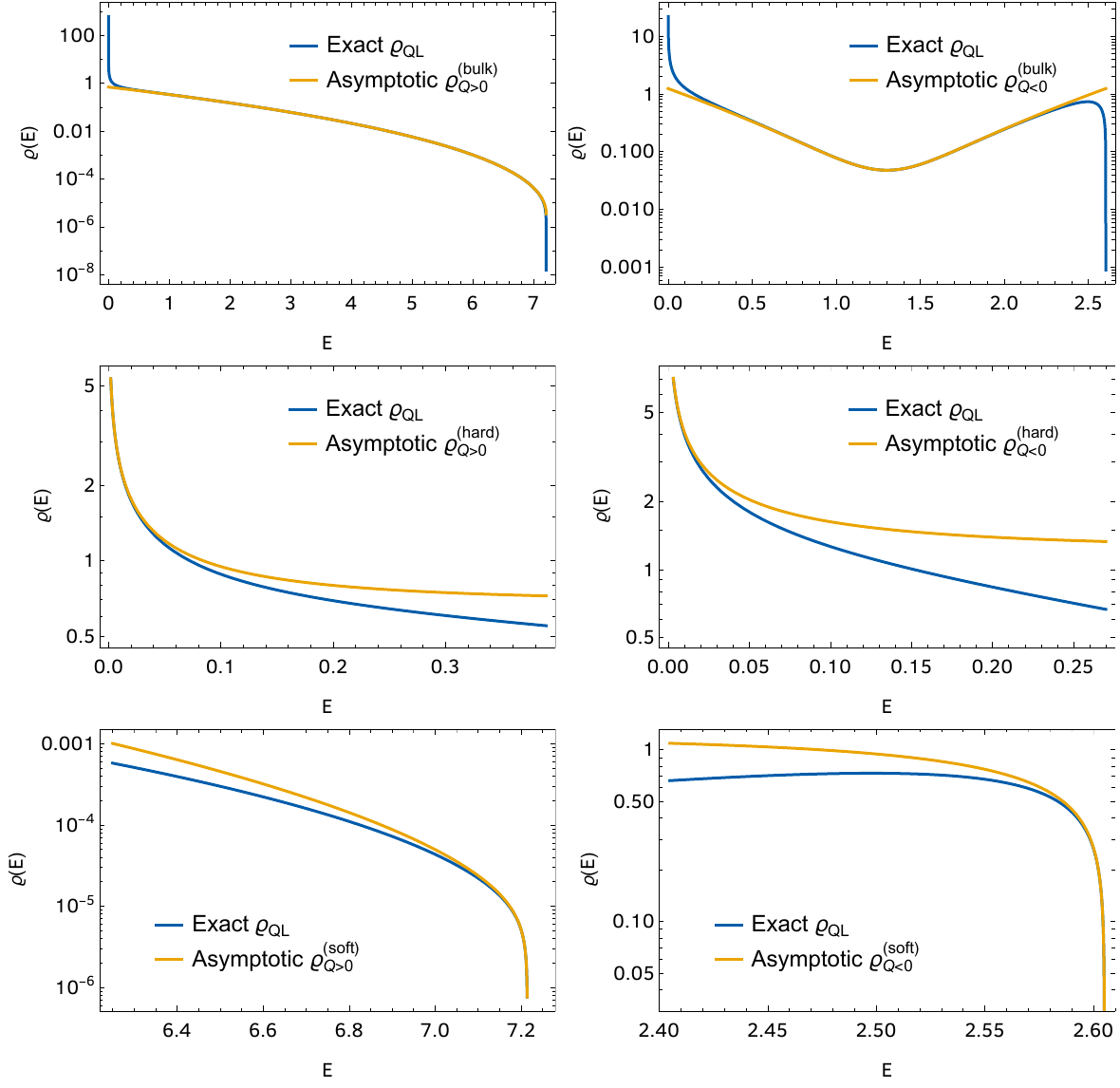}
	\caption{Comparison of the asymptotic spectral densities (orange curves) for the bulk, hard edge, and soft edge with the $Q$-Laguerre density (blue curve), in the respective domains of applicability. (Left) Densities for $Q>0$, namely, $Q=t_3(\qb=2,N=36)$. (Right) Densities for $Q<0$, namely, $Q=t_3(\qb=3,N=96)$.}
	\label{fig:M1AsymptDensity}
\end{figure} 

The spectral density close to the hard edge at $E = 0$ is
\begin{equation}\label{eq:expdenhard}
\varrho_{Q<0}^{\mathrm{(hard)}}=
C'_\abs{Q}
\coth\left[
-\frac{\pi}{\log\abs{Q}}\sqrt{\frac{E}{\EL}}
\right]
\exp\left[
-\frac{2\pi}{\log \abs{Q}}\sqrt{\frac{E}{\EL}}
\right]
\cosh\left[
\frac{\pi^2}{2\log\abs{Q}}\(1-\frac{4}{\pi}\sqrt{\frac{E}{\EL}}\)\right],
\end{equation}
which shows, besides the $1/\sqrt{E}$ divergence at the origin, a stretched exponential growth in the triple scaling limit \cite{cotler2016} corresponding to low-energy excitations slightly above the ground-state $E=0$. Interestingly, this is precisely one of the expected features of a field theory with a gravity dual. We note that the functional form is different from the result for the supersymmetric SYK model~\cite{fu2018,stanford2017,garcia2018a} whose spectrum also has a hard edge at zero energy. As was mentioned previously, the supersymmetric SYK is reduced in the low-energy limit to a super-Schwarzian action shared by JT super-gravity~\cite{forste2017}. It would be worthwhile, especially following the recently proposed classification scheme of JT supergravity \cite{johnson2021,johnson2021a} to explore whether the above spectral density, and the spectral correlation studied in next section, are related to another flavor of JT gravity. 

Finally, the negative-$Q$ soft-edge asymptotic density is given by
\begin{equation}\label{eq:asympt_final}
\begin{split}
\varrho_{Q<0}^{\mathrm{(soft)}}=
C'_\abs{Q}
&\tanh\left[
-\frac{\pi}{\log\abs{Q}}\sqrt{1-\frac{E}{\EL}}
\right]
\exp\left[
-\frac{2\pi}{\log \abs{Q}}\sqrt{1-\frac{E}{\EL}}
\right]
\\
\times
&\cosh\left[
\frac{\pi^2}{2\log\abs{Q}}\(1-\frac{4}{\pi}\sqrt{1-\frac{E}{\EL}}\)\right].
\end{split}
\end{equation}
In this limit, the comments made for the positive $Q$ also apply here. 

In Fig.~\ref{fig:M1AsymptDensity}, we compare these simple asymptotic formulas, Eqs.~(\ref{eq:asympt_initial})--(\ref{eq:asympt_final}), with the $Q$-Laguerre weight function, Eq.~(\ref{eq:spectral_density_QLaguerre}), for both positive and negative $Q$. We see excellent agreement in the respective domains of applicability, even for $\qb=2$ and relatively small $N=36$ ($Q\approx 0.45$). As $N$ increases, the bulk asymptotic formula describes the density very well almost all the way down to the soft edge. However, it does not, of course, capture the $1/\sqrt{E}$ divergence close to the origin. For odd $\qb=3$, the asymptotic formulas are also very accurate, but the asymptotic limit is only attained for much larger values of $N$ (e.g., $Q\approx -0.54$ for $N=96$). This confirms the validity of the analytical calculation and, for negative $Q$, the possible existence of a holographic duality. 

\section{Numerical spectral density of the circular WSYK model for fixed \texorpdfstring{$\qb$}{q}}
\label{sec:spectral_density_numerics}

\begin{figure}[tbp]
	\centering
	\includegraphics[width=0.55\textwidth]{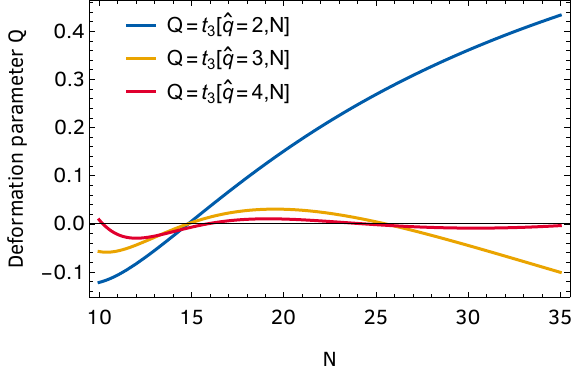}
	\caption{Dependence of the deformation parameter $Q$ on the number of Majoranas $N$ for $\qb = 2,3,4$.}
	\label{fig:QvsN}
\end{figure}

In this section, we study the accuracy of the $Q$-Laguerre approximation for the range of parameters accessible to numerical exact diagonalization, i.e., for a fixed $\qb = 2,3,4\ldots$ that does not scale with $N$. 

\begin{figure}[tbp]
	\centering
	\includegraphics[width=\textwidth]{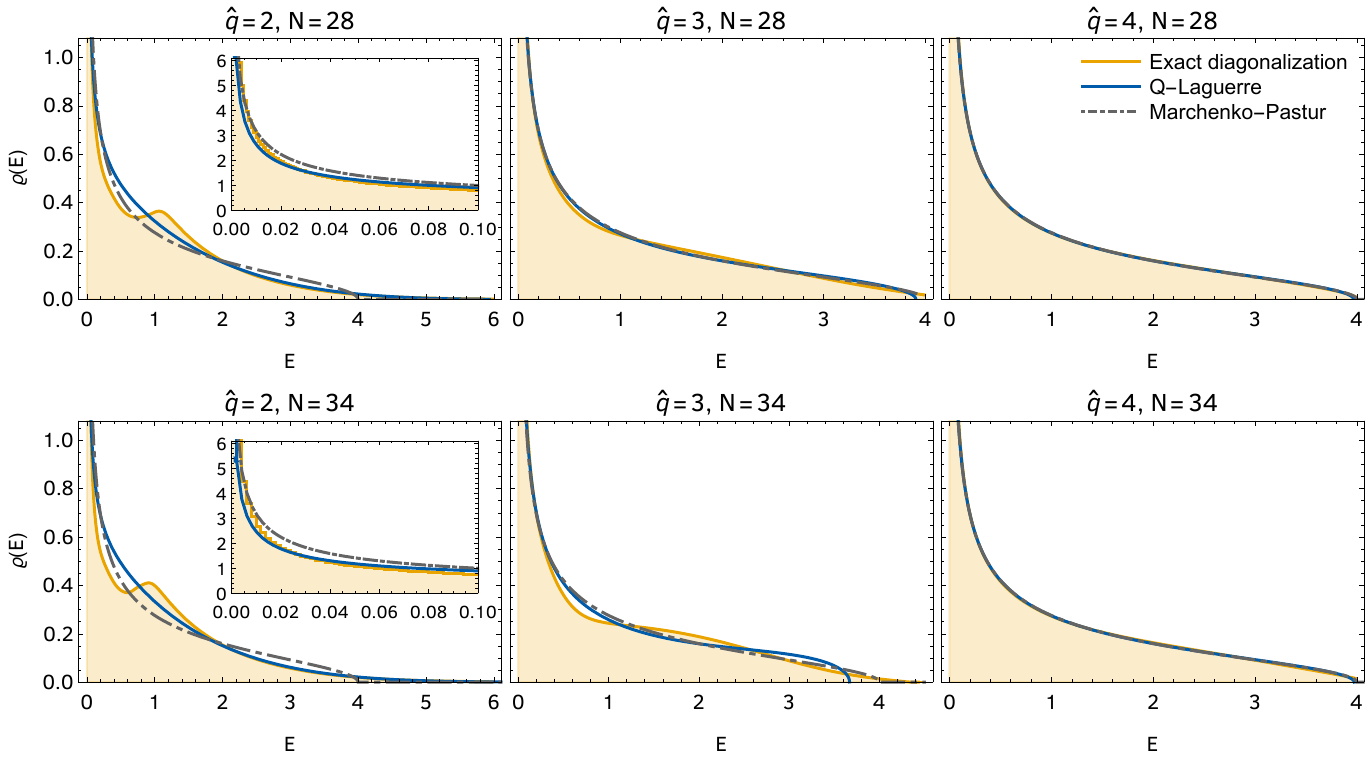}
	\caption{Spectral density of the circular ($k=1$) WSYK model with $M=1$ and $N=28$, $34$ Majorana fermions for $\qb=2$, $3$, and $4$. The orange (shaded) histograms are obtained by exact diagonalization of the WSYK Hamiltonian~(\ref{eq:def_WSYK}) for different disorder realizations totaling at least $2^{19}$ eigenvalues. The blue (full) and gray (dot-dashed) curves correspond to the $Q$-Laguerre [Eq.~(\ref{eq:spectral_density_QLaguerre})] with $Q=t_3(\qb,28)$ and Marchenko-Pastur [Eq.~(\ref{eq:Marchenko-Pastur})] predictions, respectively. We see the spectral density approaching plain random matrix results as $\qb$ increases with $N$ fixed. Insets: zoom on the hard edge for $\qb=2$. We see excellent agreement with both random matrix theory and $Q$-Laguerre near the hard edge.}
	\label{fig:M1LinearDensityq}
\end{figure}

For $\qb=3, 4,\dots$ and $N\leq34$, $Q=t_3(\qb,N)$ is very close to $0$ (e.g., we have $Q=t_3(\qb=3,N=28)\approx -0.024$ and $Q=t_3(\qb=4,N=28)\approx -0.008$), see Fig.~\ref{fig:QvsN}. Hence, deviations from plain random matrix theory are small. In Fig.~\ref{fig:M1LinearDensityq}, we compare the $Q$-Laguerre prediction~(\ref{eq:spectral_density_QLaguerre}) against numerical results obtained from exact diagonalization of the circular WSYK Hamiltonian~(\ref{eq:def_WSYK}). As can be seen in the center and right panels of Fig.~\ref{fig:M1LinearDensityq}, for these small values of $Q$, the $Q$-Laguerre density is almost indistinguishable from the random matrix result, the Marchenko-Pastur distribution, Eq.~(\ref{eq:Marchenko-Pastur}). For $\qb=3$ a modest deviation from random matrix behavior can be seen for $N=34$ in the tail of the spectrum, and there is qualitative agreement with the numerical results.

\begin{figure}[tbp]
	\centering
	\includegraphics[width=\textwidth]{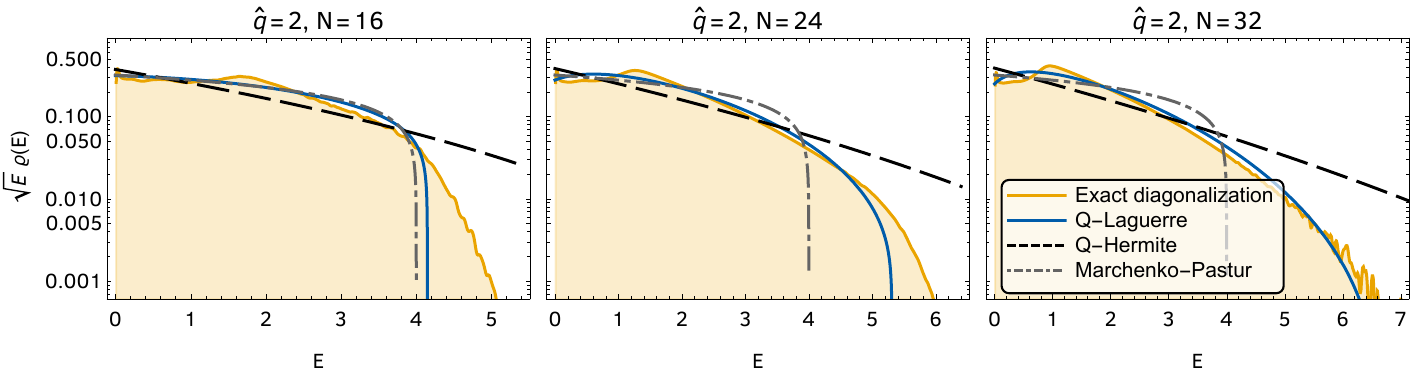}
	\caption{Spectral density (with $1/\sqrt{E}$ divergence factored out) of the circular ($k=1$) WSYK model with $M=1$ and $\qb=2$ ($q=4$) for $N=16$, $24$, and $32$ Majorana fermions. The orange (shaded) histograms are obtained by exact diagonalization of the WSYK Hamiltonian~(\ref{eq:def_WSYK}) for different disorder realizations totaling $2^{18}$ eigenvalues. The blue (full), black (dashed), and gray (dot-dashed) curves correspond to the $Q$-Laguerre [Eq.~(\ref{eq:spectral_density_QLaguerre})] with $Q=t_3(2,N)$, $Q$-Hermite [Eq.~(\ref{eq:spectral_density_linearWSYK})] with $Q=t_2(2,N)$, and Marchenko-Pastur [Eq.~(\ref{eq:Marchenko-Pastur})] predictions, respectively.}
	\label{fig:M1LaguerreDensity}
\end{figure}

For $\qb=2$, we have access to a much larger range of values of $Q=Q(2,N)$ (for instance, $Q=t_3(\qb=2,N=28)\approx 0.326$), see Fig.~\ref{fig:QvsN}. In this case, one can clearly distinguish the $Q$-Laguerre density from the random matrix one across the entire spectrum, see the left panel of Fig.~\ref{fig:M1LinearDensityq}. 

In the right half of the spectrum, $E\gtrsim 2$,
we observe an excellent agreement with the analytical $Q$-Laguerre prediction, which in this case is clearly different from both the $Q$-Hermite and the Marchenko-Pastur distribution, see Figs.~\ref{fig:M1LinearDensityq} and \ref{fig:M1LaguerreDensity}.
In particular, the $Q$-Laguerre density decays as $\exp[-E/\EL]$, while the $Q$-Hermite density as $\exp[-E/2\EL]$ (after the change of variables $E\to E^2$), and the Marchenko-Pastur distribution decays $\sqrt{1-E/\EL}$. Therefore, in this region of the spectrum, all three distributions can be clearly distinguished.
We stress that the comparison is completely determined by the analytical expression with no fitting parameters. In particular, the endpoint $\EL$ is fixed in each case: $\EL=4/(1-t_3(\qb,N))$ for $Q$-Laguerre, $\EL=4/(1-t_2(\qb,N))$ for $Q$-Hermite, and $\EL=4$ for Marchenko-Pastur. The tail of the spectrum is best seen on a logarithmic scale, see Fig.~\ref{fig:M1LaguerreDensity}. As mentioned before, to the right of the mid-spectrum bump, there is excellent agreement between the $Q$-Laguerre prediction and the numerical results. As $N$ grows, this agreement increasingly extends all the way down to the soft edge at $E=\EL$. Moreover, $Q=t_3(\qb,N)$ also increases and we start to see deviations from random matrix theory. It is also clearly visible that the numerical results are \emph{not} well described by the $Q$-Hermite density after the change of variables $E\to E^2$ (conversely, the square roots of the eigenvalues of the circular WSYK do not follow the $Q$-Hermite density of the standard SYK model).

For smaller values of the energy, still for the $\qb=2$ case, we observe a bump around the middle of the spectrum. The bump shifts slightly to lower energies as $N$ increases (not shown). Within the range of $N$ that we can reach numerically, it is unclear whether it is a finite-size effect that will go away in the large-$N$ limit or if it has another origin. In any case, it is not predicted by the $Q$-Laguerre, $Q$-Hermite, or random matrix analytical expressions. The bump is much milder for $\qb=3$ (but visible when $N=34$) and it is not observed for $\qb=4$ in the available range of $N$. A qualitatively similar nonmonotonic behavior was observed~\cite{kanazawa2017} in a SYK model with Dirac fermions and also two $\hat q = 2$ blocks.  

Near the hard edge ($E=0$), and sufficiently away from the bump, see the insets in the left panels of Fig.~\ref{fig:M1LinearDensityq}, the divergence of the spectral density is well-described by the $Q$-Laguerre density which, in this range of parameters, agrees with the Marchenko-Pastur distribution. The presence of the bump, together with the singularity at $E = 0$ and the limited range of $N$ for which $Q$ is negative, makes it difficult to test the exponential growth, once the divergence is factored out, that characterizes the $Q$-Laguerre spectral density, Eq.~(\ref{eq:expdenhard}), in this region.

\begin{figure}[tbp]
	\centering
	\includegraphics[width=0.5\textwidth]{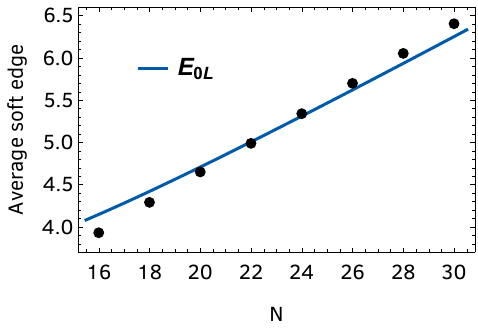}
	\caption{Spectral edge of the circular WSYK model~(\ref{eq:def_WSYK}) with $M=1$ and $\qb=2$ ($q=4$) as a function of the number of Majoranas $N$. The black dots correspond to the ensemble-averaged largest eigenvalue of $W$ obtained from exact diagonalization ($8192$ disorder realizations for $N=16$--$24$, $2048$ realizations for $N=26$, $28$, and $1472$ realizations for $N=30$), while the full line gives the $Q$-Laguerre prediction~(\ref{eq:E0_QLaguerre}) with $Q=t_3(2,N)$.}
	\label{fig:M1Edge}
\end{figure}

In Fig.~\ref{fig:M1LaguerreDensity}, there appears to be a discrepancy between numerics and the $Q$-Laguerre density at the soft (right) edge, which is more pronounced for smaller $N$. However, it is an artifact of the combination of the logarithmic scale (which amplifies small deviations), finite sampling, and relatively small values of $N$. Indeed, we know that the edge (largest eigenvalue), $E_{\mathrm{max}}$, which depends on the disorder realization, is itself a random variable whose distribution only becomes sharply peaked around its mean, $\EL$, for sufficiently large $N$. Therefore, the spectral density beyond the edge is, in reality, the distribution of the largest eigenvalues. Strictly speaking, we note that not only the largest eigenvalue contributes to the tail of the spectral density because, due to ensemble-ensemble fluctuations, it may occur that a few of the largest eigenvalues of one disorder realization are larger than the largest of another disorder realization. In order to illustrate explicitly that the discrepancy in the tail of the distribution is an artifact of the combination of a soft edge and ensemble average, we compare the largest eigenvalue of $W$ averaged over many disorder realizations with the $Q$-Laguerre prediction. Results depicted in Fig.~\ref{fig:M1Edge} show a very good agreement between this quantity and the analytical prediction, Eq.~(\ref{eq:E0_QLaguerre}), for $16\leq N\leq 30$.

\begin{figure}[tbp]
	\centering
	\includegraphics[width=\textwidth]{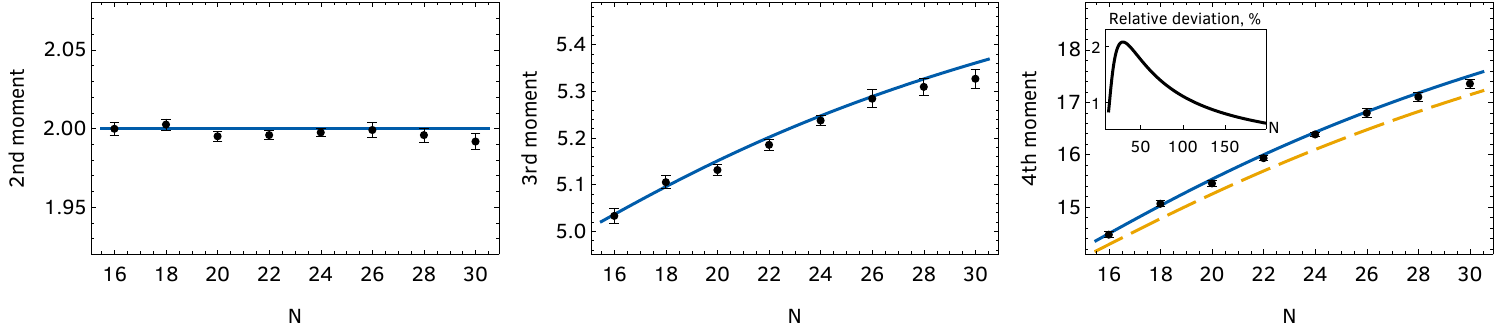}
	\caption{Low-order (normalized) moments of the circular ($k=1)$ WSYK model~(\ref{eq:def_WSYK}) with $M=1$ and $\qb=2$ ($q=4$) as a function of the number of Majoranas $N$. The black dots are obtained from ensemble-averaged exact diagonalization ($8192$ disorder realizations for $N=16$--$24$, $2048$ realizations for $N=26$, $28$, and $1472$ realizations for $N=30$), while the full blue lines gives the exact analytic expressions, Eqs.~(\ref{eq:W_moment2})--(\ref{eq:W_moment4}). For the fourth moment, the $Q$-Laguerre approximation is also plotted (dashed orange line), while the inset shows the relative deviation between the exact analytic prediction and the $Q$-Laguerre approximation. The relative deviation is maximal for the values of $N$ currently accessible, yet it is only around $2\%$.}
	\label{fig:M1Moments}
\end{figure}

We get further numerical evidence corroborating the applicability of our analytic results to systems with fixed $\qb=2$ by evaluating some low-order moments. In Fig.~\ref{fig:M1Moments}, we plot the second, third, and fourth moments as a function of the number of Majoranas $N$. As expected, there is excellent agreement between the moments obtained from exact diagonalization and the exact moments, Eqs.~(\ref{eq:W_moment2})--(\ref{eq:W_moment4}). The fourth moment is the first where there are permutation diagrams with more than one crossing and we can, therefore, compare the $Q$-Laguerre approximation against the exact result. As seen in the inset, the relative deviation between the two expressions is always below $2\%$, a value that is attained for the small values of $N$ considered here, and then decreases as $N$ increases. 

In summary, we observe a good agreement between the $Q$-Laguerre density and moments with numerical results obtained by exact diagonalization of $W$. The approximation of uncorrelated commutations and the replacement of perfect-matching crossings by permutation crossings are, therefore, well justified, at least for $\qb = 3,4$ and the values of $N$ that are numerically accessible. We again emphasize that the good agreement between numerical and analytic predictions relies crucially on the value of $Q=t_3(\qb,N)$, which could not have been obtained starting from the $Q$-Hermite (perfect-matching) combinatorics.

\section{Microscopic spectral density and level statistics of the WSYK model}\label{sec:speden}
In this section, we investigate the spectral density in the so-called microscopic limit corresponding to the smallest eigenvalues close to the hard edge of the spectrum at $E = 0$. In the context of random matrix theory~\cite{verbaarschot1994,altland1997}, this microscopic spectral density is universal~\cite{verbaarschot1998}, namely, it does not depend on the details of the Hamiltonian but only on symmetries such as the time-reversal invariance. Explicit analytic expressions for each universality class are known by using the orthogonal polynomial or the supersymmetry method~\cite{verbaarschot1993a,nagao1995,ivanov2002}. In the former, the microscopic density can be expressed in terms of the asymptotic limit of the Laguerre kernel that also describes the bulk spectral density. Previously, we have shown that the spectral density of the WSYK is well approximated by the weight function of the Al-Salam-Chihara $Q$-Laguerre polynomials~\cite{al1976} which are a variant of $Q$-Laguerre polynomials~\cite{moak1981}. It is tempting to speculate that the microscopic limit of the WSYK model will be given precisely by the random matrix results but replacing the Laguerre with $Q$-Laguerre polynomials. However, there are no reasons at all to believe that this is the case. For instance, for supersymmetric SYK models, the microscopic spectral density for some values of $N$ is similar to the chiral random matrix prediction while the average spectral density is still given \cite{garcia2018a} by the $Q$-Hermite result. We recall that, in contrast to standard random matrix theory, the relation between the average spectral density and the $Q$-Laguerre polynomials is \emph{not} through the spectral kernel of random matrix theory---which is expressed as a sum over the order of the orthogonal polynomials squared, with the sum, in turn, evaluated by the Christoffel-Darboux formula~\cite{mehta2004,guhr1998}. In our case, the moments of the spectral density are those of just the \emph{weight} of the Al-Salam-Chihara $Q$-Laguerre polynomials, which is, in principle, not directly related to the polynomials themselves. 

\begin{figure}
	\centering
	\begin{minipage}{0.5\textwidth}
		\includegraphics[width=\textwidth]{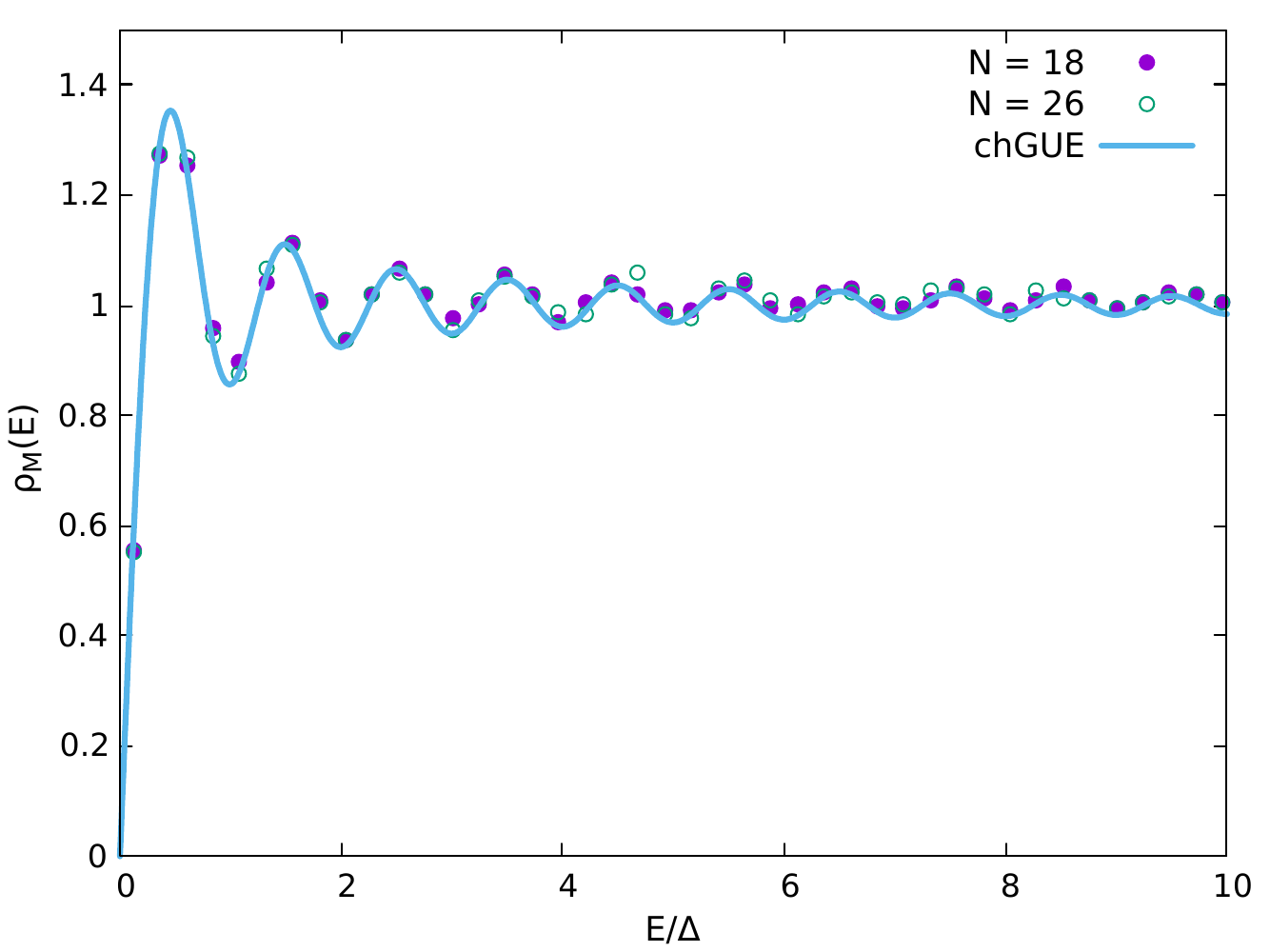}
	\end{minipage}%
	\begin{minipage}{0.5\textwidth}
		\includegraphics[width=\textwidth]{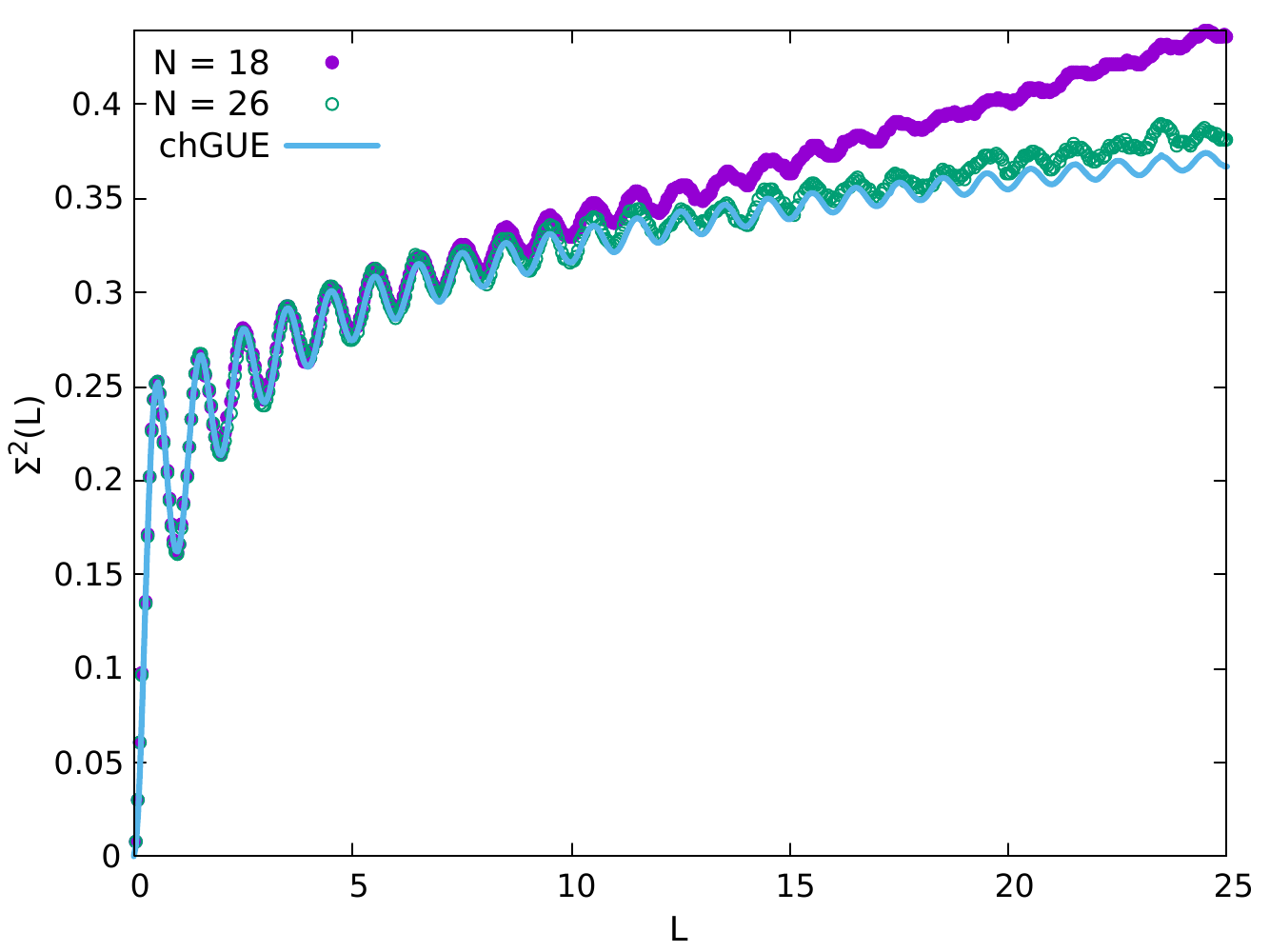}
	\end{minipage}
	\begin{minipage}{0.5\textwidth}
		\includegraphics[width=\textwidth]{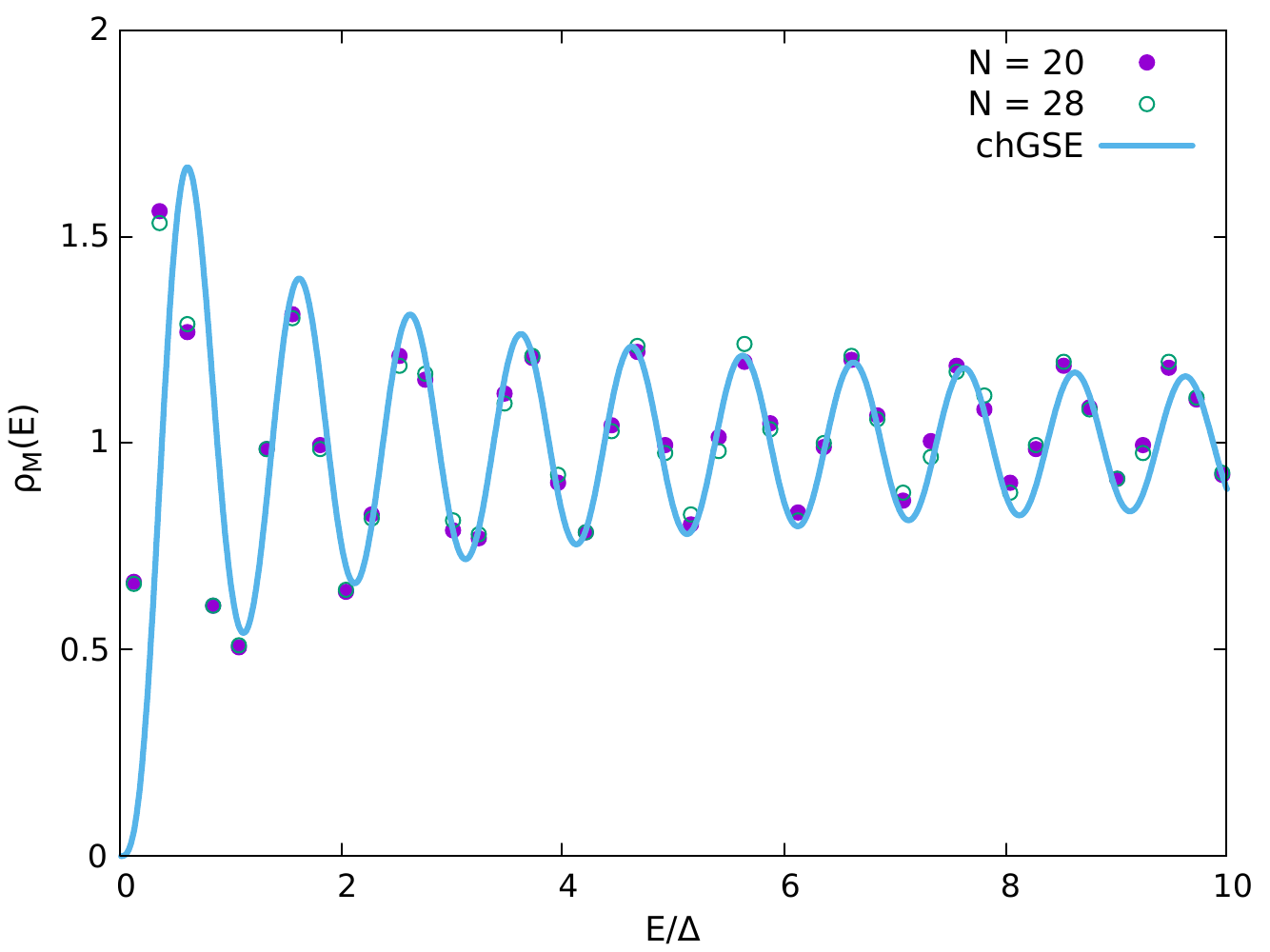}
	\end{minipage}%
	\begin{minipage}{0.5\textwidth}
		\includegraphics[width=\textwidth]{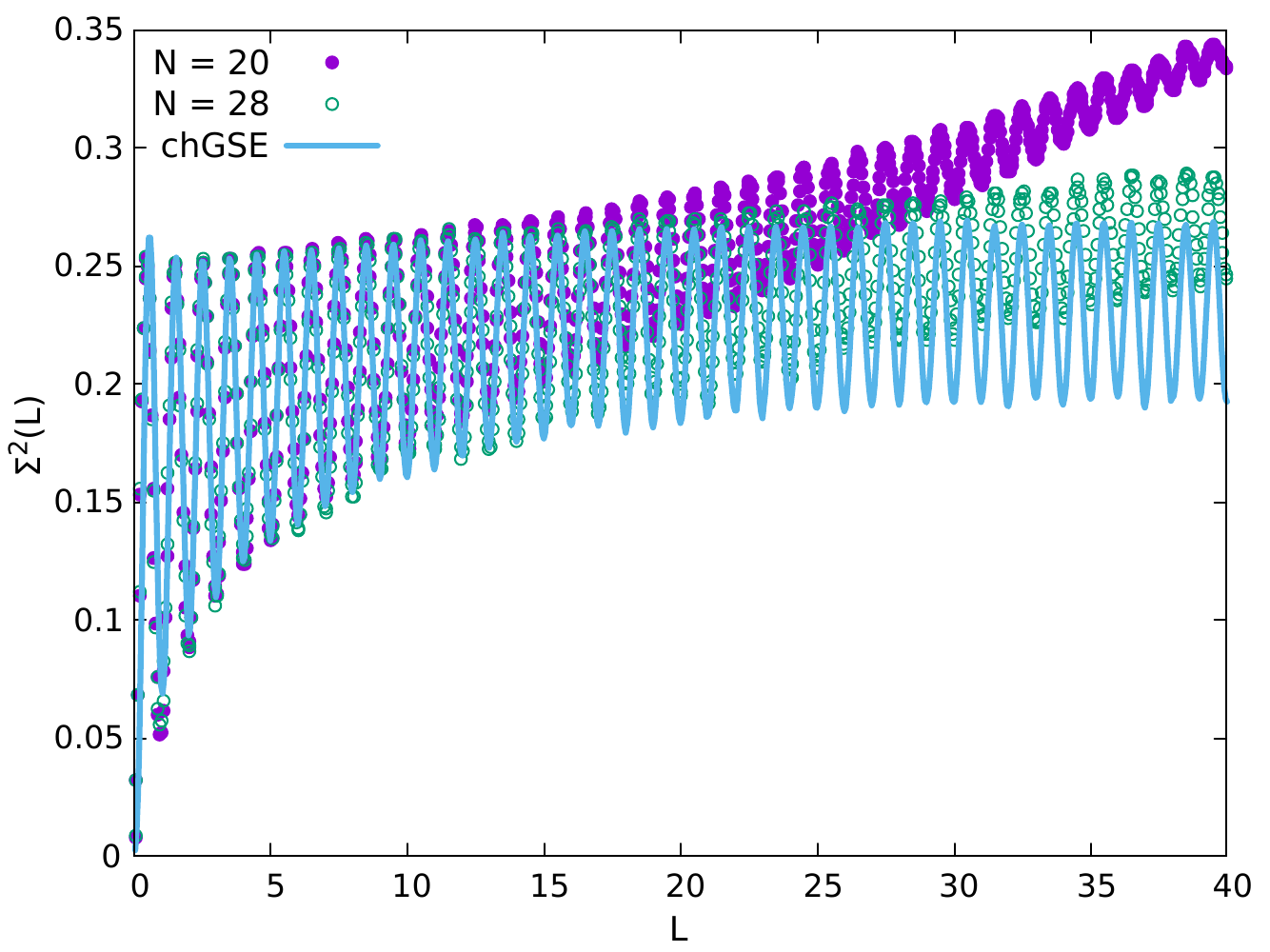}
	\end{minipage}
	\begin{minipage}{0.5\textwidth}
		\includegraphics[width=\textwidth]{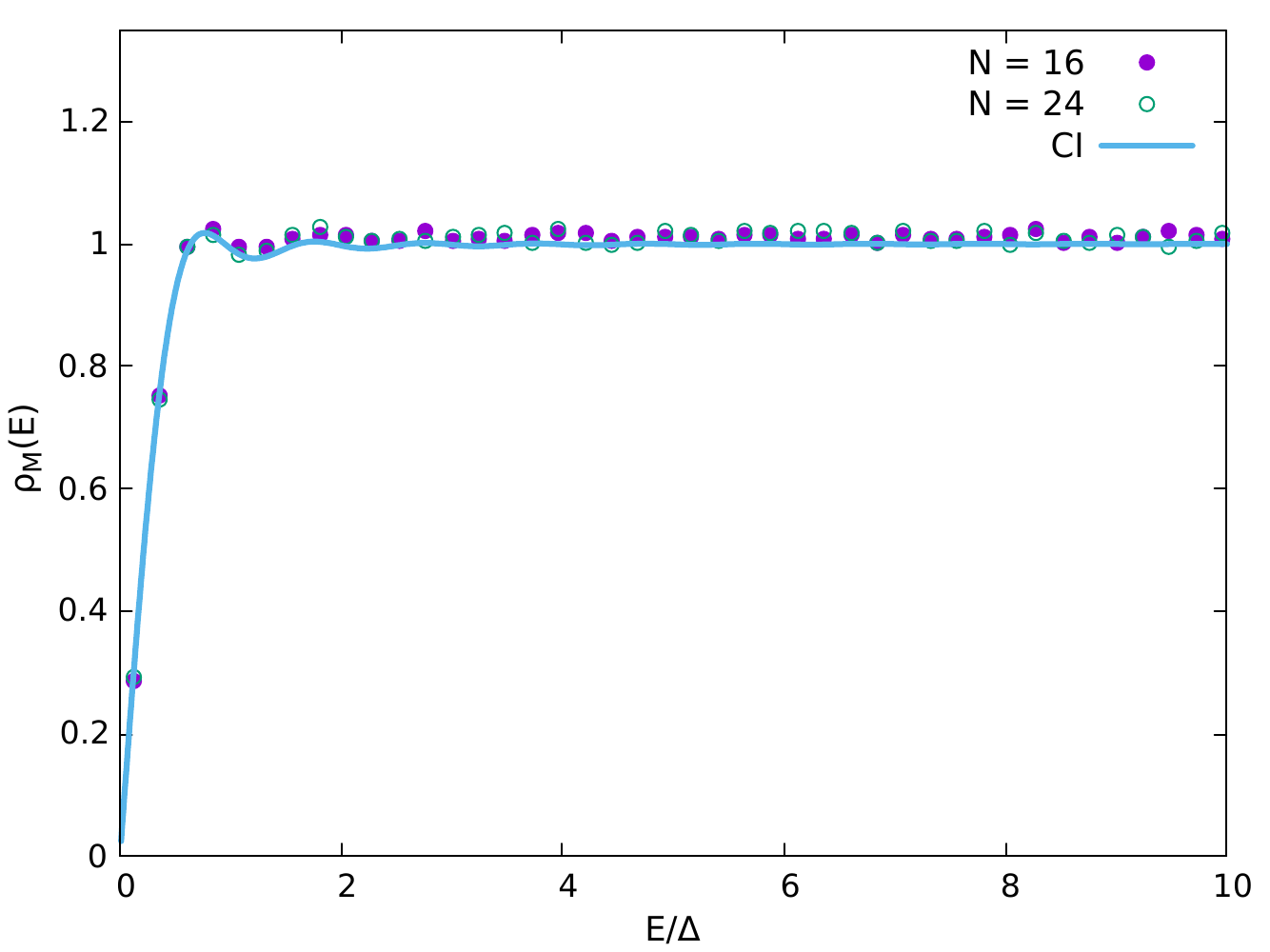} 
	\end{minipage}%
	\begin{minipage}{0.5\textwidth}
		\includegraphics[width=\textwidth]{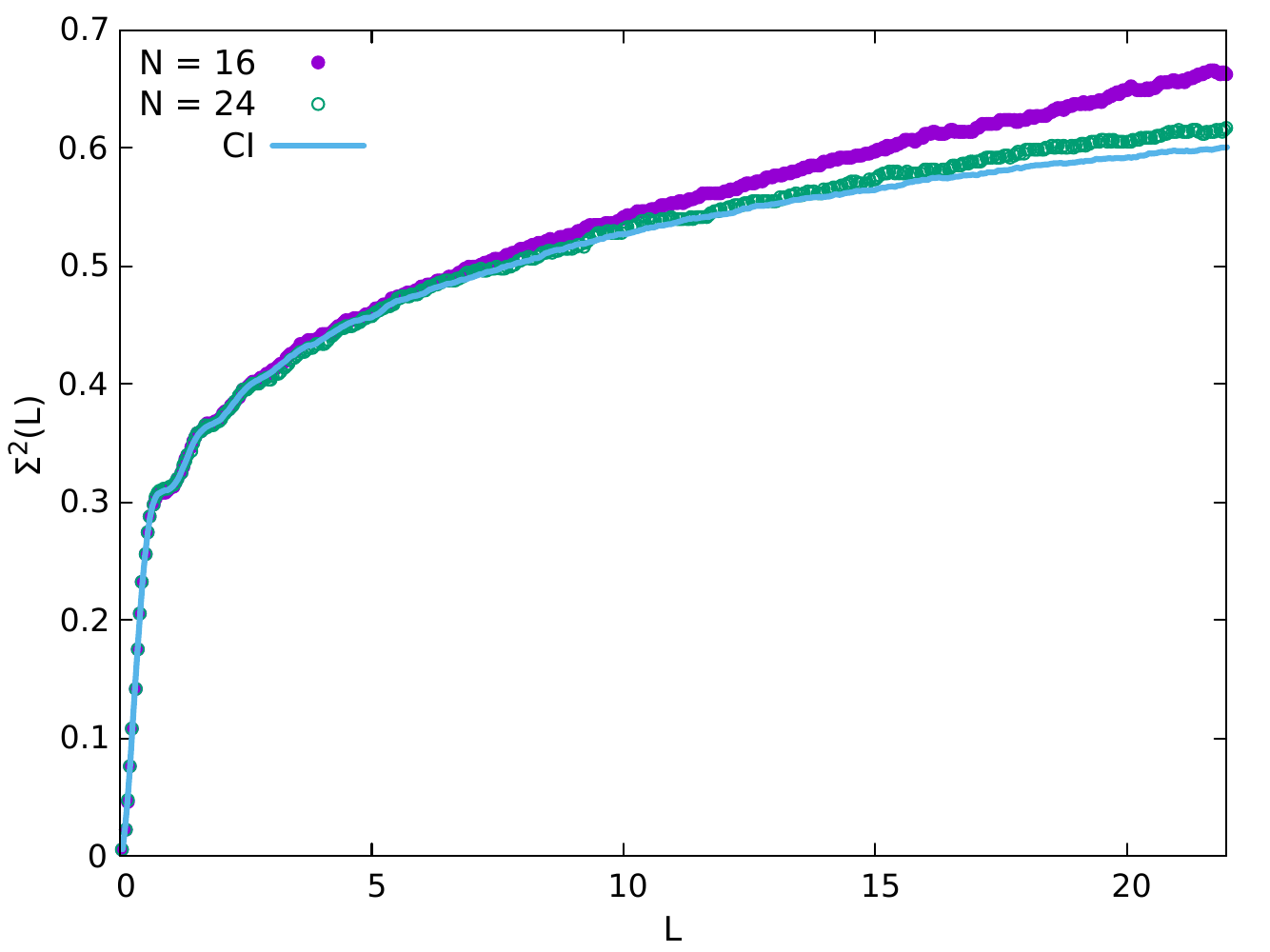} 
	\end{minipage}
	\caption{(Left) The microscopic spectral density corresponding to the first few eigenvalues closest to $E = 0$ for different $N$ and ${\hat q} = 4$. We observe a good agreement with the predictions of random matrix theory for the different universality classes. In the context of random matrix theory, this observable is universal~\cite{verbaarschot1998}, namely, it does not depend on the details of the Hamiltonian but only on its global symmetries. In order to compute this microscopic spectral density, it was necessary to extract the smooth divergent part of the average spectral density $\propto 1/E^\alpha$, with $\alpha \sim 1/2$ obtained from a single-parameter fit. The universal microscopic density can then be seen in the distribution of $E^\alpha \varrho(E)$ close to $E=0$. We do not understand well the origin of the small deviations from the chGSE prediction for $E/\Delta < 1$.
	(Right)~Number variance for different $N$ and ${\hat q} = 4$. We observe an excellent agreement with the random matrix predictions for the different universality classes even for comparatively large eigenvalues separations $\geq 10$. The range of the agreement increases with $N$. This is an indication that the dynamics is quantum chaotic even for times shorter than the Heisenberg time. }
	\label{fig:microden_nv}
\end{figure}

With these considerations in mind, in Fig.~\ref{fig:microden_nv}~(left), we depict results for the microscopic spectral density for different values of $N$. We diagonalize numerically the WSYK Hamiltonian~(\ref{eq:def_WSYK}) for $k = 1$, $M=1$, and ${\qb}=4$.
As was expected, since the global symmetries of the WSYK model depend on $N$, the results of the microscopic spectral density depend on $N$ as well.

The Bott periodicity~\cite{you2016} observed in the standard SYK model is also applicable here but for the chiral and superconducting random matrix ensembles~\cite{verbaarschot1994,verbaarschot1993a,shuryak1993a}. For instance, for $N {\rm mod}\, 8 = 0$, time-reversal symmetry is preserved and the result is close to the CI universality class in the tenfold-way classification~\cite{altland1997}. Likewise, time-reversal symmetry is broken for $N {\rm mod}\, 8 = 2, 6$ so it belongs to the chiral Gaussian Unitary Ensemble (chGUE) universality class. For $N {\rm mod}\, 8 = 4$, the universality class is that of the chiral Gaussian Symplectic Ensemble (chGSE) corresponding to systems with time-reversal invariance but no rotational invariance because of, for instance, spin-orbit coupling.
We note that, unlike the standard supersymmetric SYK model~\cite{fu2018,kanazawa2017}, where the supercharge equivalent has an odd and fixed number of Majoranas, the WSYK reproduces the chGUE universality class. This is important for any gravity interpretation where, at least for oriented surfaces, time reversal symmetry is broken in the gravity dual~\cite{yang2019}.

Overall, the agreement with the random matrix theory prediction observed in Fig.~\ref{fig:microden_nv} is excellent with very small deviations even for relatively small values of $N$.  
It would be very interesting to obtain an explicit analytical calculation of the microscopic density by combining diagrammatic and combinatorial techniques. We speculate that some $Q$-deformation of the standard random matrix results could reproduce the small deviations from the random matrix predictions as a function of $Q(\qb, N)$. 
 
As a further probe of the symmetries of the model, and the nature of the quantum dynamics, we carry out an analysis of level statistics. We first investigate an ultra-short-range correlator, the adjacent gap ratio, 
\cite{oganesyan2007,luitz2015,bertrand2016} that probes the quantum dynamics for times scales much larger than the Heisenberg time. It is defined as,
\begin{equation}
r_i = \frac{\min(\delta_i, \delta_{i+1})}{\max(\delta_i, \delta_{i+1})},
\label{eq:agr}
\end{equation}
where $\delta_i = E_i - E_{i-1}$ and the spectrum is assumed to be ordered. Its average value is sensitive to the symmetry of the system and the type of quantum dynamics. For an integrable system, $\left\langle r \right\rangle_\mathrm{P} \approx 0.38$, while for a quantum chaotic system, it is given by the random matrix prediction that depends on the symmetry class,
$\langle r \rangle_{\rm RMT}\approx 0.5307,\  0.5996,\  0.674$, for the orthogonal, unitary, and symplectic symmetry classes, respectively~\cite{atas2016}. The advantage of $\langle r \rangle$, or its distribution function, is that it is not necessary to unfold the spectrum, i.e., to rescale it such that the mean level spacing is the same for all energies. Although the main focus of the paper is the region $E \sim 0$ where, as shown for the microscopic spectral density, the impact of the nonstandard symmetry classes is stronger, we compute the average adjacent gap ratio over the full spectrum and for at least $2\times10^4$ disorder realizations for each $N$. 

The results are extremely close to the random matrix predictions: $\langle r \rangle\approx 0.5307,\  0.5995,\  0.673$ for $N = 20, 22, 24$, respectively. We note that these values are not affected by the chiral symmetries that determine the precise random matrix universality class as the average is taken over the whole spectrum and the symmetries only have effect for $E \approx 0$.
We also see that, indeed, the symmetries of the WSYK model for those $N$ are fully consistent with those found in the microscopic spectral density calculation. This excellent agreement with random matrix theory is also a strong indication that dynamics is quantum chaotic for sufficiently long times. 

In order to test the limits of applicability of the random matrix results in the region $E \approx 0$ where global symmetries are important, we investigate the number variance, a long-range correlator that probes the quantum dynamics for shorter times than the adjacent gap ratio:
\begin{equation}
\Sigma^2(L) = \left\langle {\widetilde N}^2(L)\right\rangle - \left\langle {\widetilde N}(L)\right\rangle^2.
\label{eq:nv}
\end{equation}
It is just the variance of the number of levels ${\widetilde N}$ in an energy interval of width $L$ (in units of the mean level spacing). For integrable systems, a linear growth is expected, while a signature of quantum chaos~\cite{bohigas1984} is a slow logarithmic growth for $L \gg 1$.
The results depicted in Fig.~\ref{fig:microden_nv}~(right) show a very good agreement with the random matrix prediction for the different universality classes. Deviations are only observed at relatively large energy separations, measured in units of the mean level spacing, corresponding to times substantially shorter than the Heisenberg time. As $N$ increases, the agreement with the random matrix results persists until even larger energy separations.

The numerical evaluation of the number variance was carried out by exact diagonalization of the Hamiltonian~$W$ for more than $2\times 10^4$ disorder realizations. Such a large number of disorder realizations is necessary as spectral averaging is not possible since the effect of global symmetries is restricted to the region very close to the hard edge at zero energy. For the spectral analysis, the singularity at zero energy was first removed by a smooth transformation. The resulting spectra were unfolded by a low-degree polynomial.

We also carried out a similar spectral analysis for ${\hat q}=2,3$. For ${\hat q} =3$, the results are qualitatively similar to those shown above for ${\hat q} = 4$ though, as expected because the number of Majoranas is smaller, deviations from the random matrix prediction are observed for shorter energy separations. However, for ${\hat q} = 2$, spectral correlations close to the hard edge are qualitatively different which suggests that the dynamics are not quantum chaotic in this region. In the bulk of the spectrum, short range spectral correlations are consistent with those of a quantum chaotic system though large deviations are still observed for larger separations. This is expected as the dynamics of each separate chiral block in the linear case is integrable with spectral correlations given by Poisson statistics. Although the bump observed in the spectral density is not in principle related to spectral correlations, it is intriguing that both unexpected features occur in the same range of parameters.   

Unfortunately, we could not find an analytical expression for either the microscopic spectral density or the number variance. We speculate that a $Q$-deformed random matrix ensemble, like those studied in Refs.~\cite{forrester2020,garcia2001}, has the potential to describe the deviations from the random matrix results but we could not find a suitable calculation scheme based on the combinatorial and diagrammatic techniques employed for the calculation of the spectral density.

\section{Outlook and conclusions}\label{sec:conclusions}
We enumerate a few problems that are a natural continuation of this research. We have not exhausted all the universality classes that can be obtained from the Wishart prescription. More generally, only a relatively small subset of the $38$ universality classes~\cite{bernard2002,kawabata2019} of non-Hermitian Hamiltonians has been studied in the context of SYK models. It would be interesting to find out whether a similar classification applies to non-Hermitian SYK models~\cite{garciasaverbaarschot2021} and, in each case, to identify the combinatorial problem, the relevant Touchard-Riordan expression, and the explicit form of the spectral density. 

A distinctive feature of the WSYK is the existence of a microscopic spectral density close to the hard edge of the spectrum whose features are close to the random matrix prediction. However, especially for smaller $\qb$ and not too large $N$, we expect deviations from this universal result. We could not find them analytically but we believe that the combinatorial techniques are versatile enough to provide a viable approach to this problem. It would be necessary to identify a novel scaling limit in the evaluation of the moments that likely leads to some deformation of the universal random matrix results.

We have not addressed the possible existence and properties of the gravity dual. As was mentioned earlier, for the supersymmetric SYK, it has been shown~\cite{fu2018} that the low-energy effective action is given by a super-Schwarzian, also related to a supersymmetric extension of JT gravity~\cite{forste2017}. Although the WSYK has similarities with the supersymmetric SYK, there are important differences. The charge $L_\mu$ is non-Hermitian so the eigenvalues of the Hamiltonian are not the squares of the charge ones. Also, no exact zero eigenvalues exist, albeit the ground-state tends to zero as $N\to\infty$ and there is a proliferation of low-lying excitations. Moreover, the average spectral density is different from that of the supersymmetric case, although it shares the exponential increase for low-energy excitations typical of a field theory with a gravity dual. It would be interesting to find out the effective Schwarzian for the WSYK, relate it to some flavor of (super-) JT gravity and also to give a gravity interpretation of the universal microscopic spectral density that may be related to features deep in the quantum regime where the discreteness of the spectrum is important.

Other problems that deserve further attention are the study of elliptic WSYK models where the coupling $k$ of the imaginary part is $0 < k < 1$. Combinatorially, it involves a mix of perfect matchings and permutations that may be related to other flavor of $Q$-polynomials. More generally, it would be interesting to investigate whether there is a general relation between the weight of $Q$-polynomials and combinatorial problems related to the moments of random Hamiltonians of strongly interacting systems. 

In conclusion, we have investigated a Wishart extension of the SYK model consisting of the product of two Hermitian conjugate SYK matrices, each with complex random couplings. By using combinatorial and diagrammatic techniques, we have computed analytically the low-order moments of the spectral density and have found striking similarities with those of the weight function of Al-Salam-Chihara $Q$-Laguerre polynomials~\cite{al1976,kasraoui2011AAM}, a type of $Q$-Laguerre polynomials~\cite{moak1981}. Based on these similarities, we have carried out a parameter-free comparison between the spectral density of the WSYK model computed numerically and this $Q$-Laguerre weight function where $Q(\hat q, N)$ has been computed analytically. For $\qb= 3,4$, and sufficiently large $N$, we have found good agreement between numerical and analytical results. 
For $\hat q = 4$, we have also reported that the short-range and long-range spectral correlations of the model are in good agreement with the random matrix prediction even for relatively large eigenvalues separations. This is an indication of quantum chaotic dynamics even for time scales much shorter than the Heisenberg time. Depending on $N$, the universality class corresponds to that of superconducting or chiral random matrix ensembles. In particular, we found chGUE correlations for $\hat{q}=4$ and $N\mod 8=2,6$. This was the only remaining symmetry class of the tenfold-way classification whose level statistics had not been realized in the SYK model.

\acknowledgments
We thank T.\ Prosen, P.\ Ribeiro, and S.\ Wu for fruitful discussions. LS acknowledges support by FCT through Grant No.\ SFRH/BD/147477/2019.
AMGG was supported by the NSFC Grant No.\ 11874259, the National Key R$\&$D Program of China (Project ID: 2019YFA0308603), and a Shanghai talent program.

\appendix
\renewcommand\thesubsection{\Alph{section}.\arabic{subsection}}

\section{Review of the combinatorial approach to the standard SYK model}
\label{app:review_standard_SYK}

In this Appendix, we review the computation of the spectral density of the standard SYK model~(\ref{eq:def_SYK}) using the method of moments, following Refs.~\cite{garcia2016,garcia2017,cotler2016,berkooz2019}.

Because the random variables $J_\va$ are Gaussian, the odd moments of the SYK Hamiltonian vanish, while the even moments, $\av{\Tr H^{2p}}$, are evaluated by Wick contraction, i.e., by summing over all possible pair contractions of the indices $\va$, $\vb$, $\vc$, etc. For instance, the first nontrivial moment (the fourth) can be explicitly evaluated as:
 \begin{equation}\label{eq:H_moment4_example}
 \begin{split}
 \av{\Tr H^4}
 &=\sum_{\va,\vb,\vc,\vd}\av{J_\va J_\vb J_\vc J_\vd}\Tr\(\Gamma_\va\Gamma_\vb\Gamma_\vc\Gamma_\vd\)\\
 &=\av{J^2}^2\sum_{\va,\vb}\left[
 \Tr\(\Gamma_\va\Gamma_\va\Gamma_\vb\Gamma_\vb\)+
 \Tr\(\Gamma_\va\Gamma_\vb\Gamma_\va\Gamma_\vb\)+
 \Tr\(\Gamma_\va\Gamma_\vb\Gamma_\vb\Gamma_\va\)
 \right]\\
 &=2^{N/2}\av{J^2}^2\binom{N}{q}^2\left[
 2+2^{-N/2}\binom{N}{q}^{-2}\sum_{\va,\vb}
 \Tr\(\Gamma_\va\Gamma_\vb\Gamma_\va\Gamma_\vb\)
 \right].
 \end{split}
 \end{equation}
 We can represent graphically the two terms inside square brackets in the last line of Eq.~(\ref{eq:H_moment4_example}) by introducing a diagrammatic notation for the $\Gamma$-matrices and their contractions,
 \begin{equation}
 \begin{tikzpicture}[baseline=($0.75*(a)+0.25*(x)$)]
 \begin{feynman}[inline=(a)]
 \vertex[large, dot] (a) {};
 \vertex[below=0.3cm of a] (x) {\scriptsize{$\va$}};
 \end{feynman}
 \end{tikzpicture}
 =\Gamma_\va
 \qquad \text{and} \qquad
 \begin{tikzpicture}[inner sep=2pt,baseline=(a)]
 \begin{feynman}[inline=(a)]
 \vertex (a) {};
 \vertex[right=0.6cm of a] (b) {};
 \vertex[below=0.205cm of a] (x) {\scriptsize{$\va$}};
 \vertex[below=0.18cm of b] (y) {\scriptsize{$\vb$}};
 \diagram*{(a) --[half left,min distance=0.6cm] (b)};
 \end{feynman}
 \end{tikzpicture}
 =\frac{\av{J_\va J_\vb}}{\av{J^2}}=\delta_{\va,\vb}.
 \end{equation}
 After correctly normalizing $\Tr$ by $2^{-N/2}$ and $\sum_{\va}$ by $\binom{N}{q}^{-1}$, we obviously have (we omit the labels of the dots and edges throughout)
 \begin{equation}
 t_1(q,N) \equiv 
 \begin{tikzpicture}[inner sep=2pt,baseline=(b)]
 \begin{feynman}[inline=(a)]
 \vertex[dot] (a) {};
 \vertex[dot,right=0.4cm of a] (b) {};
 \diagram*{(a) --[half left,min distance=0.4cm] (b)};
 \end{feynman}
 \end{tikzpicture}
 = 2^{-N/2} \binom{N}{q}^{-1} \sum_{\va}
 \Tr\(\Gamma_\va \Gamma_\va\)
 =1,
 \end{equation}
 while the last term in Eq.~(\ref{eq:H_moment4_example}) can be represented and evaluated as~\cite{garcia2016}
 \begin{equation}\label{eq:diagram_t2_perfect_matching}
 t_2(q,N)\equiv
 \begin{tikzpicture}[inner sep=2pt,baseline=(b)]
 \begin{feynman}[inline=(a)]
 \vertex[dot] (a) {};
 \vertex[dot,right=0.4cm of a] (b) {};
 \vertex[dot,right=0.4cm of b] (c) {};
 \vertex[dot,right=0.4cm of c] (d) {};
 \diagram*{(a) --[half left,min distance=0.4cm] (c)};
 \diagram*{(b) --[half left,min distance=0.4cm] (d)};
 \end{feynman}
 \end{tikzpicture}
 =2^{-N/2}\binom{N}{q}^{-2}\sum_{\va,\vb}\Tr\(\Gamma_\va \Gamma_\vb \Gamma_\va \Gamma_\vb\)
 =\binom{N}{q}^{-1}\sum_{s=0}^q 
 (-1)^{q+s}\binom{q}{s}\binom{N-q}{q-s}.
 \end{equation}
 To evaluate $t_2$ we have to commute $\Gamma_\va$ with $\Gamma_\vb$. Let $\Gamma_\va$ and $\Gamma_\vb$ have $s$ $\gamma$-matrices in common, i.e., $s=\abs{\va\cap\vb}$ for fixed $\va$ and $\vb$. We can then express the sum over $\va$ and $\vb$ as a sum over the $q$ indices in $\vb$ and over $s$ by allowing for all possible combinations of indices inside $\va$ and $\vb$: out of the $N$ possible $\gamma$-matrices, we fix the $q$ distinct $\gamma$-matrices in $\Gamma_\vb$ in all $\binom{N}{q}$ possible ways, choose the $s$ indices in $\va$ that coincide with the $q$ indices in $\vb$ in all $\binom{q}{s}$ ways, and allow for the $\binom{N-q}{q-s}$ distinct combinations of the remaining $(q-s)$ $\gamma$-matrices in $\Gamma_\va$ to be any of the $(N-q)$ $\gamma$-matrices still available. The commutation of $\Gamma_\va$ and $\Gamma_\vb$ gives a phase $(-1)^{q+s}$ according to Eq.~(\ref{eq:relations_Gamma}). This procedure yields exactly the factors in Eq.~(\ref{eq:diagram_t2_perfect_matching}).
 
The diagrammatic representation of higher-order moments can also be immediately written down. The $2p$th moment $\av{\tr H^{2p}}$ will have $2p$ dots ordered on a line corresponding to the $2p$ insertions of $\Gamma$-matrices in the trace. We then connect the $2p$ dots by $p$ edges in all possible ways as required by Wick's Theorem. Each diagram thus obtained corresponds to a \emph{perfect matching} $\pi$. The set of all perfect matchings of $2p$ elements is denoted by $\mathcal{M}_{2p}$ and has $(2p-1)!!$ elements. We can write the moments of the SYK model as a sum over perfect matchings $\pi\in\mathcal{M}_{2p}$,
\begin{equation}
\frac{1}{\sigmaH^{2p}}\frac{\av{\Tr H^{2p}}}{\Tr\id}
=\sum_{\pi\in\mathcal{M}_{2p}}t(\pi),
\end{equation}
where $\sigmaH=\sqrt{\av{J^2}\binom{N}{q}}$ is the energy scale of the SYK model and the weight~$t(\pi)$ is the contribution of the trace associated to the diagram of $\pi$.

With this diagrammatic notation, the first few moments of $H$ are explicitly found to be:
 \begingroup
 \allowdisplaybreaks
 \begin{align}
 \av{\Tr H^0}&=2^{N/2},
 \\
 \av{\Tr H^2}&=
 2^{N/2}\av{J^2}\binom{N}{q}
 \hspace{6pt}
 \begin{tikzpicture}[inner sep=2pt,baseline=(b)]
 \begin{feynman}[inline=(a)]
 \vertex[dot] (a) {};
 \vertex[dot,right=0.4cm of a] (b) {};
 \diagram*{(a) --[half left,min distance=0.4cm] (b)};
 \vertex[right=0.2cm of a] (x) {};
 \vertex[below=0.35cm of x] {$1$};
 \end{feynman}
 \end{tikzpicture}
 =2^{N/2}\av{J}^2\binom{N}{q},
 \\
 \begin{split}
 \av{\Tr H^4}&=
 2^{N/2}\(\av{J^2}\binom{N}{q}\)^2 
 \(
 \begin{tikzpicture}[inner sep=2pt,baseline=(b)]
 \begin{feynman}[inline=(a)]
 \vertex[dot] (a) {};
 \vertex[dot,right=0.4cm of a] (b) {};
 \vertex[dot,right=0.4cm of b] (c) {};
 \vertex[dot,right=0.4cm of c] (d) {};
 \diagram*{(a) --[half left,min distance=0.4cm] (b)};
 \diagram*{(c) --[half left,min distance=0.4cm] (d)};
 \vertex[right=0.2cm of b] (x) {};
 \vertex[below=0.35cm of x] {$1$}; 
 \end{feynman}
 \end{tikzpicture}
 +
 \begin{tikzpicture}[inner sep=2pt,baseline=(b)]
 \begin{feynman}[inline=(a)]
 \vertex[dot] (a) {};
 \vertex[dot,right=0.4cm of a] (b) {};
 \vertex[dot,right=0.4cm of b] (c) {};
 \vertex[dot,right=0.4cm of c] (d) {};
 \diagram*{(a) --[half left,min distance=0.4cm] (c)};
 \diagram*{(b) --[half left,min distance=0.4cm] (d)};
 \vertex[right=0.2cm of b] (x) {};
 \vertex[below=0.35cm of x] {$t_2$};
 \end{feynman}
 \end{tikzpicture}
 +
 \begin{tikzpicture}[inner sep=2pt,baseline=(b)]
 \begin{feynman}[inline=(a)]
 \vertex[dot] (a) {};
 \vertex[dot,right=0.4cm of a] (b) {};
 \vertex[dot,right=0.4cm of b] (c) {};
 \vertex[dot,right=0.4cm of c] (d) {};
 \diagram*{(a) --[half left,min distance=0.4cm] (d)};
 \diagram*{(b) --[half left,min distance=0.4cm] (c)};
 \vertex[right=0.2cm of b] (x) {};
 \vertex[below=0.35cm of x] {$1$}; 
 \end{feynman}
 \end{tikzpicture}
 \)
 \\
 &=2^{N/2}\(\av{J^2}\binom{N}{q}\)^2
 \(2+t_2\),
 \end{split}
 \\
 \begin{split}\nonumber
 \av{\Tr H^6}&=
 2^{N/2}\(\av{J^2}\binom{N}{q}\)^3
 \end{split}
 \\
 \begin{split}\nonumber
 \times&\left(
 \begin{tikzpicture}[inner sep=2pt,baseline=(b)]
 \begin{feynman}[inline=(a)]
 \vertex[dot] (a) {};
 \vertex[dot,right=0.4cm of a] (b) {};
 \vertex[dot,right=0.4cm of b] (c) {};
 \vertex[dot,right=0.4cm of c] (d) {};
 \vertex[dot,right=0.4cm of d] (e) {};
 \vertex[dot,right=0.4cm of e] (f) {};
 \diagram*{(a) --[half left,min distance=0.4cm] (b)};
 \diagram*{(c) --[half left,min distance=0.4cm] (d)}; 
 \diagram*{(e) --[half left,min distance=0.4cm] (f)};
 \vertex[right=0.2cm of c] (x) {};
 \vertex[below=0.35cm of x] {1};
 \end{feynman}
 \end{tikzpicture}
 +
 \begin{tikzpicture}[inner sep=2pt,baseline=(b)]
 \begin{feynman}[inline=(a)]
 \vertex[dot] (a) {};
 \vertex[dot,right=0.4cm of a] (b) {};
 \vertex[dot,right=0.4cm of b] (c) {};
 \vertex[dot,right=0.4cm of c] (d) {};
 \vertex[dot,right=0.4cm of d] (e) {};
 \vertex[dot,right=0.4cm of e] (f) {};
 \diagram*{(a) --[half left,min distance=0.4cm] (b)};
 \diagram*{(c) --[half left,min distance=0.4cm] (e)}; 
 \diagram*{(d) --[half left,min distance=0.4cm] (f)};
 \vertex[right=0.2cm of c] (x) {};
 \vertex[below=0.35cm of x] {$t_2$};
 \end{feynman}
 \end{tikzpicture}
 +
 \begin{tikzpicture}[inner sep=2pt,baseline=(b)]
 \begin{feynman}[inline=(a)]
 \vertex[dot] (a) {};
 \vertex[dot,right=0.4cm of a] (b) {};
 \vertex[dot,right=0.4cm of b] (c) {};
 \vertex[dot,right=0.4cm of c] (d) {};
 \vertex[dot,right=0.4cm of d] (e) {};
 \vertex[dot,right=0.4cm of e] (f) {};
 \diagram*{(a) --[half left,min distance=0.4cm] (b)};
 \diagram*{(c) --[half left,min distance=0.4cm] (f)}; 
 \diagram*{(d) --[half left,min distance=0.4cm] (e)};
 \vertex[right=0.2cm of c] (x) {};
 \vertex[below=0.35cm of x] {1};
 \end{feynman}
 \end{tikzpicture}
 +
 \begin{tikzpicture}[inner sep=2pt,baseline=(b)]
 \begin{feynman}[inline=(a)]
 \vertex[dot] (a) {};
 \vertex[dot,right=0.4cm of a] (b) {};
 \vertex[dot,right=0.4cm of b] (c) {};
 \vertex[dot,right=0.4cm of c] (d) {};
 \vertex[dot,right=0.4cm of d] (e) {};
 \vertex[dot,right=0.4cm of e] (f) {};
 \diagram*{(a) --[half left,min distance=0.4cm] (c)};
 \diagram*{(b) --[half left,min distance=0.4cm] (d)}; 
 \diagram*{(e) --[half left,min distance=0.4cm] (f)};
 \vertex[right=0.2cm of c] (x) {};
 \vertex[below=0.35cm of x] {$t_2$};
 \end{feynman}
 \end{tikzpicture}
 +
 \begin{tikzpicture}[inner sep=2pt,baseline=(b)]
 \begin{feynman}[inline=(a)]
 \vertex[dot] (a) {};
 \vertex[dot,right=0.4cm of a] (b) {};
 \vertex[dot,right=0.4cm of b] (c) {};
 \vertex[dot,right=0.4cm of c] (d) {};
 \vertex[dot,right=0.4cm of d] (e) {};
 \vertex[dot,right=0.4cm of e] (f) {};
 \diagram*{(a) --[half left,min distance=0.4cm] (c)};
 \diagram*{(b) --[half left,min distance=0.4cm] (e)}; 
 \diagram*{(d) --[half left,min distance=0.4cm] (f)};
 \vertex[right=0.2cm of c] (x) {};
 \vertex[below=0.35cm of x] {$t_3'$};
 \end{feynman}
 \end{tikzpicture}
 \right.
 \end{split}
 \\
 \begin{split}
 &+
 \begin{tikzpicture}[inner sep=2pt,baseline=(b)]
 \begin{feynman}[inline=(a)]
 \vertex[dot] (a) {};
 \vertex[dot,right=0.4cm of a] (b) {};
 \vertex[dot,right=0.4cm of b] (c) {};
 \vertex[dot,right=0.4cm of c] (d) {};
 \vertex[dot,right=0.4cm of d] (e) {};
 \vertex[dot,right=0.4cm of e] (f) {};
 \diagram*{(a) --[half left,min distance=0.4cm] (c)};
 \diagram*{(b) --[half left,min distance=0.4cm] (f)}; 
 \diagram*{(d) --[half left,min distance=0.4cm] (e)};
 \vertex[right=0.2cm of c] (x) {};
 \vertex[below=0.35cm of x] {$t_2$};
 \end{feynman}
 \end{tikzpicture}
 +
 \begin{tikzpicture}[inner sep=2pt,baseline=(b)]
 \begin{feynman}[inline=(a)]
 \vertex[dot] (a) {};
 \vertex[dot,right=0.4cm of a] (b) {};
 \vertex[dot,right=0.4cm of b] (c) {};
 \vertex[dot,right=0.4cm of c] (d) {};
 \vertex[dot,right=0.4cm of d] (e) {};
 \vertex[dot,right=0.4cm of e] (f) {};
 \diagram*{(a) --[half left,min distance=0.4cm] (d)};
 \diagram*{(b) --[half left,min distance=0.4cm] (c)}; 
 \diagram*{(e) --[half left,min distance=0.4cm] (f)};
 \vertex[right=0.2cm of c] (x) {};
 \vertex[below=0.35cm of x] {$1$};
 \end{feynman}
 \end{tikzpicture}
 +
 \begin{tikzpicture}[inner sep=2pt,baseline=(b)]
 \begin{feynman}[inline=(a)]
 \vertex[dot] (a) {};
 \vertex[dot,right=0.4cm of a] (b) {};
 \vertex[dot,right=0.4cm of b] (c) {};
 \vertex[dot,right=0.4cm of c] (d) {};
 \vertex[dot,right=0.4cm of d] (e) {};
 \vertex[dot,right=0.4cm of e] (f) {};
 \diagram*{(a) --[half left,min distance=0.4cm] (d)};
 \diagram*{(b) --[half left,min distance=0.4cm] (e)}; 
 \diagram*{(c) --[half left,min distance=0.4cm] (f)};
 \vertex[right=0.2cm of c] (x) {};
 \vertex[below=0.35cm of x] {$t_3$};
 \end{feynman}
 \end{tikzpicture}
 +
 \begin{tikzpicture}[inner sep=2pt,baseline=(b)]
 \begin{feynman}[inline=(a)]
 \vertex[dot] (a) {};
 \vertex[dot,right=0.4cm of a] (b) {};
 \vertex[dot,right=0.4cm of b] (c) {};
 \vertex[dot,right=0.4cm of c] (d) {};
 \vertex[dot,right=0.4cm of d] (e) {};
 \vertex[dot,right=0.4cm of e] (f) {};
 \diagram*{(a) --[half left,min distance=0.4cm] (d)};
 \diagram*{(b) --[half left,min distance=0.4cm] (f)}; 
 \diagram*{(c) --[half left,min distance=0.4cm] (e)};
 \vertex[right=0.2cm of c] (x) {};
 \vertex[below=0.35cm of x] {$t_3'$};
 \end{feynman}
 \end{tikzpicture}
 +
 \begin{tikzpicture}[inner sep=2pt,baseline=(b)]
 \begin{feynman}[inline=(a)]
 \vertex[dot] (a) {};
 \vertex[dot,right=0.4cm of a] (b) {};
 \vertex[dot,right=0.4cm of b] (c) {};
 \vertex[dot,right=0.4cm of c] (d) {};
 \vertex[dot,right=0.4cm of d] (e) {};
 \vertex[dot,right=0.4cm of e] (f) {};
 \diagram*{(a) --[half left,min distance=0.4cm] (e)};
 \diagram*{(b) --[half left,min distance=0.4cm] (c)}; 
 \diagram*{(d) --[half left,min distance=0.4cm] (f)};
 \vertex[right=0.2cm of c] (x) {};
 \vertex[below=0.35cm of x] {$t_2$};
 \end{feynman}
 \end{tikzpicture}
 \end{split}
 \\
 \begin{split}\nonumber
 &+
 \left.
 \begin{tikzpicture}[inner sep=2pt,baseline=(b)]
 \begin{feynman}[inline=(a)]
 \vertex[dot] (a) {};
 \vertex[dot,right=0.4cm of a] (b) {};
 \vertex[dot,right=0.4cm of b] (c) {};
 \vertex[dot,right=0.4cm of c] (d) {};
 \vertex[dot,right=0.4cm of d] (e) {};
 \vertex[dot,right=0.4cm of e] (f) {};
 \diagram*{(a) --[half left,min distance=0.4cm] (e)};
 \diagram*{(b) --[half left,min distance=0.4cm] (d)}; 
 \diagram*{(c) --[half left,min distance=0.4cm] (f)};
 \vertex[right=0.2cm of c] (x) {};
 \vertex[below=0.35cm of x] {$t_3'$};
 \end{feynman}
 \end{tikzpicture}
 +
 \begin{tikzpicture}[inner sep=2pt,baseline=(b)]
 \begin{feynman}[inline=(a)]
 \vertex[dot] (a) {};
 \vertex[dot,right=0.4cm of a] (b) {};
 \vertex[dot,right=0.4cm of b] (c) {};
 \vertex[dot,right=0.4cm of c] (d) {};
 \vertex[dot,right=0.4cm of d] (e) {};
 \vertex[dot,right=0.4cm of e] (f) {};
 \diagram*{(a) --[half left,min distance=0.4cm] (e)};
 \diagram*{(b) --[half left,min distance=0.4cm] (f)}; 
 \diagram*{(c) --[half left,min distance=0.4cm] (d)};
 \vertex[right=0.2cm of c] (x) {};
 \vertex[below=0.35cm of x] {$t_2$};
 \end{feynman}
 \end{tikzpicture}
 +
 \begin{tikzpicture}[inner sep=2pt,baseline=(b)]
 \begin{feynman}[inline=(a)]
 \vertex[dot] (a) {};
 \vertex[dot,right=0.4cm of a] (b) {};
 \vertex[dot,right=0.4cm of b] (c) {};
 \vertex[dot,right=0.4cm of c] (d) {};
 \vertex[dot,right=0.4cm of d] (e) {};
 \vertex[dot,right=0.4cm of e] (f) {};
 \diagram*{(a) --[half left,min distance=0.4cm] (f)};
 \diagram*{(b) --[half left,min distance=0.4cm] (c)}; 
 \diagram*{(d) --[half left,min distance=0.4cm] (e)};
 \vertex[right=0.2cm of c] (x) {};
 \vertex[below=0.35cm of x] {$1$};
 \end{feynman}
 \end{tikzpicture}
 +
 \begin{tikzpicture}[inner sep=2pt,baseline=(b)]
 \begin{feynman}[inline=(a)]
 \vertex[dot] (a) {};
 \vertex[dot,right=0.4cm of a] (b) {};
 \vertex[dot,right=0.4cm of b] (c) {};
 \vertex[dot,right=0.4cm of c] (d) {};
 \vertex[dot,right=0.4cm of d] (e) {};
 \vertex[dot,right=0.4cm of e] (f) {};
 \diagram*{(a) --[half left,min distance=0.4cm] (f)};
 \diagram*{(b) --[half left,min distance=0.4cm] (d)}; 
 \diagram*{(c) --[half left,min distance=0.4cm] (e)};
 \vertex[right=0.2cm of c] (x) {};
 \vertex[below=0.35cm of x] {$t_2$};
 \end{feynman}
 \end{tikzpicture}
 +
 \begin{tikzpicture}[inner sep=2pt,baseline=(b)]
 \begin{feynman}[inline=(a)]
 \vertex[dot] (a) {};
 \vertex[dot,right=0.4cm of a] (b) {};
 \vertex[dot,right=0.4cm of b] (c) {};
 \vertex[dot,right=0.4cm of c] (d) {};
 \vertex[dot,right=0.4cm of d] (e) {};
 \vertex[dot,right=0.4cm of e] (f) {};
 \diagram*{(a) --[half left,min distance=0.4cm] (f)};
 \diagram*{(b) --[half left,min distance=0.4cm] (e)}; 
 \diagram*{(c) --[half left,min distance=0.4cm] (d)};
 \vertex[right=0.2cm of c] (x) {};
 \vertex[below=0.35cm of x] {$1$};
 \end{feynman}
 \end{tikzpicture}
 \right)
 \end{split}
 \\
 \begin{split}\nonumber
 &=
 2^{N/2}\(\av{J^2}\binom{N}{q}\)^3
 \(5+6t_2+3t_3'+t_3\),
 \end{split}
 \end{align}
 \endgroup
 where below each diagram we wrote its value. Note that different diagrams can yield the same value because of the cyclic property of the trace. In the evaluation of the sixth moment, two new diagrams arise, $t_3$ and $t_3'$. We can see that the diagram $t_3'$ corresponds to two copies of $t_2$ glued together---i.e., with fixed $\vb$, one can independently commute $\Gamma_\va$ and $\Gamma_\vc$ with $\Gamma_\vb$---and, therefore, we find
 \begin{equation}
 t_3'\equiv
 \begin{tikzpicture}[inner sep=2pt,baseline=(b)]
 \begin{feynman}[inline=(a)]
 \vertex[dot] (a) {};
 \vertex[dot,right=0.4cm of a] (b) {};
 \vertex[dot,right=0.4cm of b] (c) {};
 \vertex[dot,right=0.4cm of c] (d) {};
 \vertex[dot,right=0.4cm of d] (e) {};
 \vertex[dot,right=0.4cm of e] (f) {};
 \diagram*{(a) --[half left,min distance=0.4cm] (c)};
 \diagram*{(b) --[half left,min distance=0.4cm] (e)}; 
 \diagram*{(d) --[half left,min distance=0.4cm] (f)};
 \end{feynman}
 \end{tikzpicture}
 =2^{-N/2}\binom{N}{q}^{-3}\sum_{\va,\vb,\vc} \Tr\(
 \Gamma_\va \Gamma_\vb \Gamma_\va \Gamma_\vc \Gamma_\vb \Gamma_\vc
 \)
 =t_2^2.
 \end{equation}
 On the other hand, the diagram $t_3$ cannot be reduced to a product of independent lower-order diagrams. It is instead given by~\cite{garcia2016,garcia2018}:
 \begin{equation}\label{eq:diagram_t3_perfect_matching}
 \begin{split}
 t_3(q,N)&\equiv
 \begin{tikzpicture}[inner sep=2pt,baseline=(b)]
 \begin{feynman}[inline=(a)]
 \vertex[dot] (a) {};
 \vertex[dot,right=0.4cm of a] (b) {};
 \vertex[dot,right=0.4cm of b] (c) {};
 \vertex[dot,right=0.4cm of c] (d) {};
 \vertex[dot,right=0.4cm of d] (e) {};
 \vertex[dot,right=0.4cm of e] (f) {};
 \diagram*{(a) --[half left,min distance=0.4cm] (d)};
 \diagram*{(b) --[half left,min distance=0.4cm] (e)}; 
 \diagram*{(c) --[half left,min distance=0.4cm] (f)};
 \end{feynman}
 \end{tikzpicture}
 = 2^{-N/2} \binom{N}{q}^{-3} \sum_{\va,\vb,\vc} \Tr\(
 \Gamma_\va \Gamma_\vb \Gamma_\vc \Gamma_\va \Gamma_\vb \Gamma_\vc \)
 \\
 &=\binom{N}{q}^{-2} 
 \sum_{s=0}^q \sum_{r=0}^q \sum_{m=0}^r 
 (-1)^{q+s+m}
 \binom{q}{s}\binom{N-q}{q-s}\binom{s}{r-m}
 \binom{2(q-s)}{m}\binom{N-2q+s}{q-r}.
 \end{split}
 \end{equation}
 The computation of this diagram is as follows. We have to first commute $\Gamma_\va$ with the product $\Gamma_\vb\Gamma_\vc$ and then commute $\Gamma_\vb$ with $\Gamma_\vc$. Let $\Gamma_\vb$ and $\Gamma_\vc$ have $s$ factors in common and $\Gamma_\va$ have $r$ factors in common with the product $\Gamma_\vb\Gamma_\vc$, of which $(r-m)$ are in common with both $\Gamma_\vb$ and $\Gamma_\vc$. The sum over $\va$, $\vb$, and $\vc$ can then be expressed as a sum over the $q$ indices in $\vb$---which yields a factor $\binom{N}{q}$---and over $s$, $r$, and $m$. Now, with all indices in $\vb$ fixed we proceed as before. First, we fix the indices in $\vc$, choosing the $s$ indices in common with $\vb$ out of the $q$ possible values and the remaining $(q-s)$ indices in $\vc$ from the unused $(N-q)$ $\gamma$-matrices. Then, we fix the indices in $\va$, choosing the $(r-m)$ indices in common with both $\vb$ and $\vc$ out of the $s$ indices common to $\vb$ and $\vc$, the $m$ indices in common with only one of $\vb$ or $\vc$ from the $2(q-s)$ indices that $\vb$ and $\vc$ do not share, and the remaining $(q-r)$ indices in $\va$ from the spare $(N-2q+s)$ $\gamma$-matrices.
 Finally, the commutation of $\Gamma_\va$ with $\Gamma_\vb\Gamma_\vc$ gives a factor $(-1)^{2q+m}$, while the commutation of $\Gamma_\vb$ with $\Gamma_\vc$ gives a factor $(-1)^{q+s}$.
 
 In principle, one can compute $t(\pi)$ exactly for every diagram and for all $p$ as done for $t_2$ and $t_3$ in Eqs.~(\ref{eq:diagram_t2_perfect_matching}) and (\ref{eq:diagram_t3_perfect_matching}), respectively. However, the computations quickly become intractable (e.g., there are $7!!=105$ diagrams at the next order involving up to six independent commutations of $\Gamma$-matrices). Alternatively, one can, with certain approximations, give a simple combinatorial interpretation to the weight $t(\pi)$ of each diagram at leading and next-to-leading order in $1/N$. The latter becomes exact when $q\propto\sqrt{N}$~\cite{erdos2014,feng2019}.
 
 To leading order in $1/N$ and fixed $q$, different $\Gamma$-matrices have no common $\gamma$-matrices. We thus ignore the commutations of the $\Gamma$-matrices altogether and replace all the traces by $1$. We then simply count the number of allowed diagrams (i.e., pair contractions) at each order. Now, these are exactly the moments of the normal distribution,
 \begin{equation}
 \frac{1}{\sigmaH^{2p}}\frac{\av{\Tr H^{2p}}}{\Tr\id}
 =(2p-1)!!,
 \end{equation}
 and, therefore, to first-order in $1/N$, the spectral density of the SYK model is Gaussian.
 
 To next-to-leading order---or exactly when $q\propto\sqrt{N}$---we take the commutations of $\Gamma$-matrices into account but ignore their correlations. The number of required commutations in a trace equals the number of crossings in the corresponding perfect-matching diagram. Since the trace with a single permutation was evaluated in Eq.~(\ref{eq:diagram_t2_perfect_matching}) and corresponds to diagram $t_2$, a diagram $\pi$ with $\mathrm{cross}(\pi)$ crossings is approximately given by $t_2^{\mathrm{cross}(\pi)}$.\footnote{The number of crossings of a perfect matching corresponds precisely to the number of crossings of lines in its diagram. This is no longer true for permutation diagrams.}%
 \footnote{The number of perfect matchings in $\mathcal{M}_{2p}$ with $k$ crossings can be explicitly computed for arbitrary $(p,k)$ and is tabulated as sequence A067311 in the Online Encyclopedia of Integer Sequences (OEIS)~\cite{OEIS_A067311}.}
 By summing over all diagrams, the moments are given by
 \begin{equation}\label{eq:moments_combinatorics_SYK}
 \frac{1}{\sigmaH^{2p}}\frac{\av{\Tr H^{2p}}}{\Tr\id}
 =\sum_{\pi\in \sM_{2p}} t_2^{\mathrm{cross}(\pi)}.
 \end{equation}
 The sum on the right-hand side of Eq.~(\ref{eq:moments_combinatorics_SYK}) can be evaluated by the Touchard-Riordan formula~\cite{riordan1975}. Alternatively, it can be recognized as the $2p$th moment of the orthogonality weight of the $Q$-Hermite polynomials with $Q=t_2(q,N)$~\cite{ismail1987}:
\begin{equation}\label{eq:spectral_density_QHermite}
\varrho_\mathrm{QH}(E;Q)
=(Q;Q)_\infty(-Q;Q)_\infty^2\frac{2}{\pi \EH}
\sqrt{1-E^2/\EH^2}
\prod_{k=1}^\infty
\(1-\frac{4E^2/\EH^2}{2+Q^{k}+Q^{-k}}\),
\end{equation}
 supported on $-\EH\leq E\leq \EH$, where $\EH$ is the (dimensionless) ground-state energy of the SYK model given by
 \begin{equation}\label{eq:E0_QHermite}
 \EH=\frac{2}{\sqrt{1-Q}},
 \end{equation}
 and $(a;Q)_\infty=\prod_{k=0}^\infty \(1-aQ^k\)$ is the $Q$-Pochhammer symbol. Note that the spectral density is of the form of the Wigner semicircle distribution times a $Q$-dependant multiplicative correction.

\section{Spectral density of the linear WSYK model with \texorpdfstring{$M=1$}{M=1}}
\label{app:linear_WSYK}
 
In this Appendix, we consider the $M=1$ linear WSYK model, defined in Eq.~(\ref{eq:def_WSYK}), which is equivalent to the $\mathcal{N}=1$ supersymmetric SYK model. Its spectral density can be trivially obtained from the standard SYK model by a change of variables. Indeed, because the couplings are real we have $W=L^2$ and, hence, the eigenvalues of $W$ are the squares of the eigenvalues of $L$. Effecting the change of variables $E\to E^2$ and multiplying by the associated Jacobian $1/\sqrt{E}$, we obtain the following spectral density, which has to be evaluated at $Q=t_2(\qb,N)$ (recall that each $\Gamma$-matrix now only has $\qb=q/2$ $\gamma$-matrices):
\begin{equation}\label{eq:spectral_density_linearWSYK}
\begin{split}
\varrho(E;Q)&=\frac{1}{\sqrt{E}}\varrho_{\mathrm{QH}}(\sqrt{E};Q)\\
&=(Q;Q)_\infty(-Q;Q)_\infty^2\frac{2}{\pi \EL}
\sqrt{\frac{1-E/\EL}{E/\EL}}
\prod_{k=1}^\infty
\(1-\frac{4E/\EL}{2+Q^{k}+Q^{-k}}\),
\end{split}
\end{equation}
supported on $0\leq E\leq \EL$, where the (dimensionless) spectral edge of the WSYK model is
\begin{equation}\label{eq:E0_linear_WSYK}
\EL=\EH^2=\frac{4}{1-Q}.
\end{equation}
As would be expected, the spectral density assumes the form of the single-channel Marchenko-Pastur distribution times the $Q$-Hermite multiplicative correction.

\section{Simple asymptotic formulas for the WSYK spectral density}
\label{app:asymptotics}

In this Appendix, we derive the simple asymptotic formulas presented in the main text for the $Q$-Laguerre density~(\ref{eq:spectral_density_QLaguerre}) in the different regimes, namely, the bulk, $0\ll E\ll \EL$, the hard edge, $E\to0$, and the soft edge, $E\to\EL$, Eqs.~(\ref{eq:asympt_initial})--(\ref{eq:asympt_final}). The same calculation was done for the standard and supersymmetric SYK models in Refs.~\cite{cotler2016,garcia2017} and~\cite{garcia2018a}, respectively. The computation has to be performed separately for positive and negative $Q$. For large enough $N$---the limit we are mostly interested in---$Q=t_3(\qb,N)$ is positive (resp.\ negative) for even (resp.\ odd) $\qb$.

\paragraph*{Positive \texorpdfstring{$Q$}{Q} (even \texorpdfstring{$\qb$}{qhat})}

We start by rewriting Eq.~(\ref{eq:spectral_density_QLaguerre}) as
\begin{equation}
\begin{split}
\log\varrho_{\mathrm{QL}}(E;Q)
&=\log C_Q+\frac{1}{2}\log\(\frac{1-E/\EL}{E/\EL}\)+
\sum_{k=0}^{+\infty}\log\left[
\frac{1-\frac{4\(1-2E/\EL\)^2}{\(Q^{k/2}+Q^{-k/2}\)^2}}{\(1-\frac{2\(1-2E/\EL\)}{Q^k+Q^{-k}}\)^2}
\right]
\\
&=\log C_Q+\frac{1}{2}
\sum_{k=-\infty}^{+\infty}\log\left[
1-\frac{\(1-2E/\EL\)^2}{\cosh^2\(k\log Q/2\)}
\right]
-\sum_{k=-\infty}^{+\infty}\log\left[
{1-\frac{1-2E/\EL}{\cosh\(k\log Q\)}}
\right],
\end{split}
\end{equation}
with
\begin{equation}\label{eq:C_Q} C_Q=\frac{(Q;Q)_\infty^2(-Q;Q)_\infty^2}{(-Q^2;Q^2)_\infty^2}
\frac{2}{\pi \EL}.
\end{equation}

Performing a Poisson resummation, we have
\begin{equation}
\begin{split}
\log\varrho_{\mathrm{QL}}(E;Q)&=
\log C_Q+
\sum_{n=-\infty}^{+\infty}\int_{0}^{\infty}\d x\, \cos\(2\pi n x\)\log\left[
1-\frac{\(1-2E/\EL\)^2}{\cosh^2\(x \log Q /2\)}
\right]\\
&-\sum_{n=-\infty}^{+\infty}\int_{0}^{\infty}\d x\, \cos\(2\pi n x\)\log\left[
1-\frac{1-2E/\EL}{\cosh\(x\log Q\)}
\right].
\end{split}
\end{equation}
Both integrals can be evaluated analytically~\cite{oberhettinger2012}, yielding
\begin{equation}
\begin{split}
&\log\varrho_{\mathrm{QL}}(E;Q)=\log C_Q\\
&-\frac{1}{2}\sum_{n=-\infty}^{+\infty}\frac{
	1-\cosh\left[
	\frac{4\pi n}{\log Q}\arcsin\(1-\frac{2E}{\EL}\)
	\right]-2\cosh\(\frac{\pi^2 n}{\log Q}\)+2\cosh\left[
	\frac{2\pi n}{\log Q}\arccos \(\frac{2E}{\EL}-1\)
	\right]	
}{n\sinh\(\frac{2\pi^2n}{\log Q}\)}.
\end{split}
\end{equation}
We evaluate the $n=0$ term in the sum by taking the limit of the summand as $n\to0$. Then, using the fact that the summand is an even function of $n$, we can rewrite the spectral density as:
\begin{equation}\label{eq:spectral_density_Poisson_resum}
\begin{split}
&\varrho_{\mathrm{QL}}(E;Q)=
C'_Q\exp\left[
\frac{2\arcsin^2\(1-\frac{2E}{\EL}\)-\arccos^2\(\frac{2E}{\EL}-1\)}{\log Q}
\right]\\
&\times\exp\left[-\sum_{n=1}^{\infty}\frac{
	1-\cosh\left[
	\frac{4\pi n}{\log Q}\arcsin\(1-\frac{2E}{E_0}\)
	\right]-2\cosh\(\frac{\pi^2 n}{\log Q}\)+2\cosh\left[
	\frac{2\pi n}{\log Q}\arccos \(\frac{2E}{E_0}-1\)
	\right]	
}{n\sinh\(\frac{2\pi^2n}{\log Q}\)}\right],
\end{split}
\end{equation}
where $C'_Q=C_Q\exp[\pi^2/(4\log Q)]$.

Up to this point, the computation is exact and Eqs.~(\ref{eq:spectral_density_QLaguerre}) and (\ref{eq:spectral_density_Poisson_resum}) are equivalent. For large $N$, we have $Q\to 1^-$ and we can replace the hyperbolic functions by single exponents, i.e., $\cosh(x/\log Q)\approx (1/2)\exp(-\abs{x}/\log Q)$ and $\sinh(x/\log Q)\approx -(1/2)\exp(-\abs{x}/\log Q)$. In this limit, and using the Taylor expansion of $\log(1-x)$, Eq.~(\ref{eq:spectral_density_Poisson_resum}) reads as
\begin{equation}\label{eq:spectral_density_Laguerre_approx}
\begin{split}
&\varrho_{\mathrm{QL}}(E;Q)
\\&\approx C''_Q 
\exp\left[
\frac{2\arcsin^2\(1-\frac{2E}{\EL}\)-\arccos^2\(\frac{2E}{\EL}-1\)}{\log Q}\right]
\frac{
	1-\exp\left[
	\frac{4\pi}{\log Q}
	\arccos\abs{\(1-\frac{2E}{\EL}\)}
	\right]}{\(
	1-\exp\left[
	\frac{2\pi}{\log Q}
	\arccos\(1-\frac{2E}{\EL}\)
	\right]\)^2},
\end{split}
\end{equation}
with
\begin{equation}
C''_Q=
\frac{C'_Q}{\(1+\exp[\pi^2/\log Q]\)^2}
=\frac{C_Q\exp\left[\pi^2/\(4\log Q\)\right]}{\(1+\exp[\pi^2/\log Q]\)^2}.
\end{equation} Equation~(\ref{eq:spectral_density_Laguerre_approx}) approximates the $Q$-Laguerre spectral density~(\ref{eq:spectral_density_QLaguerre}) extremely well, even for relatively small system sizes. For instance, the two expressions are within $0.5\%$ of one another throughout their support for $\qb=2$ and $N=16$, i.e., $Q=t_3(2,16)$, while their relative deviation is below $10^{-10}\%$ for $Q=t_3(2,40)$. 

We can further simplify Eq.~(\ref{eq:spectral_density_Laguerre_approx}), depending on the value of $E/\EL$. For $E$ well inside the bulk, $0\ll E\ll\EL$, we can safely ignore the second term in Eq.~(\ref{eq:spectral_density_Laguerre_approx}) and we obtain the asymptotic formula for the bulk density:
\begin{equation}\label{eq:spectral_density_bulk}
\varrho_{Q>0}^{(\mathrm{bulk})}(E;Q)=C''_Q 
\exp\left[
\frac{2\arcsin^2\(1-\frac{2E}{\EL}\)-\arccos^2\(\frac{2E}{\EL}-1\)}{\log Q}\right].
\end{equation}

Close to the hard edge, $E=0$, we have to expand $\arcsin x\approx \pi/2-\sqrt{2(1-x)}$ around $x=1$ and $\arccos x\approx \pi-\sqrt{2(1+x)}$ around $x=-1$. Inserting these expansions in Eq.~(\ref{eq:spectral_density_Laguerre_approx}), we obtain the asymptotic hard-edge density,
\begin{equation}\label{eq:spectral_density_hard}
\varrho_{Q>0}^{(\mathrm{hard})}(E;Q)=
C''_Q 
\exp\left[-\frac{\pi^2}{2\log Q}\right]
\coth\left[-\frac{2\pi}{\log Q}\sqrt{\frac{E}{\EL}} \right].
\end{equation}

Finally, near the soft edge, $E=\EL$, we have to expand $\arcsin x\approx -\pi/2+\sqrt{2(1+x)}$ around $x=-1$ and $\arccos x\approx \sqrt{2(1-x)}$ around $x=1$. Inserting these expansions in Eq.~(\ref{eq:spectral_density_Laguerre_approx}), we obtain the asymptotic soft-edge density:
\begin{equation}\label{eq:spectral_density_soft}
\varrho_{Q>0}^{(\mathrm{soft})}(E;Q)= 
C''_Q\exp\left[\frac{\pi^2}{2\log Q}\right]
\,2\sinh\left[-\frac{4\pi}{\log Q}\sqrt{1-\frac{E}{\EL}}\right].
\end{equation}

\paragraph*{Negative \texorpdfstring{$Q$}{Q} (odd \texorpdfstring{$\qb$}{qhat})}

We now turn to negative $Q$. The computation proceeds similarly to positive $Q$, but one has to treat separately the product factors with even and odd $k$ in the $Q$-Laguerre spectral density. For negative $Q$, $Q=-\abs{Q}$ and we can rewrite Eq.~(\ref{eq:spectral_density_QLaguerre}) as
\begin{equation}
\begin{split}
\log\varrho_{\mathrm{QL}}(E;&Q)
\\
=\log C_Q
&+\frac{1}{2}\sum_{k=-\infty}^{+\infty}\log\left[
\frac{1-\frac{4\(1-2E/\EL\)^2}{\(\abs{Q}^{k}+\abs{Q}^{-k}\)^2}}{\(1-\frac{2\(1-2E/\EL\)}{\abs{Q}^{2k}+\abs{Q}^{-2k}}\)^2}
\right]
+\frac{1}{2}\sum_{k=-\infty}^{+\infty}\log\left[
\frac{1+\frac{4\(1-2E/\EL\)^2}{\(\abs{Q}^{k-1/2}-\abs{Q}^{-k+1/2}\)^2}}{\(1+\frac{2\(1-2E/\EL\)}{\abs{Q}^{2k-1}+\abs{Q}^{-2k+1}}\)^2}
\right]
\\
=\log C_Q
&+\frac{1}{2}\sum_{k=-\infty}^{+\infty}\log\left[
1-\frac{\(1-2E/\EL\)^2}{\cosh^2\(k\log \abs{Q}\)}
\right]
+\frac{1}{2}\sum_{k=-\infty}^{+\infty}\log\left[
1+\frac{\(1-2E/\EL\)^2}{\sinh^2\((k-1/2)\log \abs{Q}\)}
\right]
\\
&-\sum_{k=-\infty}^{+\infty}\log\left[
{1-\frac{1-2E/\EL}{\cosh\(2k\log \abs{Q}\)}}
\right]
-\sum_{k=-\infty}^{+\infty}\log\left[
{1+\frac{1-2E/\EL}{\cosh\((2k-1)\log \abs{Q}\)}}
\right],
\end{split}
\end{equation}
where $C_Q$ is again given by Eq.~(\ref{eq:C_Q}). As before, we perform a Poisson resummation, obtaining:
\begin{equation}
\begin{split}
\log&\varrho_{\mathrm{QL}}(E;Q)
=\log C_Q
+\sum_{n=-\infty}^{+\infty}\int_0^{\infty}\d x
\cos\(2\pi n x\)
\log\left[
1-\frac{\(1-2E/\EL\)^2}{\cosh^2\(x\log \abs{Q}\)}
\right]
\\
&+\sum_{\substack{n=-\infty\\n\neq 0}}^{+\infty} (-1)^n
\int_0^{\infty}\d x \cos\(2\pi n x\)
\log\left[
1+\frac{\(1-2E/\EL\)^2}{\sinh^2\(x\log \abs{Q}\)}
\right]
+\int_0^{\infty}\d x \log\left[
1+\frac{\(1-2E/\EL\)^2}{\sinh^2\(x\log \abs{Q}\)}
\right]
\\
&-2\sum_{n=-\infty}^{+\infty}\int_0^{\infty}\d x
\cos\(2\pi n x\)\log\left[
{1-\frac{1-2E/\EL}{\cosh\(2x\log \abs{Q}\)}}
\right]
\\
&-2\sum_{n=-\infty}^{+\infty}(-1)^n
\int_0^{\infty}\d x \cos\(2\pi n x\)\log\left[
{1+\frac{1-2E/\EL}{\cosh\(2x\log \abs{Q}\)}}
\right].
\end{split}
\end{equation}
The first, fourth, and fifth integrals were already performed for positive $Q$\footnote{Note that the change of variables $x\to2x$ can be effected by changing $n\to n/2$ in the result of the integration.}, while the second and third can also be evaluated analytically~\cite{garcia2018a}. We also isolated the $n=0$ term from the second integral, as its values cannot be obtained from the general-$n$ result as the limit $n\to0$. Performing the integrations, the spectral density is exactly rewritten as:
\begin{equation}\label{eq:spectral_density_Poisson_resum_negativeQ}
\begin{split}
&\varrho_{\mathrm{QL}}(E;Q)
\\
&=C'_\abs{Q}
\exp\left[
\frac{2\arcsin^2\(1-\frac{2E}{\EL}\)-\frac{1}{2}\arccos^2\(1-\frac{2E}{\EL}\)-\frac{1}{2}\arccos^2\(\frac{2E}{\EL}-1\)-\pi \abs{\arcsin\(1-\frac{2E}{\EL}\)}}{\log \abs{Q}}
\right]
\\
&\times\exp\left[-\sum_{n=1}^{\infty}\frac{
	1-\cosh\left[
	\frac{2\pi n}{\log \abs{Q}} \arcsin\(1-\frac{2E}{E_0}\)
	\right]-2\cosh\(\frac{\pi^2 n}{2\log \abs{Q}}\)+2\cosh\left[
	\frac{\pi n}{\log \abs{Q}}
	\arccos \(\frac{2E}{E_0}-1\)
	\right]	
}{n\sinh\(\frac{\pi^2n}{\log \abs{Q}}\)}\right]
\\
&\times\exp\left[
2\sum_{n=1}^{\infty}(-1)^n\frac{\cosh\(\frac{\pi^2 n}{2\log \abs{Q}}\)-\cosh\left[
	\frac{\pi n}{\log \abs{Q}}
	\arccos \(1-\frac{2E}{E_0}\)
	\right]}{n\sinh\(\frac{\pi^2n}{\log \abs{Q}}\)}
\right]
\\
&\times\exp\left[
\sum_{n=1}^{\infty}\frac{(-1)^n}{n}\(
1-\exp\left\{\frac{2\pi n}{\log \abs{Q}}\abs{\arcsin\(1-\frac{2E}{\EL}\)}
\right\}\)\right],
\end{split}
\end{equation}
with $C'_\abs{Q}=C_Q\exp[\pi^2/(4\log\abs{Q})]$. Next, in the large-$N$ limit, we again replace the hyperbolic functions by single exponents and use the Taylor expansion of $\log(1-x)$ to approximate the $Q$-Laguerre density by
\begin{equation}\label{eq:spectral_density_Laguerre_approx_Qneg}
\begin{split}
\varrho_{\mathrm{QL}}(E;Q)\approx C'_\abs{Q}
&\cosh\left[
\frac{\pi}{\log\abs{Q}}
\abs{\arcsin\(1-\frac{2E}{\EL}\)}
\right]
\\
\times&\exp\left[
\frac{2\arcsin^2\(1-\frac{2E}{\EL}\)-\frac{1}{2}\arccos^2\(1-\frac{2E}{\EL}\)-\frac{1}{2}\arccos^2\(\frac{2E}{\EL}-1\)}{\log \abs{Q}}
\right]
\\
\times&\frac{
1-\exp\left[
\frac{2\pi}{\log\abs{Q}}\arccos\abs{1-\frac{2E}{\EL}}
\right]}{
\(
1-\exp\left[
\frac{\pi}{\log\abs{Q}}\arccos\(1-\frac{2E}{\EL}\)
\right]\)^2
\(
1-\exp\left[
\frac{\pi}{\log\abs{Q}}\arccos\(\frac{2E}{\EL}-1\)
\right]\)^2
}.
\end{split}
\end{equation}
This approximate formula once again describes extremely well the $Q$-Laguerre density. For example, for $\qb=3$ and $N=64$, Eqs.~(\ref{eq:spectral_density_QLaguerre}) and (\ref{eq:spectral_density_Laguerre_approx_Qneg}) are within $0.01\%$ of one another.

Finally, we give the negative-$Q$ asymptotic spectral densities for the bulk, hard edge, and soft edge. As before, the bulk spectral density is well approximated by dropping the last term in Eq.~(\ref{eq:spectral_density_Laguerre_approx_Qneg}):
\begin{equation}
\begin{split}
\varrho_{Q<0}^{\mathrm{(bulk)}}=
C'_\abs{Q}
&\cosh\left[
\frac{\pi}{\log\abs{Q}}
\abs{\arcsin\(1-\frac{2E}{\EL}\)}
\right]
\\
\times&\exp\left[
\frac{2\arcsin^2\(1-\frac{2E}{\EL}\)-\frac{1}{2}\arccos^2\(1-\frac{2E}{\EL}\)-\frac{1}{2}\arccos^2\(\frac{2E}{\EL}-1\)}{\log \abs{Q}}
\right].
\end{split}
\end{equation} 
Expanding the $\arcsin$ and $\arccos$ around $E\approx 0$ and $E\approx \EL$, we obtain the hard- and soft-edge asymptotic densities, respectively:
\begin{align}
\varrho_{Q<0}^{\mathrm{(hard)}}=
C'_\abs{Q}
&\coth\left[
-\frac{\pi}{\log\abs{Q}}\sqrt{\frac{E}{\EL}}
\right]
\exp\left[
-\frac{2\pi}{\log \abs{Q}}\sqrt{\frac{E}{\EL}}
\right]
\cosh\left[
\frac{\pi^2}{2\log\abs{Q}}\(1-\frac{4}{\pi}\sqrt{\frac{E}{\EL}}\)\right],
\\
\begin{split}
\varrho_{Q<0}^{\mathrm{(soft)}}=
C'_\abs{Q}
&\tanh\left[
-\frac{\pi}{\log\abs{Q}}\sqrt{1-\frac{E}{\EL}}
\right]
\exp\left[
-\frac{2\pi}{\log \abs{Q}}\sqrt{1-\frac{E}{\EL}}
\right]
\\
\times
&\cosh\left[
\frac{\pi^2}{2\log\abs{Q}}\(1-\frac{4}{\pi}\sqrt{1-\frac{E}{\EL}}\)\right].
\end{split}
\end{align}

\section{Ansatz for the spectral density of the circular WSYK model with \texorpdfstring{$M>1$}{M>1}}
\label{app:M>1}

In this Appendix, we consider the circular WSYK model with $M>1$. Proceeding as before, we will first give the leading-order spectral density in the strict limit $N\to\infty$ with fixed $\qb$. Contrary to the $M=1$ case, we were not able to find a density that is exact to next-to-leading order, as we did not find the $Q$-orthogonal polynomials whose moments exactly reproduce the moments of our model. Nevertheless, we propose an ansatz in terms of generalized Al-Salam-Chihara $Q$-Laguerre polynomials that describes the numerical results for large $M$ (i.e., scaling with $N$) to very high accuracy.

The leading-order moments (ignoring commutations) are obtained by setting, in Eq.~(\ref{eq:moments_W_sum_permutations}), $t(\sigma)=1$ for all $\sigma\in\mathcal{S}_p$,
\begin{equation}\label{eq:M>1_leading_order}
\frac{1}{\sigmaL^p}\frac{\av{\Tr W^p}}{\Tr \id}
=\sum_{\sigma\in\mathcal{S}_p} M^{\mathrm{cyc}(\sigma)}=\sum_{k=1}^p c(p,k)M^k,
\end{equation}
where we recall that $\mathrm{cyc}(\sigma)$ is the number of cycles in the permutation $\sigma$ (number of closed loops in the respective diagram) and we rewrote the sum in terms of the number of permutations of $p$ elements with $k$ cycles, which is known as the unsigned Stirling number of the first kind, $c(p,k)$.\footnote{The unsigned Stirling numbers of the first kind are tabulated as sequence A130534 in the OEIS~\cite{OEIS_A130534}.} 
Now, it is known that $c(p,k)$ are also the coefficients of the polynomial $M(M+1)(M+2)\cdots(M+p-1)=(M+p-1)!/(M-1)!$, expanded in increasing powers of $M$, i.e., $M(M+1)(M+2)\cdots(M+p-1)=\sum_{k=1}^p c(p,k)M^k$. Thus, we conclude that the traces of $W$ are given by
\begin{equation}
\frac{1}{\sigmaL^p}\frac{\av{\Tr W^p}}{\Tr \id}
=\frac{(M+p-1)!}{(M-1)!}
\end{equation}
and are immediately identified as the moments of a $\chi^2$ distribution with $2M$ degrees of freedom. To leading order in the limit $N\to\infty$ we thus find the spectral density of the $M>1$ circular WSYK model to be given by $\varrho(E)=E^{M-1}\exp(-E)/(M-1)!$.

Let us now address the next-to-leading-order, where we consider only uncorrelated commutations.
In Sec.~\ref{sec:spectral_density_WSYK}, we saw that the relevant orthogonal polynomials for the $M=1$ circular WSYK model are the Al-Salam-Chihara $Q$-Laguerre polynomials. Furthermore, in random matrix theory, the relevant (classical) orthogonal polynomials for the multichannel Wishart-Laguerre ensemble are the generalized Laguerre polynomials $L_n^{(\alpha)}(x)$. It is, therefore, natural to conjecture that the relevant orthogonal polynomials for the circular WSYK model with $M>1$ are the generalized Al-Salam-Chihara $(Q,y)$-Laguerre polynomials $L_{n}^{(\alpha)}(x;Q,y)$, recently introduced in Ref.~\cite{pan2020SLC}. These polynomials are orthonormal with respect to the weight function
\begin{equation}\label{eq:spectral_density_gen_QyLaguerre}
\begin{split}
\varrho_{\mathrm{QL}}^{(\alpha)}(E;Q,y)=
&\frac{(Q;Q)_\infty (Q^{\alpha+1};Q)_\infty}{(-Q^2/y;Q^2)_\infty(-Q^{2(\alpha+1)}y;Q^2)_\infty}
\frac{1-Q}{2\pi E}\sqrt{\(E_+-E\)\(E-E_-\)}\\
&\times\prod_{k=1}^\infty 
\frac{
	1-\frac{4v^2(E)}{(1+Q^k)(1+Q^{-k})}}{
	\(1-\frac{2v(E)}{Q^k/\sqrt{y}+Q^{-k}\sqrt{y}}\)
	\(1-\frac{2v(E)}{Q^{k+\alpha}\sqrt{y}+Q^{-k-\alpha}/\sqrt{y}}\)},
\end{split}
\end{equation}
supported on $E_-<E<E_+$, where $-1\leq Q\leq 1$ as before, $y\geq 1$, $\alpha$ is a positive integer, the left and right edges are
\begin{equation}\label{eq:E0_gen_QyLaguerre}
E_\pm=\frac{(\sqrt{y}\pm 1)^2}{1-Q},
\end{equation} 
and $v(E)$ is the recentered and rescaled energy,
\begin{equation}
v(E)=\frac{\bar{E}-E}{\Delta E/2},
\end{equation}
with 
\begin{equation}
\bar{E}=\frac{E_++E_-}{2}=\frac{y+1}{1-Q}
\quad\text{and}\quad
\Delta E=E_+-E_-=\frac{4\sqrt{y}}{1-Q}.
\end{equation}
If we set $y=1$ and $\alpha=0$, we recover the spectral density~(\ref{eq:spectral_density_QLaguerre}) with	 $E_+=\EL$ and $E_-=0$. The spectral density~(\ref{eq:spectral_density_gen_QyLaguerre}) is of the form of the multichannel Marchenko-Pastur distribution times a multiplicative correction. 
The combinatorial interpretation of its $p$th moment, $\mu_p$, was found in Ref.~\cite{pan2020SLC} to be
\begin{equation}
\mu_p=\sum_{\sigma\in\mathcal{S}_p} 
\beta^{\mathrm{rec}(\sigma)}
y^{\mathrm{wex}(\sigma)}
Q^{\mathrm{cross}({\sigma})},
\end{equation}
where $\beta=[\alpha+1]_Q:=(1-Q^{\alpha+1})/(1-Q)$ is the $Q$-analog of the integer $\alpha+1$, $\mathrm{rec}(\sigma)$ is the number of records in the permutation $\sigma$ (in terms of diagrams, the number of edges drawn above the dots with no other edges on top), and $\mathrm{wex}(\sigma)$ is its number of weak excedances (the total number of edges drawn above the dots). The lowest moments explicitly read as
\begin{align}
\label{eq:gen_QyLaguerre_mu1}
\mu_1&=\beta y,
\\
\label{eq:gen_QyLaguerre_mu2}
\mu_2&=\beta^2 y^2+\beta y,
\\
\label{eq:gen_QyLaguerre_mu3}
\mu_3&=\beta^3 y^2 +\beta^2 y^2\(2+Q\)+\beta y\(1+y\),
\\
\label{eq:gen_QyLaguerre_mu4}
\mu_4&=\beta^4 y^4 +\beta^3 y^3 \(3+2Q+Q^2\)+
\beta^2y^2\(\(3+2y\)\(1+Q\)+Q^2\)+
\beta y \(1+\(3+Q\)y+y^2\).
\end{align}

It remains to determine $Q$, $y$, and $\alpha$ in terms of the physical parameters $t_3$ and $M$ (under the approximation of uncorrelated commutations). As mentioned before, we were not able to \emph{derive} the correct parameters, but propose an ansatz that describes the numerical results for large $M$ to very high accuracy. We propose that
\begin{equation}\label{eq:ansatz_ybeta}
\alpha=M-1 \Leftrightarrow \beta=[M]_Q
\quad\text{and}\quad
y=\frac{M}{[M]_Q},
\end{equation}
while $Q$ should be the solution of the equation
\begin{equation}\label{eq:ansatz_Q}
Q=\frac{t_3}{[M]_Q}.
\end{equation}
Conversely, we can use Eqs.~(\ref{eq:ansatz_ybeta}) and (\ref{eq:ansatz_Q}) to specify the physical parameters $M$ and $t_3$ as
\begin{equation}\label{eq:ansatz_Mt3}
M=\beta y\quad\text{and}\quad
t_3=\beta Q.
\end{equation}

\begin{figure}[tbp]
	\centering
	\includegraphics[width=0.45\textwidth]{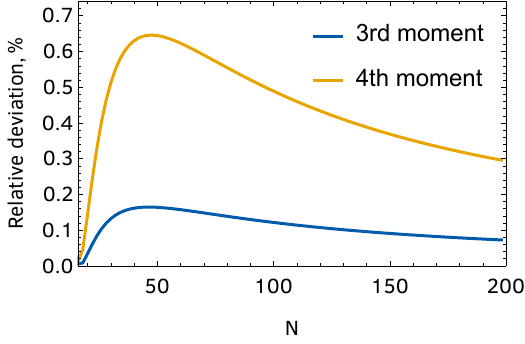}
	\caption{Relative deviation between the exact analytical prediction for the third and fourth moments of $W$, Eqs.~(\ref{eq:W_moment3}) and (\ref{eq:W_moment4}), and the generalized $(Q,y)$-Laguerre ansatz, Eqs.~(\ref{eq:gen_QyLaguerre_mu3}), (\ref{eq:gen_QyLaguerre_mu4}), (\ref{eq:ansatz_ybeta}), and (\ref{eq:ansatz_Q}), as a function of $N$, for $\qb=2$ ($q=4$) and $M=N$. The relative deviation is maximal (yet very small, below $1\%$) for the numerically accessible system sizes and decreases with increasing $N$.}
	\label{fig:MNRelativeDev}
\end{figure}

This ansatz for $(Q,y,\alpha)$ has several desirable features: (i) it reproduces exactly the two lowest moments (with no crossings); (ii) in the random matrix limit ($Q\to0$), it recovers the $M$-channel Marchenko-Pastur distribution, since $y\to M$ and $\beta\to1$; (iii) in the limit $Q\to 1$, we have $\alpha\to M-1$ and $y\to 1$ and, because of the the combinatorial interpretation of the moments in this limit~\cite{hwang2020EJC},
\begin{equation}
\mu_p=\sum_{\sigma\in\mathcal{S}_p}y^{\mathrm{wex}(\sigma)}(\alpha+1)^{\mathrm{cyc}(\sigma)},
\end{equation}
$M$ counts the number of cycles in each permutation, in agreement with our leading-order result, Eq.~(\ref{eq:M>1_leading_order}); and (iv) when $M=1$, we find $Q\to t_3$, $y\to1$, and $\alpha\to0$, such that we recover the results of Sec.~\ref{sec:spectral_density_WSYK}. Note that our ansatz is not the only one with these properties. It is also important to note that the third and fourth moments of the weight of generalized $(Q,y)$-Laguerre polynomials, given by Eqs.~(\ref{eq:gen_QyLaguerre_mu3}) and (\ref{eq:gen_QyLaguerre_mu4}), \emph{do not match exactly} the (normalized) moments of $W$ obtained by approximating $t_4\approx t_3^2$ and inserting the ansatz~(\ref{eq:ansatz_Mt3}) into Eqs.~(\ref{eq:W_moment3}) and (\ref{eq:W_moment4}). However, in the case $M=N$, the two expressions are within around $1\%$ of each other and the relative deviation decreases with increasing $N$, see Fig.~\ref{fig:MNRelativeDev}. This result justifies, \emph{a posteriori}, the validity of the approximations and our ansatz, for large $M$. For smaller $M$, e.g., $M=2$, the relative deviation plateaus at a finite value for large $N$ and the value of this plateau in turn decreases as $M$ increases (for instance, for $M=5$ it is already below $2\%$). It is not clear to us, at this point, the reason why our ansatz is effective only for large $M$. Our results also indicate that large $M$ here means \emph{scaling with $N$}. Indeed, we see a decrease of the relative deviation with $N$ not only for $M=N$ but also, e.g., for $M=\sqrt{N}$. On the other hand, even for fixed $M=60$---which is larger than the $M=N$ for the numerically-accessible $N$---we see the plateau of nonvanishing relative deviation (although at a very small value).

\begin{figure}[tbp]
	\centering
	\includegraphics[width=\textwidth]{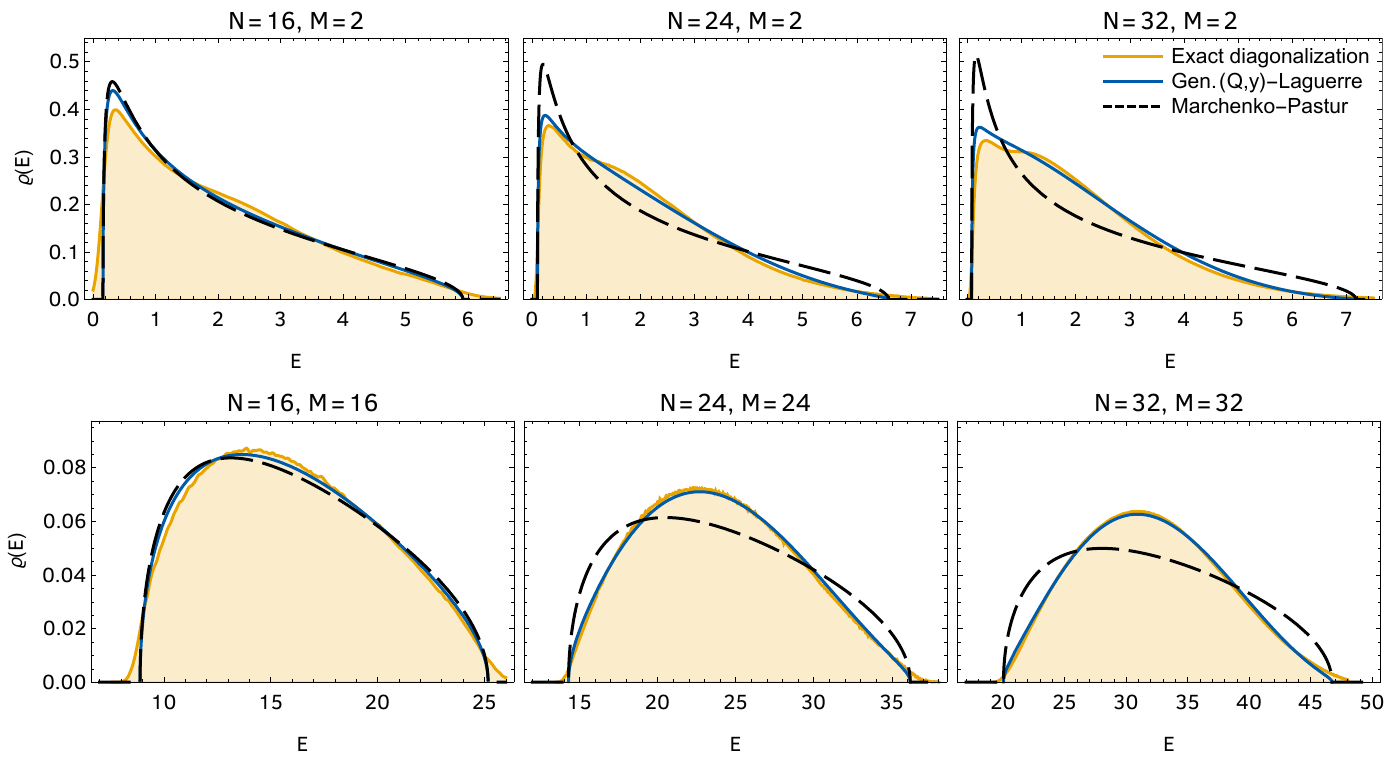}
	\caption{Spectral density of the circular WSYK model with $\qb=2$ ($q=4$) for small, $M=2$, and large, $M=N$, number of charges and three different number of Majoranas, $N=16$, $24$, and $32$. The orange (shaded) histograms are obtained from numerical exact diagonalization of the Hamiltonian~(\ref{eq:def_WSYK}). The blue (full) curves correspond to the generalized Al-Salam-Chihara $(Q,y)$-Laguerre weight [Eq.~(\ref{eq:spectral_density_gen_QyLaguerre})] with parameter ansatz~(\ref{eq:ansatz_ybeta}) and (\ref{eq:ansatz_Q}), while the black (dashed) curve is given by the multichannel Marchenko-Pastur distribution.}
	\label{fig:MNLaguerreDensity}
\end{figure}

\begin{figure}[tbp]
	\centering
	\includegraphics[width=\textwidth]{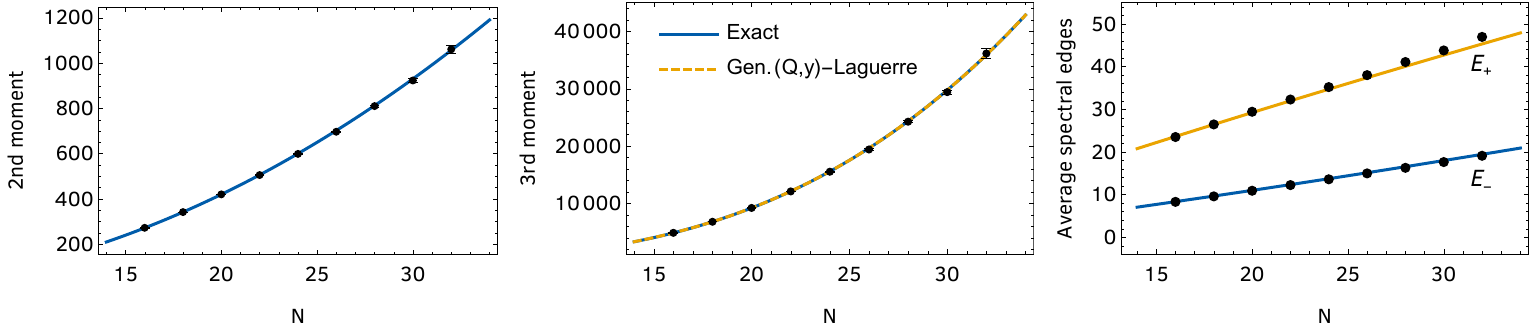}
	\caption{Low-order (normalized) moments and spectral edges of the circular WSYK model with $M=N$ and $\qb=2$ ($q=4$) as a function of the number of Majoranas $N$. The black dots are obtained form ensemble-averaged exact diagonalization ($2^{18}$ eigenvalues). For the moments (left and center panel) the blue line gives the exact analytic expression, Eqs.~(\ref{eq:W_moment2}) and (\ref{eq:W_moment3}), respectively. For the third moment, the generalized Al-Salam-Chihara $(Q,y)$-Laguerre prediction [Eqs.~(\ref{eq:gen_QyLaguerre_mu3}), (\ref{eq:ansatz_ybeta}), and (\ref{eq:ansatz_Q})] is also plotted (dashed orange line) and is seen to perfectly overlap with the exact result (and the numerics). For the spectral edge, the curves are given by the generalized Al-Salam-Chihara $(Q,y)$-Laguerre prediction, Eq.~(\ref{eq:E0_gen_QyLaguerre}).}
	\label{fig:MNMomentsEdge}
\end{figure}

Figure~\ref{fig:MNLaguerreDensity} shows the spectral density of the circular WSYK model with $\qb=2$ ($q=4$) and different $N$, for $M=2$ and $M=N$. As expected from the previous discussion, there is excellent quantitative agreement between the numerical exact diagonalization results and the generalized $(Q,y)$-Laguerre ansatz, for large $M=N$. (It also shows that, once again, the Marchenko-Pastur distribution does not give the correct density). For small and fixed $M=2$, while there are noticeable deviations, our ansatz still captures the main qualitative features of the spectral density (much better than standard random matrix theory, in any case). Finally, to further highlight the very accurate description our ansatz gives of the numerical results for $M=N$, we plot, in Fig.~\ref{fig:MNMomentsEdge}, the second and third moments and the left and right (soft) spectral edges of $W$.

\vfill

\addcontentsline{toc}{section}{References}
\bibliography{10_14lucaswishart2021.bbl}

\end{document}